\documentclass{aa}  

\usepackage{graphicx}
\usepackage{txfonts}
\usepackage{color}
\newcommand{\imin}{i_{\rm min}}
\newcommand{\imax}{i_{\rm max}}

\newcommand{\jmax}{j_{\rm max}}
\newcommand{\xionedim}{\xi_{\rm 1d}}
\newcommand{\lamrf}{\lambda_{\rm RF}}
\newcommand{\lamlya}{\lambda_{\rm Ly\alpha}}

\newcommand{\xicosmo}{\xi^{\rm ph}}
\newcommand{\xilya}{\xi_{\rm Ly\alpha}}
\newcommand{\xismooth}{\xi_{\rm smooth}}
\newcommand{\xipl}{\xi^{\rm pl}}
\newcommand{\xipeak}{\xi_{\rm peak}}
\newcommand{\xif}{\xi}

\newcommand{\xifhat}{\hat{\xi}}

\newcommand{\fqlam}{f_q(\lambda)}
\newcommand{\dqlam}{\delta_q(\lambda)}
\newcommand{\dhat}{\hat\delta}
\newcommand{\dhatqlam}{\hat\delta_q(\lambda)}
\newcommand{\dtildeqlam}{\tilde\delta_q(\lambda)}
\newcommand{\apar}{\alpha_\parallel}
\newcommand{\aperp}{\alpha_\perp}
\newcommand{\hMpc}{h^{-1}{\rm Mpc}}
\newcommand{\lya}{Ly$\alpha$}
\newcommand{\Lya}{Ly$\alpha$~}
\newcommand{\Lyb}{Ly$\beta$~}
\newcommand{\om}{\Omega_M}
\newcommand{\oc}{\Omega_C}
\newcommand{\ob}{\Omega_B}
\newcommand{\on}{\Omega_\nu}
\newcommand{\ol}{\Omega_\Lambda}

\newcommand{\DA}{D_A}
\newcommand{\DM}{D_M}
\newcommand{\DHub}{D_H}
\newcommand{\DV}{D_V}
\newcommand{\rperp}{r_\perp}
\newcommand{\rpar}{r_\parallel}
\newcommand{\rperppseudo}{\tilde{r}_\perp}
\newcommand{\rparpseudo}{\tilde{r}_\parallel}
\newcommand{\lcdm}{$\Lambda$CDM~}

\newcommand{\betalya}{\beta_{\rm Ly\alpha}}
\newcommand{\blya}{b_{\rm Ly\alpha}}
\newcommand{\betaciv}{\beta_{\rm CIV}}
\newcommand{\bciv}{b_{\rm CIV}}
\newcommand{\betaq}{\beta_{\rm q}}
\newcommand{\bqso}{b_{\rm q}}
\newcommand{\betahcds}{\beta_{\rm HCD}}
\newcommand{\bhcds}{b_{\rm HCD}}
\newcommand{\Lhcds}{L_{\rm HCD}}
\newcommand{\Apeak}{A_{\rm peak}}
\newcommand{\NHI}{N_{\rm HI}}
\newcommand{\lamuv}{\lambda_{\rm UV}}

\begin{document}

   \title{Measurement of BAO correlations at $z=2.3$ with SDSS DR12 
\lya-Forests}

   \author{
Julian~E.~Bautista\inst{1},
Nicol\'as~G.~Busca\inst{2}, 
Julien~Guy\inst{3},
James~Rich\inst{4},
Michael~Blomqvist\inst{5},
H\'elion~du~Mas~des~Bourboux\inst{4},
Matthew~M.~Pieri\inst{5},
Andreu~Font-Ribera\inst{6,7},
Stephen~Bailey\inst{6},
Timoth\'ee~Delubac\inst{8}, 
David~Kirkby\inst{9},
Jean-Marc~Le Goff\inst{4},
Daniel~Margala\inst{9},
An\v{z}e~Slosar\inst{10},
Jose Alberto Vazquez\inst{10},
Joel~R.~Brownstein\inst{1},
Kyle~S.~Dawson\inst{1},
Daniel~J.~Eisenstein\inst{11}
Jordi~Miralda-Escud\'{e}\inst{12,13},
Pasquier~Noterdaeme\inst{14},
Nathalie~Palanque-Delabrouille\inst{4},
Isabelle~P\^aris\inst{5},
Patrick~Petitjean\inst{14},
Nicholas~P.~Ross\inst{15}
Donald~P.~Schneider \inst{16,17},
David~H.~Weinberg\inst{18},
Christophe~Y\`eche\inst{4,6}
          }

\institute{
Department of Physics and Astronomy, University of Utah, 115 S 1400 E, Salt Lake City, UT 84112, USA
\and
APC, Universit\'{e} Paris Diderot-Paris 7, CNRS/IN2P3, CEA,
Observatoire de Paris, 10, rue A. Domon \& L. Duquet,  Paris, France
\and
LPNHE,  CNRS/IN2P3,  Universit\'e  Pierre  et  Marie  Curie
Paris 6, Universit\'e Denis Diderot Paris 7, 4 place Jussieu, 75252
Paris CEDEX, France
\and
IRFU, CEA, Universit\'e Paris-Saclay,  F-91191 Gif-sur-Yvette, France
\and
Aix Marseille Universit\'e, CNRS, LAM (Laboratoire d’Astrophysique de Marseille)
 UMR 7326, F-13388, Marseille, France
\and
Lawrence Berkeley National Laboratory, 1 Cyclotron Road, Berkeley, CA 94720, USA
\and
Department of Physics and Astronomy, University College London, Gower Street, London, United Kingdom 
\and
Laboratoire d'Astrophysique, Ecole polytechnique Fed\'erale de Lausanne, CH-1015
 Lausanne, Switzerland
\and
Department of Physics and Astronomy, University of California, Irvine, 
CA 92697, USA
\and
Brookhaven National Laboratory, 2 Center Road,  Upton, NY 11973, USA
\and
Harvard-Smithsonian Center for Astrophysics, 60 Garden St., Cambridge, MA 02138, USA
\and
Instituci\'{o} Catalana de Recerca i Estudis  Avan\c{c}ats, Barcelona, Catalonia
\and
Instituci\'{o} de Ciències del Cosmos, Universitat de Barcelona (UB-IEEC), Catalonia
\and 
Universit\'e Paris 6 et CNRS, Institut d'Astrophysique de Paris, 98bis blvd. Ara
go, 75014 Paris, France
\and
Institute for Astronomy, University of Edinburgh, Royal Observatory, Edinburgh, EH9 3HJ, United Kingdom
\and
Department of Astronomy and Astrophysics, The Pennsylvania State University, University Park, PA 16802
\and
Institute for Gravitation and the Cosmos, The Pennsylvania State University, University Park, PA 16802
\and
Department of Astronomy, Ohio State University, 140 West 18th Avenue, Columbus, 
OH 43210, USA
}

  \abstract
{
  We have used flux-transmission correlations in \Lya forests to measure
  the imprint of 
  baryon acoustic oscillations (BAO).
The study uses spectra of
157,783 quasars in the redshift range $2.1\le z \le 3.5$
from the Sloan Digital Sky Survey (SDSS) Data Release 12 (DR12).
Besides the statistical improvements on our previous studies
using SDSS DR9 and DR11, we have implemented
numerous improvements in the analysis procedure, allowing
us to construct a physical model of the correlation function and to
investigate potential systematic errors in the determination of
the BAO peak position.
The Hubble distance, $\DHub=c/H(z)$, relative to the sound
horizon is 
$\DHub(z=2.33)/r_d=9.07  \pm 0.31$.
The best-determined combination of 
comoving angular-diameter distance, $\DM$, and the Hubble distance
is found to be $\DHub^{0.7}\DM^{0.3}/r_d=13.94\pm0.35$. This value
is $1.028\pm0.026$ times the
prediction of the flat-\lcdm model consistent with the
cosmic microwave background (CMB) anisotropy spectrum.
The errors include marginalization over the effects
of unidentified high-density absorption systems and fluctuations
in ultraviolet ionizing radiation.
Independently of the CMB measurements, the combination of our results
and other BAO observations
determine  the open-\lcdm density parameters to be
$\om=0.296 \pm 0.029$, 
$\ol=0.699 \pm 0.100$ and
$\Omega_k = -0.002 \pm 0.119$.
}


\authorrunning{J. Bautista et al.}
\titlerunning{BAO at $z=2.3$}

   \maketitle
%

\section{Introduction}

The sound waves that propagated in the pre-recombination universe
produced a pronounced peak in the two-point 
correlation function of the cosmological-density field
\citep{1970ApJ...162..815P,1970Ap&SS...7....3S}.
This ``baryon-acoustic oscillation'' (BAO) peak, first
observed by \citet{2005ApJ...633..560E} and \citet{2005MNRAS.362..505C},
is centered on a
comoving distance equal to 
the  sound horizon at the drag epoch, $r_d$.
Observed  at a redshift $z$, the BAO peak position in 
the transverse (angular) direction determines the ratio
$\DM(z)/r_d$, where $\DM(z)=(1+z)\DA(z)$ is the ``comoving''
angular-diameter distance  ($\DA$ is the ``traditional''
angular-diameter distance).
In the radial (redshift) direction, the peak position determines 
$\DHub(z)/r_d$, where $\DHub(z)=c/H(z)$ is the Hubble distance.
Since $\DM$, $\DHub$ and $r_d$ all have simple dependencies on the
cosmological parameters, observation of the BAO feature constrains
those parameters, especially when combined
with cosmic microwave background
(CMB) data \citep{2014A&A...571A..16P,2016A&A...594A..13P,2015PhRvD..92l3516A}.
In particular, one can constrain models beyond the flat-\lcdm model
that describes CMB data, deriving constraints on
cosmological curvature and the dark-energy equation of state.
Furthermore,
the shape of the spectrum of CMB anisotropies can be used to calculate
the value of $r_d$ to percent-level precision,
$r_d\sim147$~Mpc
and the use of this value allows one to 
derive  $\DM(z)$ and $\DHub(z)$ from BAO measurements.
These absolute distances can be 
combined with relative distances determined with type Ia supernovae
\citep{2014A&A...568A..22B},
to extrapolate to $z=0$, yielding a
``top-down'' measurement of $H_0$ \citep{2015PhRvD..92l3516A}.

Most studies of the BAO peak have used 
galaxies at redshifts $z<0.8$ as tracers of the density.
The first observations used $z\sim 0.35$ data from the 
Sloan Digital Sky Survey (SDSS) \citep{2005ApJ...633..560E} 
and $z\sim0.2$ data from the
Two-Degree Field Redshift Survey 
\citep{2005MNRAS.362..505C},
and the combination of
these sets \citep{2007MNRAS.381.1053P,2010MNRAS.401.2148P}.
Since then,
results of increasingly higher precision have
been obtained, most significantly in the redshift range  $0.35<z<0.65$ from
the Baryon Oscillation Spectroscopy Survey (BOSS) of SDSS-III 
\citep{2012MNRAS.427.3435A,2014MNRAS.439...83A,2014MNRAS.441...24A}
with the results of the complete survey being summarized in
\citet{2016arXiv160703155A}. 
Results at other redshifts have been obtained by
6dFGRS at $z\sim0.11$ \citep{2011MNRAS.416.3017B},
WiggleZ at $0.4<z<0.8$ \citep{2011MNRAS.415.2892B}, 
SDSS-I  at $z\sim0.35$ \citep{2012MNRAS.427.2132P,2012MNRAS.427.2168M,2012MNRAS.426..226C,2013MNRAS.431.2834X} and
SDSS-I at
$z\sim0.15$ \citep{2015MNRAS.449..835R}.
There is an impressive agreement of the results of these studies 
with the expectations
of flat-\lcdm models based on CMB data, as emphasized
by \citet{2016A&A...594A..13P}.

At higher redshifts, BAO correlations can be seen using
quasars and their Lyman-$\alpha$ (\lya) forests as mass
tracers \citep{2007PhRvD..76f3009M}.
The correlations in the \lya-forest flux-transmission field 
of  BOSS quasars were first studied in 
\citet{2011JCAP...09..001S} and the BAO peak was seen in 
SDSS data release DR9 
\citep{2013A&A...552A..96B,2013JCAP...04..026S,2013JCAP...03..024K}
and DR11 \citep{2015A&A...574A..59D}.
Cross-correlations of the \lya\ absorption with the distribution of quasars
were detected in DR9 \citep{2013JCAP...05..018F}, and the first BAO detection
was presented in \citet{2014JCAP...05..027F}. 

In this paper, we study the auto-correlation function, $\xif$,
of the \Lya flux-transmission field
using the SDSS data
release DR12 and we update the  cosmological constraints
reported in \citet{2015A&A...574A..59D}.
Our study of the  quasar-forest cross-correlation function
will be presented in a future publication
(du Mas des Bourboux et al., in preparation). 
In addition to using the $15\%$  increase of survey area of DR12 over DR11, 
the following improvements over the analysis of \citet{2015A&A...574A..59D}
have been implemented:

\begin{itemize}
\item Spectroscopic pipeline improvements (Sect. \ref{samplesec})
that includes  a new algorithm for the extraction of spectra from 
the CCD images that results in a more linear flux response.
This modification allows us to correct for 
the mean distortion of the flux-transmission 
field due to imperfect spectral modeling
of standard stars.
We also 
correct for the differential positioning of  quasar and stellar fibers in the
focal plane due to optimization at different wavelengths
\citep{2016ApJ...831..157M}.

\item The use of mock spectra with improved modeling of metal absorbers 
(Sect. \ref{mocksec}),
including both \lya-metal and metal-metal correlations.

\item Modeling of the distortions of the correlation function
due to quasar-continuum fitting
(Sect. \ref{xisec}).  This allows us to construct a physical
model of the correlation function over the range $10<r<200~\hMpc$.

\item Modeling of spurious correlations introduced by the pipeline 
(Sect. \ref{instcorrsec}), calculation of their effect
on the correlation function, and searches for unidentified
spurious correlations  using the Carbon-IV (CIV) forest,
$142.0<\lamrf<152.0$~nm.

\item Fits of the data (Sect. \ref{fitssec}) that
marginalize over the contributions to the correlation function
of metals, unidentified high-column-density systems, 
and fluctuations of ionizing UV flux.

\end{itemize}
None of these improvements induce significant
changes in the derived values of $\DHub/r_d$ and $\DM/r_d$; the differences
(Sect. \ref{cosmosec}) with those of
\citet{2015A&A...574A..59D} are consistent with statistical
fluctuations induced by the increased sample size.

This paper is organized as follows.
Section \ref{samplesec} describes the DR12 data used in this
analysis. Section \ref{mocksec} gives a brief description of the
mock spectra used to test the analysis procedure, with
a more detailed description being found in \citet{2015JCAP...05..060B}.
Section \ref{xisec} presents our method of estimating 
the flux-transmission field, 
its
correlation function,
and the associated
covariance matrix.
Section \ref{instcorrsec} studies spurious correlations induced by the pipeline.
In Section \ref{fitssec} we fit the mocks and data to derive the BAO
peak position parameters, $\DM(z=2.33)/r_d$ and $\DHub(z=2.33)/r_d$.
Section \ref{systsec} investigates possible systematic errors
in the measurement.
In Section \ref{cosmosec} we compare our measured peak position
with the predictions of \lcdm models and derive constraints
on the associated cosmological parameters.
Section \ref{conclusionssec} presents a brief summary.

\begin{figure*}[htbp]
\begin{center}
\includegraphics[width=.90\textwidth]{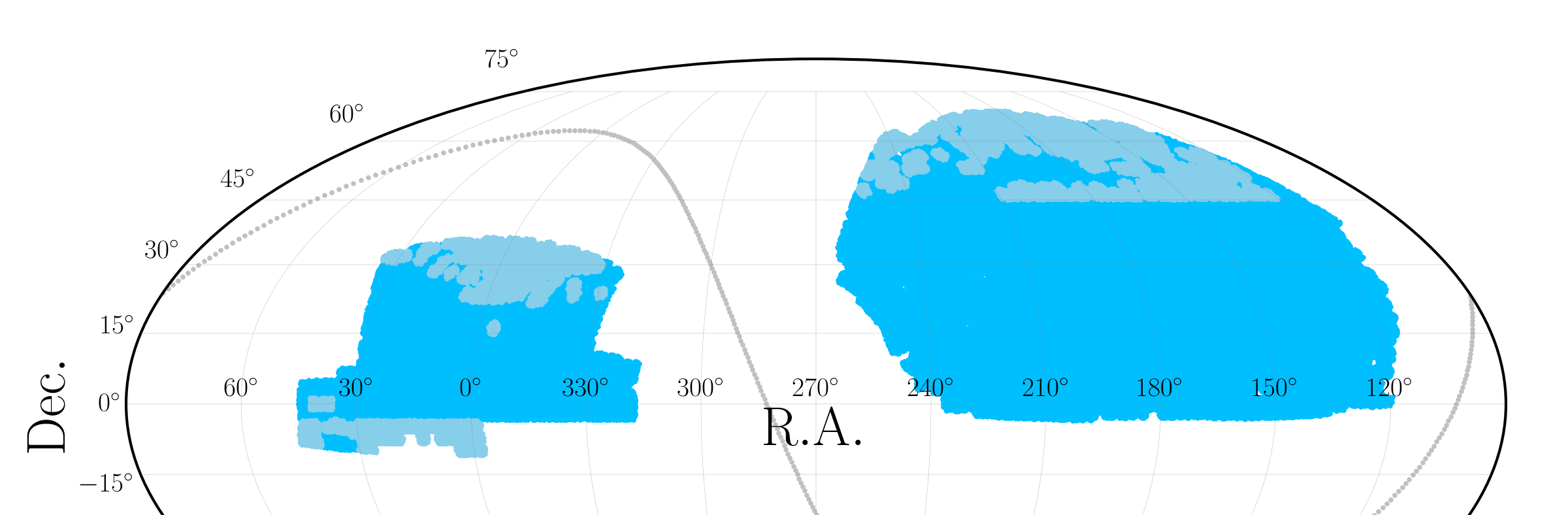}
\caption{SDSS DR12 footprint 
  (in J2000 equatorial coordinates) used in this work.
  The survey covers one quarter of the sky ($10^4{\rm deg^2}$).
  The light blue regions are those added beyond the area covered
  by our previous study \citep{2015A&A...574A..59D}.
  The dotted line is the Galactic plane.
}
\label{fig:skysectors}
\end{center}
\end{figure*}

\begin{figure}
\begin{center}
\includegraphics[width=\columnwidth]{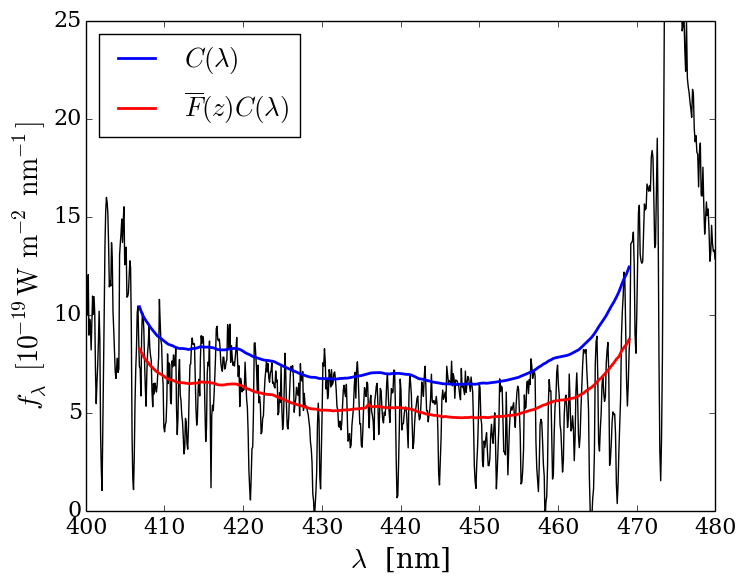}
\caption{
Example of a BOSS quasar spectrum of 
redshift 2.91 (smoothed to the width of analysis pixels).
The red and blue lines cover the
forest region used in our analysis, $104.0<\lamrf<120.0$~nm.
This region is sandwiched between the quasar's \Lyb and \Lya
emission lines, respectively at 
400.9 and 475.4~nm (restframe 102.572 and 
121.567~nm).
The blue line is
the model of the continuum, $C_q(\lambda)$;
the red line is the
product of the continuum and the
mean transmission, $C_q(\lambda)\times\overline{F}(z)$,
as calculated by the method described in Sect. \ref{xisec}. 
}
\label{spectrumfig}
\end{center}
\end{figure}

\begin{figure}
\includegraphics[width=\columnwidth]{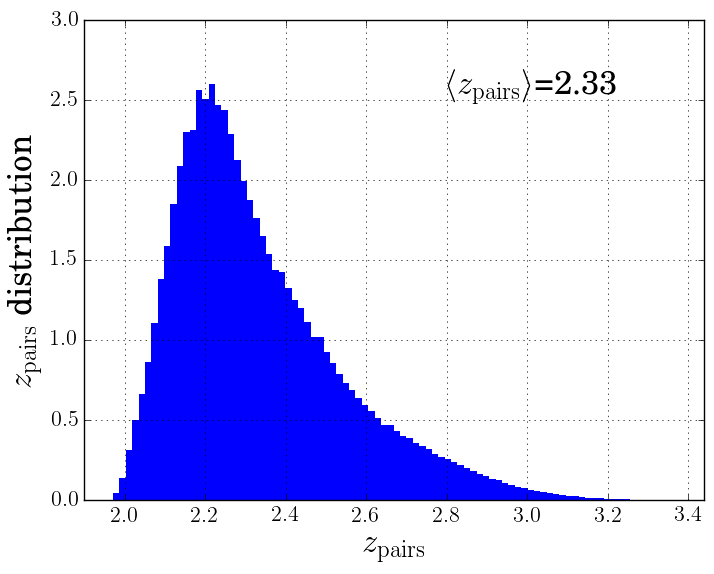}
\caption{Weighted redshift distribution of pairs
  of \lya~forest pixels. The mean is $\langle z\rangle=2.33$.
  Included in the distribution are the $\sim5\times10^{10}$ pairs
  within $20~\hMpc$ of the center of the BAO peak.
}
\label{zdistfig}
\end{figure}

\section{The BOSS quasar sample and data reduction}
\label{samplesec}

The quasar sample  was gathered over a five-year period
by the SDSS-III Collaboration
\citep{2011AJ....142...72E, 
1998AJ....116.3040G, 
2006AJ....131.2332G, 
2013AJ....146...32S
}.
We use the data from  the twelfth Data Release (DR12) of SDSS
as presented in 
\citet{2015ApJS..219...12A}.
The associated quasar catalog is described in \citet{2016arXiv160806483P}.
Most of the quasar spectra were obtained by
the Baryon Oscillation Spectroscopic Survey, BOSS \citep{2013AJ....145...10D},
but DR12 also includes six months of data from the SEQUELS 
\footnote{http://www.sdss.org/dr12/algorithms/ancillary/boss/sequels/} program.
The DR12 celestial footprint covering
$\sim\pi~\rm{sr}\sim10^4~{\rm deg^2}$
is displayed in 
Fig.~\ref{fig:skysectors}.
An example of a quasar spectrum in the \lya-forest region is shown
in Fig.~\ref{spectrumfig}.
The redshift distribution of measurement pairs in the forest is shown
in Fig.~\ref{zdistfig}.

The quasar target selection used in BOSS, summarized in
\citet{2012ApJS..199....3R},  combines different targeting methods described
in \citet{2010A&A...523A..14Y}, \citet{2011ApJ...743..125K}, and
\citet{2011ApJ...729..141B}.
The selection algorithms use SDSS photometry and, when available,
data from
the GALEX survey \citep{2005ApJ...619L...1M} in the UV;
the UKIDSS survey \citep{2007MNRAS.379.1599L} in the NIR,
and the FIRST survey \citep{1995ApJ...450..559B} in the radio.

The DR12 data were  processed using a new software package that differs
from the standard DR12 SDSS-III pipeline 
\citep{2012AJ....144..144B} 
and which has become the standard
pipeline for SDSS DR13 \citep{2016arXiv160802013S}.
For each object,
both pipelines provide a flux calibrated spectrum, $f(\lambda)$, errors,
and an object classification (galaxy, quasar, star).
A model spectrum is fit to $f(\lambda)$ providing  
a redshift estimate.
For this study,
we use the ``coadded'' spectra constructed from
typically four exposures of 15 minutes resampled at
wavelength pixels of width 
$\Delta\log_{10}\lambda=10^{-4}$ 
($c\Delta\lambda/\lambda\sim{\rm 69~km\,s^{-1}}$).
For the small number of quasars with repeated observations,
the coadded spectra can include exposures widely separated
in time.

An important difference with respect to the DR12 pipeline
is that pixels on the CCD image are combined to give a flux $f(\lambda)$
with pixel-weights determined only by the CCD readout noise.
While this method is sub-optimal because it ignores photo-electron Poisson
noise, compared to the DR12 method
it yields an unbiased estimate of $f(\lambda)$ 
since the weights do not depend
on the observed CCD counts which are needed to estimate Poisson noise.
A more detailed description of the changes to the extraction pipeline 
is given in Appendix~\ref{sec:new_pipeline}.

\begin{figure}
\begin{center}
\includegraphics[width=\columnwidth]{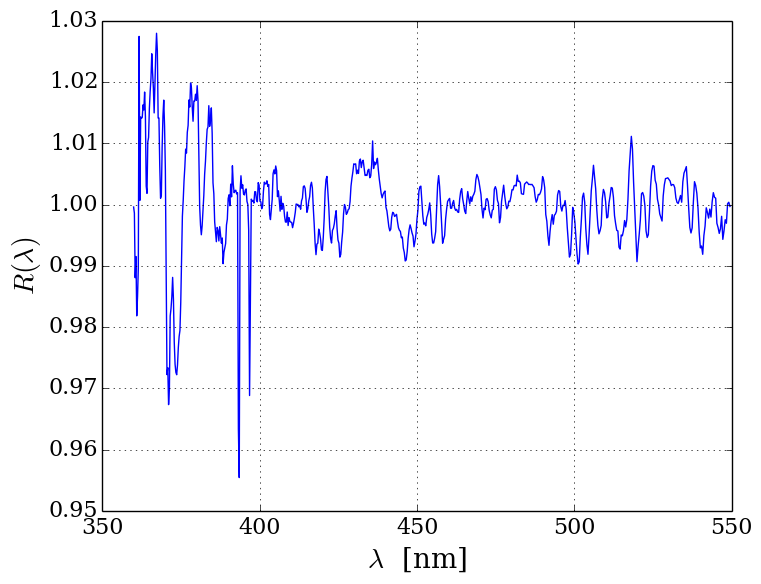}
\caption{
  Mean ratio, $R(\lambda)$, of observed flux to pipeline-model
  flux as a function of 
observed wavelength for quasar spectra to the red of the \Lya emission
line ($\lambda_{\rm RF}>130~{\rm nm}$).
(The mean is calculated by weighting each measurement by the
inverse of the pipeline variance.) 
In this mostly unabsorbed region of quasar spectra, the percent-level 
wavelength-dependent deviations from
unity are due to  imperfect modeling of calibration stars and
to the calcium H and K lines (393.4 and 396.9~nm) due to Galactic absorption.
}
\label{meanfluxfig}
\end{center}
\end{figure}

The ratio of observed flux to model flux, averaged over all spectra,
is an important
diagnostic of pipeline systematic errors.
Figure~\ref{meanfluxfig} shows the average ratio, $R(\lambda)$,
as a function of observed wavelength for quasar spectra
on the red side of \Lya emission.
Since model imperfections for individual quasar spectra are
averaged over in this figure (because of the range of quasar redshifts),
the deviations from unity
reflect imperfections in the spectrograph flux calibration.
The figure reveals
percent-level deviations that are
mostly due to 
imperfect modeling of  photo-spectroscopic standard stars.
The calcium H and K lines from interstellar absorption are also visible.
In the analysis to be presented here,
the $f(\lambda)$ are  divided by $R(\lambda)$ to correct
on average for these artifacts.
This procedure is  effective only with the fluxes from the DR13 pipeline
since the non-linearities present in the DR12 pipeline make
the correction flux-dependent.
We show in Section \ref{instcorrsec} that after this global correction,
remaining calibration artifacts (due to their time-dependence) are
sufficiently small to have a negligible effect on the measurement
of the correlation function.

The spectra of all quasar targets were visually inspected
\citep{2012A&A...548A..66P,2014A&A...563A..54P,2016arXiv160806483P} 
to correct for misidentifications, to flag broad absorption lines (BALs),
and to  establish the definitive quasar redshift.
Damped \Lya troughs (DLAs)
were visually flagged, but also identified and characterized
automatically \citep{2012A&A...547L...1N}. 
The visual inspection of DR12 confirmed
297,301 quasars, of which 
181,719 are in the redshift range appropriate for
this study, 2.1~$\leq$~$z_q$~$\leq$~3.5. 
We discarded
quasars with visually identified BALs  (to avoid the necessity
of modeling their profiles in the forest)
leaving 160,868  quasars. 
A further cut requiring a minimum number of
unmasked forest pixels (50 ``analysis pixels''; see below) 
yielded a sample of 157,922 quasars.
Finally, 139 spectra failed the
continuum-fitting procedure (Sect. \ref{deltaestimationsec}),
leaving
157,783 spectra compared to 
137,562  in the \citet{2015A&A...574A..59D} investigation.

For the measurement of the flux transmission, we use the
rest-frame wavelength interval
\begin{equation}
104.0\,<\lamrf<120.0~{\rm nm} \; .
\label{rflambdarange}
\end{equation}
As illustrated in Fig.~\ref{spectrumfig},
this range is 
bracketed by the  emission lines $\lambda_{\rm Ly\beta}=102.572$~nm and 
$\lamlya=121.567$~nm.  This region
was chosen as the maximum range that avoids the large pixel
variances on the slopes of the two lines due to  quasar-to-quasar
diversity of line-emission strength and profile.
The absorber redshift, 
$z=\lambda/\lamlya-1$,
is required to lie in the range 
$1.96<z<3.44$. The lower limit
is set by the requirement that the observed wavelength be greater
than 360~nm, 
below which the system throughput is less than 10\% of its peak value. 
The upper limit is produced by the maximum quasar
redshift of 3.5, beyond which the BOSS surface density of quasars
is not sufficient to be useful for this study. 
The weighted distribution of redshifts of absorber pairs
near the BAO peak position 
is shown in Fig. \ref{zdistfig}.
The distribution has a mean of $\overline{z}=2.33$.

For the determination of the correlation function, we
use analysis pixels 
that are the inverse-variance-weighted flux average
over three adjacent pipeline pixels.
Throughout the rest of this paper, ``pixel'' refers to analysis
pixels unless otherwise stated.
The width of these pixels is $207~{\rm km~s^{-1}}$
corresponding at $z\sim2.33$ to an 
observed-wavelength width $\sim0.28~{\rm nm}$ and a
comoving radial distance of $\sim2.0~\hMpc$.
The total sample of 157,783
spectra thus provides $\sim 3\times10^7$ measurements
of \Lya absorption
over an effective volume of $\sim 50~{\rm Gpc}^3$.

\begin{figure}[htbp]
\begin{center}
\includegraphics[width=\columnwidth]{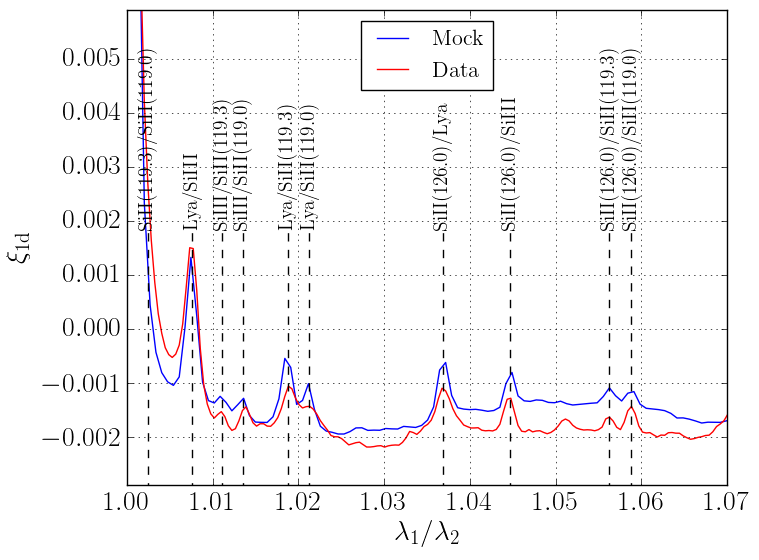}
\caption{
  One-dimensional flux-correlation function, $\xionedim$,
  for BOSS quasars showing
correlations of $\dqlam$ within the same forest.
The correlation function is shown as a function of wavelength
ratio for the data and for the mocks (procedure Met1).
Prominent peaks due to \lya-metal and metal-metal correlations
are indicated.
The peak at $\lambda_1/\lambda_2\sim1.051$, which is seen in the
data but not in the mocks, is due to  CII(133.5)-SiIV(140.277)
  at $z\sim1.85$, outside the redshift range covered by the mocks.
}
\label{fig:xi1d}
\end{center}
\end{figure}
	
\begin{table}
  \centering
\caption{Parameters of the flat \lcdm cosmological model used
for the production and analysis of the mock spectra and
the CMB-based flat \lcdm model from \citet{2016A&A...594A..13P}
used for the analysis of the data.
The models are defined by the cold dark matter, baryon, and massive
neutrinos densities, the Hubble constant, the number of light
neutrino species, and the inhomogeneity parameters, $\sigma_8$ and $n_s$.
The sound horizon, $r_d$, is calculated using
CAMB \citep{2000ApJ...538..473L}.
} 
\label{modtable}
\begin{tabular}{l c c }
 & Mocks & Planck    \\
  &          &  (TT+lowP)    \\
\hline \hline
\noalign{\smallskip}
$\om h^2$  & 0.1323 & 0.1426  \\
$=\oc h^2$   & 0.1096  & 0.1197  \\
$\;+\ob h^2$   & 0.0227 & 0.02222    \\
$\;+\on h^2$    & 0 & 0.0006  \\
$h $    & 0.7 & 0.6731  \\
$N_\nu$  & 3  & 3  \\
$\sigma_8$ & 0.795 & 0.829 \\
$n_s$ & 0.97  & 0.9655 \\
\hline
$\om$    &0.27  & 0.3147  \\
$r_d$ [Mpc]   & 149.7 & 147.33   \\
    & (104.80~$h^{-1}$) & (99.17~$h^{-1}$)   \\
$\DM(2.33)/r_d$   & 38.63 & 39.15  \\
$\DHub(2.33)/r_d$   & 8.744 & 8.612  \\
\end{tabular}
\end{table}

\section{Mock quasar spectra}
\label{mocksec}

In order to test the analysis procedure and
investigate statistical and possible systematic errors, we created 100 sets
of mock spectra that
reproduce the essential
physical and instrumental characteristics of the BOSS spectra.
The basic method for the production of the mocks with \Lya absorption
is described
in \citet{2012JCAP...01..001F}.
Except for the implementation of absorption due to metals and
high column-density systems (HCDs, i.e., damped \lya-systems and Lyman-limit
systems),
the mocks used in this study are identical to the
DR11 mocks \citep{2015JCAP...05..060B}
that were  used in \citet{2015A&A...574A..59D}.
The DR11 mocks therefore cover $\sim15\%$ less solid angle than the DR12 data.

For each set of spectra, the background
quasars were assigned  the angular positions and
redshifts of the DR11 quasars.
The unabsorbed spectra (continua)
of the quasars were generated using the Principal
Component Analysis  eigenspectra of \citet{2005ApJ...618..592S}.
The amplitudes for each eigenspectrum were randomly drawn from Gaussian
distributions with a dispersion equal to that of
the corresponding eigenvalues in
Table 1 of \citet{2006ApJS..163..110S}.
The overall normalization was chosen by fitting the  mock spectrum to the
corresponding observed spectrum.

  The \Lya absorption field was generated 
  using the technique described in \citet{2012JCAP...01..001F}
  in which a
  Gaussian random field, $\delta_G$, is first
  defined at the positions of the $\sim10^7$ observed forest pixels.
  Flux transmissions were defined by the non-linear transformation
  \begin{displaymath}
    F(\delta_G) =
    \exp\left[-a(z)e^{b(z)\delta_G}
      \right].
  \end{displaymath}
  The two functions, a(z) and b(z) are chosen so that the resulting mean
  transmission and  variance are near those observed in the data.
  The correlations of the field $\delta_G$ are chosen so that
  the correlations of $F(\delta_G)$
  follow the linear correlations of the ``mock'' cosmology
  of Table \ref{modtable} modified by non-linear effects \citep{2003ApJ...585...34M}.
  For pixels randomly distributed in space, this procedure would involve
  inverting a $10^7\times10^7$ matrix.  To reduce the problem
  to a manageable size, use was made
  of the fact that the forest pixels are nearly parallel, allowing
  a separate treatment of radial and transverse coordinates.

In the final step, the spectra were modified to include the effects of 
the BOSS spectrograph point spread function (PSF), 
readout noise, photon noise, and flux systematic errors.

For each of the 100 mock data sets,
four types of spectra were produced and analyzed.
The first type consists simply of \Lya flux-transmission
fraction, $F_{\rm Ly\alpha}(\lambda)$, modified for the  wavelength resolution  but
without multiplication by a quasar continuum spectrum, $C(\lambda)$.
Analysis of this
mock type allowed us to study the recovery of the BAO peak position
under the most favorable conditions.
With the introduction of the quasar continuum,
the second type consists of more realistic spectra,
$F_{\rm Ly\alpha}(\lambda)C(\lambda)$.
Analysis of this type tests our ability to fit the quasar
continuum and to model the resulting distortion of the correlation
function.

The final two types of spectra add to $F_{\rm Ly\alpha}(\lambda)$
absorption that is
due to HCDs and to metals.
Following  \citet{2012JCAP...07..028F},
HCDs are placed at randomly chosen
pixels where the  optical depth was above a chosen
threshold. 
The neutral-hydrogen column densities
were drawn randomly with  $\NHI> 10^{17.2}~{\rm cm^{-2}}$,
assuming an intrinsic power-law distribution corrected for
self-shielding and normalized 
to match the observations of
\citet{2005ApJ...635..123P}.
Because the HCDs are placed in redshift space,
the resulting HCDs have correlations that trace
the underlying density field but with a redshift-space distortion
parameter
that is not necessarily equal to that of physical HCDs.

Absorption due to metal transitions
was also simulated by adding absorption 
in proportion to the \Lya absorption at the same redshift.
The important transitions can be seen 
in the ``one dimensional''
correlation function, $\xionedim(\lambda_1/\lambda_2)$ 
for fluxes at two wavelengths within the same forest,
as shown in Fig.~\ref{fig:xi1d}.
In this figure, the peaks correspond to absorption by two different
transitions by material at the same physical position but a different 
wavelength.
The most prominent features are due to pairs comprising \Lya and one
metal transition, but there are also features due to metal-metal pairs.
More detailed information can be obtained by stacking spectra
around strong absorbers in the forest 
\citep{2010ApJ...724L..69P,2014MNRAS.441.1718P}.
Table \ref{metaltable} lists the transitions included in the
mock spectra.

We have tested two procedures for adding metal absorption
to the primary \Lya absorption.
The methods are defined by the assumed relation between
\Lya absorption and metal absorption at the same physical position.
For each wavelength, $\lambda$, with \Lya absorption
$A_{\rm Ly\alpha}(\lambda)=1-F_{\rm Ly\alpha}(\lambda)$,
an appropriate metal absorption $A_m(\lambda+\Delta\lambda_m)$
for each transition, $m$, is chosen.
A simple procedure would be
to choose $A_m\propto A_{\rm Ly\alpha}$ with a proportionality
constant chosen to reproduce the \lya-metal features
in the observed $\xionedim$.
However, this procedure does not produce significant 
features corresponding to metal-metal correlations.
To do this it is necessary to add, for a randomly chosen small fraction of
wavelengths, a much larger metal absorption.

The first procedure (hereafter Met1) 
aims to match the observed line-of-sight correlation function $\xionedim$,
in particular the amplitudes of the peaks associated 
to \lya-metal and metal-metal correlations.
We first transform the pure mock \Lya flux into ``optical depth'' 
$\tau_{\rm Ly\alpha}$ and define the metal absorption of transition $m$ 
as $\tau_m = a_m \tau_{\rm Ly\alpha}$. 
A quadratic term in 
$\tau_{\rm Ly\alpha}$ 
is added for a fraction of the strong \Lya
absorbers to simulate strong metal absorption. 
The parameters $a_m$, the quadratic terms and rates of strong absorption are 
set to match the observed amplitudes in the data $\xionedim$.

The second procedure (hereafter Met2)
aims to match the observed stack of high signal-to-noise ratio \Lya
absorbers, while maintaining consistency with the $\xionedim$
in the resulting mock $\xionedim$.
Following \citet{2014MNRAS.441.1718P},
stacks were produced using the DR12 \Lya forest sample.
As a function of the \Lya transmission,
we measured amplitudes for the metal absorption features.
The obtained “flux decrements” caused by metals were treated as
target average flux decrements. Metals are implemented in
Met2 mocks as a mix of weak and strong metal absorption,
generating a mock $\xionedim$ consistent with the observed
$\xionedim$ and  consistent with the absorption frame measurements of
\citet{2014MNRAS.441.1718P}.

The metal absorption added by these methods is due to the
presence of metals at
the same physical position as HI.
Because our mocks provide the HI density only for redshifts $z>1.9$,
we cannot generate absorption at lower redshifts that
nevertheless
appears in the \Lya forest because of
transition wavelengths much greater than that of \lya.
An important example is
CIV doublet ($\overline\lambda=154.9$~nm) where the absorption at an observed wavelength of 400~nm is
due to material at $z\sim1.6$.
  Fortunately this absorption has little effect on the
  correlation function, as we will see in Sect. \ref{fitssec}.

The statistical properties of metal absorption in the mocks are determined
by the underlying density field. 
However, the analysis
procedure interprets absorption at a given wavelength as 
absorption due to the \Lya transition.
Because of this behavior, the metal contribution to
the measured correlation function is shifted and deformed in $(\rperp,\rpar)$
space. 
In particular, the large correlation due to HI and metals
at the same physical position is seen at  $(\rperp\sim0,\rpar)$
with 
$\rpar=(1+z)\DHub(z)\Delta\lambda/\lambda$ where $\Delta\lambda/\lambda$
is the relative wavelength separation of the metal feature with respect to \lya.
This leads to a large correlation at this $\rpar$ (and $\rperp\sim0$), 
so if the amplitude is significant, it can be measured.
Table \ref{metaltable} lists
the apparent $\rpar$ corresponding to vanishing \lya-metal separation.

\begin{table}
\centering
\caption{
Transitions contributing to absorption by metals in the mock spectra.
The second column shows
$\rpar= (1+\bar{z})\DHub(\bar{z})(\lambda_{met}-\lamlya/)\lambda$,
the reconstructed separation
corresponding to 
correlations of \Lya and metal absorption at the same
physical position. 
} 
\label{metaltable}
\begin{tabular}{l c   }
  transition &  $\rpar$ [$\hMpc$]\\
\hline \hline
\noalign{\smallskip}
SiIII(120.7~nm)  & 21    \\
SiIIa(119.0~nm)  &  64   \\
SiIIb(119.3~nm)  & 56    \\
SiIIc(126.0~nm)  & 111    \\
\end{tabular}
\end{table}%

\begin{figure}[tb]
\begin{center}
\includegraphics[width=\columnwidth]{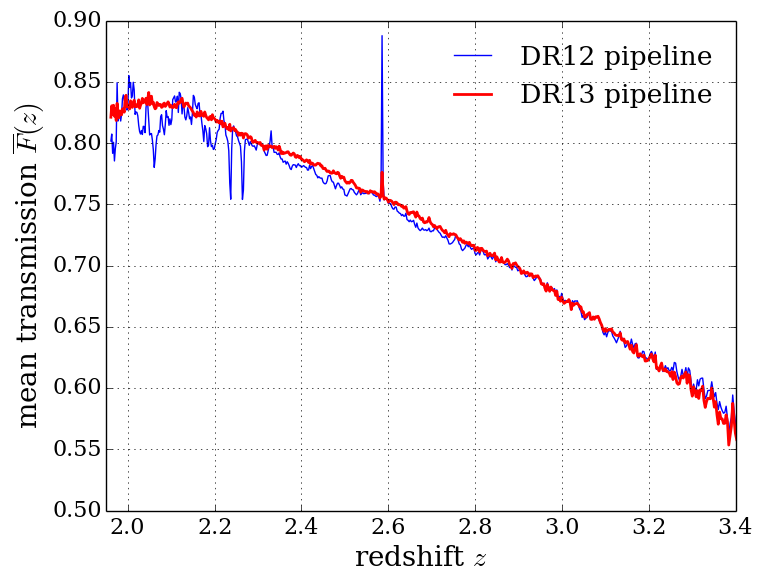}
\caption{
  Mean transmission, $\overline{F}(\lambda)$, calculated
  using the DR12 pipeline
  of our previous investigations, and
  using the DR13 pipeline applied to DR12 data of this analysis.  The DR13 pipeline allows
  us to divide forest spectra by the correction factor
  (Fig. \ref{meanfluxfig})
  derived from the
  spectra on the red side of \Lya emission.
  This procedure eliminates most of the small scale structure
  found with the old pipeline that is due to artifacts of stellar
  modeling, Galactic calcium H and K absorption, and sky lines.
}
\label{meantransfig}
\end{center}
\end{figure}

\section{Measurement of flux-transmission field and its correlation function}
\label{xisec}

In this Section we describe the measurement of the
correlation function of the transmitted flux fraction:
\begin{equation}
\dqlam=\frac{\fqlam}{ C_q(\lambda)\overline{F}(z)}-1 
\;\; .
\label{delta:def}
\end{equation}
Here, $\fqlam$ is the observed flux density for quasar $q$
at observed wavelength $\lambda$,
$C_q(\lambda)$ is the unabsorbed flux density 
(the so-called ``continuum'') and
 $\overline{F}(z)$ is the mean transmitted fraction
 at the absorber redshift, $z(\lambda)=\lambda/\lamlya-1$. 
Measurement of the flux-transmission field $\dqlam$ requires estimates
of the product $C_q(\lambda)\overline{F}(z)$ for each quasar.
An example is shown in Fig.~\ref{spectrumfig}.
The estimation procedures, 
described in this Section, differ slightly from 
those of our previous studies \citep{2013A&A...552A..96B,2015A&A...574A..59D}.
One important modification is that
we now calculate the distortion of the correlation
function due to continuum-fitting
(Sect. \ref{distortionsec}).

\subsection{Estimation of $\dqlam$}
\label{deltaestimationsec}

As in our previous investigations,
we assume the quasar continuum, $C_q(\lambda)$, 
is the product of  a universal
function of the rest-frame wavelength, $\lamrf=\lambda/(1+z_q)$, and
a linear function of $\log\lambda$, included to account for 
quasar spectral diversity:
\begin{equation}
C_q(\lambda) 
= C(\lamrf)
(a_q + b_q\log(\lambda))
\label{templatewarpeq}
\end{equation}
with $C(\lamrf)$ being normalized so that its integral over the forest
is equal to unity.
The $(a_q,b_q)$ and $C(\lamrf)$ are determined by maximizing
the likelihood function given by
\begin{equation}
L = \prod_{q,\lambda} P(\, \fqlam\;|\;C_q(\lambda)\,) \;.
\end{equation} 
Here $P(\fqlam\, | \,C_q(\lambda))$ is the 
probability to observe a flux $\fqlam$
for a given continuum found by convolving the intrinsic probability,
$D(F=\fqlam/C_q(\lambda),z)$,
with the observational resolution assumed to be Gaussian: 
\begin{equation}
 P(\, \fqlam\;|\;C_q(\lambda)\,)
\propto
\int_0^1 dF D(F,z)\exp\left[
\frac{-(C_qF-\fqlam)^2}{2\sigma_q(\lambda)^2}
\right]\;.
\end{equation}
Here, $\sigma_q(\lambda)^2$ is the variance due to readout noise and
photon statistics.

A continuum-determination method is defined by the assumed form of
$D(F,z)$.
For this work, it is taken to be 
the log-normal model of absorption used to generate the mock data,
corresponding to ``method 2'' of our previous studies
\citep{2015A&A...574A..59D}.
As a check, we also use ``method 1'', which is equivalent
to using a narrow Gaussian for  $D(F,z)$, thereby producing
only the product $C_q(\lambda)\overline{F}(\lambda)$ for each forest.

In practice, we maximize the likelihood iteratively by 
assuming a $C(\lamrf)$ to determine the $(a_q,b_q)$.
The mean absorption $\overline{F}(\lambda)$ is then calculated
by an appropriately weighted average of $\fqlam/C_q(\lambda)$ 
(for fixed $\lambda$) after which
$C(\lamrf)$ is recalculated as 
an average of $\fqlam/\overline{F}(z)$
(for fixed $\lamrf$).
The procedure is stopped after ten iterations,
at which point a stability of $\sim10^{-4}$ is reached
for $C(\lamrf)$ and $\overline{F}(\lambda)$.
Figure~\ref{spectrumfig} shows a spectrum with
its $C_q(\lambda)$  
and $C_q\overline{F}$.

The function $\overline{F}(z)$ to be used in 
Eq. \ref{delta:def} to calculate $\dqlam$ is 
the mean of  $\fqlam/C_q(\lambda)$ and can thus differ slightly
from   $\overline{D(F,z)}$ used in the model to estimate the $C_q(\lambda)$.
The use of the mean of $\fqlam/C_q(\lambda)$
ensures that the mean of $\delta$ is zero at each
redshift (or wavelength).

Figure~\ref{meantransfig} displays the  calculated $\overline{F}(z)$,
both for the DR13 pipeline used here and the DR12 pipeline used
previously.
The use of the new pipeline removes most of the artifacts
in the old analysis.
We emphasize, however, that the derived $\overline{F}(z)$
is dependent on the assumed
form of $D(F,z)$ and should therefore
not be considered as a measurement of the mean absorption.
For example,
the flattening of  $\overline{F}(z)$ for $z<2.1$ ($\lambda<377~{\rm nm}$)
suggests that we have slightly underestimated  $\overline{F}(z)$ at
these redshifts.  Since, by construction, the mean $\delta$ vanishes
at each redshift, this implies  
that our procedure makes a compensating  overestimate
of the $C_q(\lambda)$.  Since it is the product $\overline{F}(z)C_q(\lambda)$
that determines $\dqlam$, 
the  measured correlation function is therefore not strongly affected.

Those forests with identified DLAs are given a special treatment.
All pixels where the absorption due to the DLA is higher than 20\% are not used.
The absorption in the wings is corrected  using a 
Voigt profile following
the procedure of \citet{2012A&A...547L...1N}.  

We denote as $\dtildeqlam$ the estimate of $\dqlam$ using
the relation (\ref{delta:def}).
Because forest data is used to fit for $C_q(\lambda)$ and 
$\overline{F}(z)$
the measured $\dtildeqlam$ is not equal to the original $\dqlam$.
We can identify two effects.
First, the use of $\overline{F}(\lambda)$ is equivalent to the
transformation
\begin{equation}
\dtildeqlam = \dqlam
-\overline{\delta(\lambda)}
\label{dhatdef1}
\end{equation}
where the over-bar refers to the average over forests at fixed $\lambda$.
Second,
the fitting of $(a_q,b_q)$ with the data biases toward zero
the  mean $\dtildeqlam$
and mean $(\log\lambda-\overline{\log\lambda} )\dtildeqlam$.
To simplify this effect and facilitate its correction
we explicitly subtract
the mean and first moment for each forest
\begin{equation}
\dhatqlam = \dtildeqlam
- \overline{ \tilde\delta_q} 
- (\Lambda -\overline{\Lambda_q}) 
\frac{\overline{(\Lambda-\overline{\Lambda_q})\tilde\delta}}
{\overline{(\Lambda-\overline{\Lambda_q})^2}}
\hspace*{5mm}
\Lambda\equiv\log\lambda
\label{dhatdef}
\end{equation}
where the over-bars refer to averages within a given forest using
the weights defined in the next Section.
The accompanying distortion of the correlation function is non-negligible,
as we will demonstrate in Sect. \ref{mockanalysissec} with the mock spectra.

The transformation (\ref{dhatdef}) has two interesting
effects on the measured flux-transmission field.
Most importantly, it makes it simple to calculate the
distortion of the correlation function (Sect. \ref{distortionsec}) and
thus simplifies the relation between the underlying physical model
and the measured correlation function.
Second, it nearly eliminates the  difference between
the correlations functions calculated with the 
two continuum fitting methods used in \citet{2015A&A...574A..59D}
with the r.m.s. difference in the two $\xi(\rperp,\rpar)$
being 0.056 of the r.m.s. uncertainty per $(\rperp,\rpar)$ bin.

\subsection{Estimation of the correlation functions}
\label{xiestimationsec}

For the estimator of the  flux auto-correlation function, we adopt 
a simple weighted sum of products of the $\dhat$:
 \begin{equation}
 \xifhat_A=\frac{\sum_{ij\in A}w_iw_j\dhat_i\dhat_j}{\sum_{ij\in A}w_iw_j}
\; ,
\label{xiautoestimator}
 \end{equation}
where the $w_i$ are weights (see below)
and each $i$ (or $j$) indexes
a measurement on a quasar $q_i$ at wavelength $\lambda_i$.
The sum over $(i,j)$ is understood to run over 
all pairs of pixels 
within 
a bin $A$ in the space of pixel separations, 
$(A)\rightarrow(\rperp,\rpar)$.
We exclude pairs of pixels from the same quasar 
to avoid the correlated errors in $\dhat_i$ and $\dhat_j$ 
arising from the estimate of $C_q(\lambda)$
for the spectrum of the  quasar.
The bins $A$ are  defined by a range of width $4~\hMpc$
of the components perpendicular and parallel to the line of sight,
$\rperp$ and $\rpar$.  
We use 50 bins in each component, spanning the range
from $0$ to $200~\hMpc$; the total number of bins used for 
evaluating the correlation function is therefore 2500.
Separations in observational pixel coordinates
(RA,Dec,$\lambda$) are transformed
to $(\rperp,\rpar)$ in units of $\hMpc$ by
assuming that absorption is due to the \Lya transition and 
using the cosmological parameters from Table \ref{modtable}
(Planck cosmology for the data
and the mock cosmology for the mocks).

As described in  \citet{2015A&A...574A..59D},
the weights, $w_i$, are chosen so as to account for both 
Poisson noise  in the flux measurement and for
the intrinsic fluctuations in $\dhat_i$ due to cosmological
large-scale structure.
The weights are set to zero for 
pixels flagged by the pipeline as having problems due, for example 
to sky emission lines or cosmic rays.
To reduce the pipeline systematics discussed in Sect. \ref{instcorrsec},
we also do not use pairs of pixels that have
nearly the same wavelength ($\rpar<4~\hMpc$) and that were
taken on the same exposures.

\subsection{The distortion matrix}
\label{distortionsec}

The transformations (\ref{dhatdef1}) and (\ref{dhatdef}) mix
pixels so that the correlation 
$\langle \dhatqlam \hat\delta_{q^\prime}(\lambda^\prime) \rangle$
is equal to the original 
$\langle \delta_q(\lambda) \delta_{q^\prime}(\lambda^\prime) \rangle$
plus a linear combination of the correlations of
other pixel-pairs
$\langle  \delta_{q^{\prime\prime}}(\lambda^{\prime\prime})
\delta_{q^{\prime\prime\prime}}(\lambda^{\prime\prime\prime}) \rangle$.
This approach means that the measured correlation function $\xifhat$
is a ``distorted'' version of the true correlation function $\xif$.
Since the transformations (\ref{dhatdef1}) and (\ref{dhatdef}) are
linear, the relation between measured and true correlation functions
is given by a distortion matrix $D_{AB}$:
\begin{equation}
  \xifhat_A  = \sum_{B}D_{AB}\xif_{B}
\label{DABfirstuseeq}
\end{equation}
where $A$ and $B$ refer to  bins in pixel separation space.
Writing $\hat\delta_i=\sum_{i^\prime}\eta_{ii^\prime}\delta_{i^\prime}$ produces
\begin{equation}
D_{AB} = w_A^{-1}
\sum_{i,j\in A}w_iw_j 
\left[
\sum_{i^\prime,j^\prime \in B}\eta_{ii^\prime}\eta_{jj^\prime} 
\right]
\label{DABeq}
\end{equation}
where $w_A=\sum_{i,j\in A}w_iw_j$.
We ignore the small effect of transformation  (\ref{dhatdef1}),
in which case 
$\eta_{ii^\prime}=0$ unless the pixels $i$ and $i^\prime$ are in the
same forest:
\begin{equation}
\eta_{ii^\prime}= \delta_{ii^\prime}- \frac{w_{i^\prime}}{\sum_{k}w_k}
-
\frac{(\Lambda_i-\overline{\Lambda})w_{i^\prime}(\Lambda_{i^\prime}-\overline{\Lambda})}
{\sum_{k}w_k(\Lambda_k-\overline{\Lambda})^2}
\hspace*{5mm}
\Lambda=\log\lambda
\end{equation}
where the two sums over $k$ include only pixels in the same forest as
that of $i$ and $i^\prime$ and the over-bars refer to averages in that forest.
The matrix $D_{AB}$ thus depends only on the geometry and weights of the survey.
We will see its effect on the mock correlation function
in Sect. \ref{mockanalysissec} (Fig.~\ref{stackedmocksfig}).

  Previous analyses
  \citep{2013A&A...552A..96B,2013JCAP...04..026S,2015A&A...574A..59D,2015JCAP...11..034B}
have dealt with the distortions introduced by continuum-fitting in
different ways.
\citet{2013A&A...552A..96B} and \citet{2015A&A...574A..59D}
model it as an
additive power law in $r$ and $\mu$ with 12 free parameters.
\citet{2015JCAP...11..034B}
assumed that continuum-fitting reduces the observed
amplitude of long-wavelength modes in the direction parallel to the
line of sight. They then model it as a multiplicative function
of $k_\parallel$ that
tends to zero at large scales (with $k_\parallel\lesssim 2\pi/L$,
where $L\sim380~\hMpc$ is the typical length of a forest)
and to one at small scales ($k_\parallel\gg2\pi/L$).
They tune the shape of the function using simulated data to ultimately
reduce the number of free parameters to one.

In our approach we do not introduce free parameters to account for the
effects of continuum-fitting. Instead, we follow the assumption,
first proposed by \citet{2013JCAP...04..026S}, that at each line of sight
the continuum-fit delta field differs from the true delta field
by a linear function in $\Lambda$.
\citet{2013JCAP...04..026S} then de-weight these “linear modes” in their
covariance matrix.
Alternatively, we use the transformation from $\delta$ to $\hat\delta$
to remove the “linear modes” from the model via the distortion matrix.

\begin{figure*}[t]
\centering
\includegraphics[width=\columnwidth]{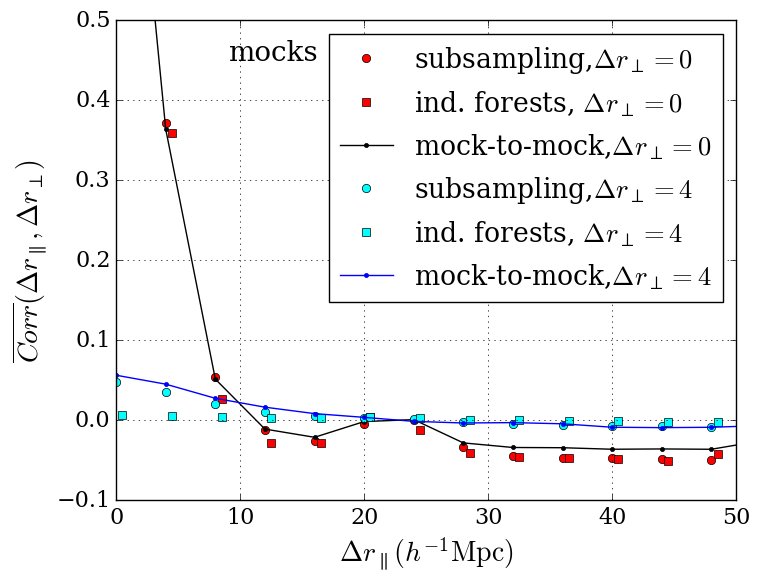}
\includegraphics[width=\columnwidth]{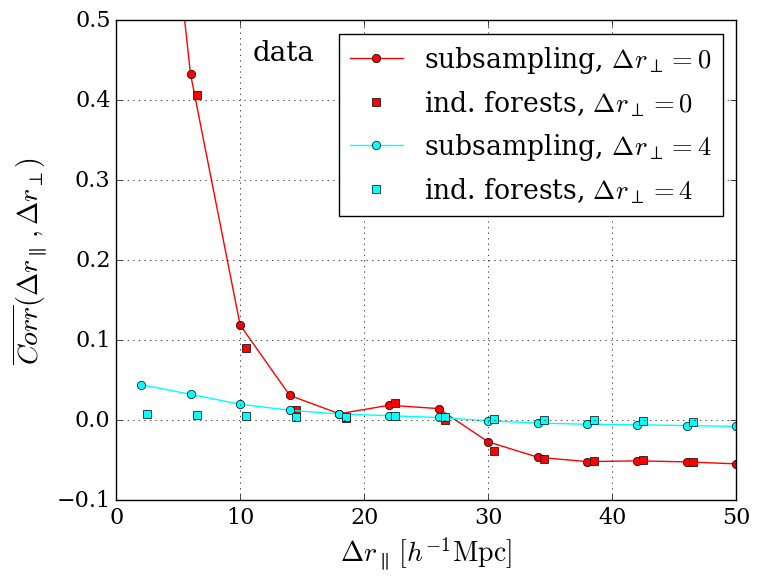}
\caption{
  Correlation,
  $Corr_{AB}=C_{AB}/\sqrt{C_{AA}C_{BB}}$, plotted vs.
$\Delta\rpar=|\rpar^A-\rpar^B|$ for the two lowest  
  intervals of $\Delta\rperp=|\rperp^A-\rperp^B|$.
  The correlation is averaged over $(\rperp^B,\rpar^B)$.
  The left panel is the  Metal (Met1) mocks and the right panel is
  for the data.
  As labeled,
the correlation is calculated by subsampling (eqn. \ref{subsamplingcov}),
by assuming ``independent forests''
(eqn. \ref{xi1dcov}, offset by $0.5\hMpc$ for clarity),
or by the mock-to-mock variation (eqn. \ref{mocktomockcov}).
Good agreement between the methods is seen, though
the independent-forest method necessarily underestimates
the correlation for $\Delta\rperp\neq0$.
The differences between the mocks and the data reflect the
differences in $\xionedim$ (Fig. \ref{fig:xi1d}).
}
\label{corrdfteq0fig}
\end{figure*}

\subsection{The covariance matrix}
\label{covmatrixsec}

The covariance matrix associated
with $\xifhat$ is:
\begin{equation}
C_{AB} =
\frac{1}{W_AW_B}
\sum_{ij\in A}\sum_{kl\in B} w_iw_j w_kw_l [
\langle \dhat_i\dhat_j \dhat_k \dhat_l \rangle
-\langle\dhat_i\dhat_j\rangle\langle\dhat_k\dhat_l\rangle
]
\label{autoautocov}
\end{equation}
where $W_A=\sum_{ij\in A}w_iw_j$.
Following \citet{2015A&A...574A..59D}, we evaluate $C_{AB}$ by dividing
the BOSS footprint into sub-samples and measuring
$\xi^s_A$ and $\xi^s_B$ in each sub-sample $s$.
Neglecting the small correlations between sub-samples,
and replacing the four-point function by the measured product
of correlations in subsamples,
the covariance is given by
\begin{equation}
C_{AB} =
\frac{1}{W_AW_B}
\sum_s W_A^sW_B^s[ \hat\xi^s_A \hat\xi^s_B - \hat\xi_A\hat\xi_B] \;,
\label{subsamplingcov}
\end{equation}
where the $W_A^s$ are the sums of weights in the sub-sample $s$.
As in \citet{2015A&A...574A..59D}, the SDSS plates  define
the sub-samples.

As a check of the subsampling method,
the sum (\ref{autoautocov})
can also be estimated by 
neglecting inter-forest correlations, 
in which case the four-point function vanishes unless
the four pixels are drawn from just two spectra:
\begin{equation}
C_{AB} =
\frac{1}{W_AW_B}
\sum_{ij\in A}\sum_{kl\in B} w_iw_j w_kw_l 
\xionedim(\lambda_i/\lambda_k)\xionedim(\lambda_j/\lambda_l)
\label{xi1dcov}
\end{equation}
The sum can then be estimated
from a random sample of forest pairs.
  Because neighboring forests are nearly parallel, the sum
  necessarily gives $C_{AB}=0$ unless $\rperp^A\sim\rperp^B$.

Finally,
for the analysis of the mock data, the covariance matrix can
also by calculated from the mock-to-mock variations of $\xi$:
\begin{equation}
C_{AB} =
\overline{ \hat\xi_A \hat\xi_B} -
\overline{\hat\xi_A} \; \overline{\hat\xi_B}
  \;,
\label{mocktomockcov}
\end{equation}
where the over-bar refers to averages over the set of 100 mocks.
Given the approximations used in the calculation of $C_{AB}$
via eqns. \ref{subsamplingcov} or \ref{xi1dcov},
it is of great importance that the mock-to-mock variations
confirm the accuracy of the other two methods when applied
to mock data.
This comparison is shown in Fig. \ref{corrdfteq0fig}.

The $2500\times2500$ element matrix, $C_{AB}$, has a relatively 
simple structure.
By far the most important elements are the diagonal elements
which are, to good approximation, inversely proportional to
the number of pixel pairs used in the calculation of the correlation
function; the number of pairs is roughly proportional to $\rperp$:
\begin{equation}
C(\rperp,\rpar,\rperp,\rpar) \;\sim\;\frac{0.043}{N_{pairs}}
\sim \frac{2.5\times10^{-10}}{\rperp/100~\hMpc}
\; .
\label{varianceauto}
\end{equation}
The variance is about twice as large as what one would calculate assuming
all (analysis) pixels used to calculate $\xi(\rperp,\rpar)$ are independent.
This decrease in the effective number of pixels
is due to the physical and instrumental correlations between neighboring
pixels in a given forest (eqn. \ref{xi1dcov}).

The off-diagonal elements of the covariance matrix also have
a simple structure.
As previously noted,
the covariance is mostly due to the two-forest part
of the four-point function which, because neighboring
forests are nearly parallel, only 
contribute to the covariance matrix elements with
$\rperp=\rperp^\prime$.
This behavior is illustrated in Fig.~\ref{corrdfteq0fig},
which displays
the mean values of the 
correlation matrix elements
as a function of $\rpar-\rpar^\prime$ 
for the smallest values of $\rperp-\rperp^\prime$.

The statistical precision of the sub-sampling calculation is
$\sim0.02$ for
individual elements of the correlation matrix.
Figure~\ref{corrdfteq0fig} reveals that only correlations with 
$\Delta\rperp\sim0$ and $\Delta\rpar<20~\hMpc$ are greater than
the statistical precision and therefore sufficiently large 
for individual matrix elements to be
measured accurately by sub-sampling.
As in \citet{2015A&A...574A..59D},
we therefore  average the sub-sampling correlation matrix
over $(\rperp^\prime,\rpar^\prime)$ and use the resulting
covariance matrix  that is a function
of $(\rperp-\rperp^\prime,\rpar-\rpar^\prime)$.

\section{Correlations introduced by the optics and data pipeline}
\label{instcorrsec}

Spurious correlations in $\dqlam$  are introduced
by the telescope and spectrometer optics, 
and by the pipeline reductions of the data. 
These correlations are superimposed on the physical $\xi(\rperp,\rpar)$
that we wish to measure;  
in this Section we estimate the various contributions.

\subsection{Optical cross-talk}

At the optical level, correlations are introduced by signals
from one object ``scattering'' into the spectra of other objects, that is
through optical cross-talk.
There is negligible cross-talk introduced in the telescope focal plane
by photons from one object entering into the fiber of another object. 
However, there can be measurable cross-talk between
neighboring fibers downstream of the wavelength dispersion 
where they are focused on the CCD.
This contamination arises
through imperfect modeling of the point-spread function
when transforming the  two-dimensional CCD image into a series 
of one-dimensional spectra, $f(\lambda)$.
We have measured the cross-talk as a function of fiber separation
by fitting the signal in sky fibers to a sky model and a weighted
sum of the spectra of the objects (quasars or galaxies) in neighboring
fibers.
Consistent with the results of \citet{2016MNRAS.457.3541C}, we find
the cross-talk for the neighboring fibers
is $\sim0.2\%$ and
$\sim0.05\%$ for fibers separated by two or three rows respectively,
and is consistent with zero for larger separations.

The cross-talk directly introduces correlations between pixels
at the same wavelength but these pixel pairs are, at any rate, not
used in the analysis.
The cross-talk introduced between pixels $\lambda_1$ and $\lambda_2$ of 
two quasars
is proportional to the product of the cross-talk amplitude
and to $\xionedim(\lambda_1/\lambda_2)$.
We have verified that these correlations are insignificant compared to
the measured $\xif$.  This fact is in part due to the fiber-assignment
strategy which avoided placing two quasar candidates
on neighboring fibers. 

\subsection{Pipeline-induced correlations}

The pipeline treatment of the spectrometer data
transforms flat-field corrected  CCD counts, $f_q^{\rm m}(\lambda)$,
to fluxes, $\fqlam$.  This process requires subtracting a 
sky contribution, $B_q(\lambda)$,
and multiplying by a calibration vector, $A_q(\lambda)$,
that corrects for the wavelength-dependent throughput of
the system:
\begin{equation}
\fqlam = A_q(\lambda)\;[f_q^{\rm m}(\lambda) - B_q(\lambda)] \;.
\end{equation}
Both $A_q(\lambda)$ and  $B_q(\lambda)$ are determined  from
spectra taken simultaneously with the data: spectra of
spectro-photometric standard stars for $A_q$, and spectra
of ``sky fibers''  (fibers pointing to empty sky regions) for $B_q$.
The spectra from the 1000 fibers
of a given exposure are treated independently for
the two 500-fiber spectrographs because
of possible differential variations of instrumental properties
such as throughput, PSF, and scattered light. 
(See \citet{2013AJ....146...32S} for the relevant details of
the spectrometer construction.)
Hereafter, the collection of fibers of a 
given plate assigned to one of the two spectrographs will be 
referred to as a ``half-plate.''
Inaccuracies in the 
determinations of $A_q$ and $B_q$ will lead directly to correlated
inaccuracies in the $\fqlam$ of a given half-plate.

Errors in the sky subtraction, $B_q(\lambda)$,  are simple to model because they
are mostly due to well-understood Poisson fluctuations of 
photo-electron counts in the 
$\sim50$  
sky fibers per half-plate.
The pipeline interpolates these
measurements in the focal plane to produce the sky background to be subtracted 
from a given quasar.  
The limited number of sky-fibers results in 
correlations between $\fqlam$ and 
$f_{q^\prime}(\lambda^\prime)$ for $q\neq q^\prime$ and $\lambda=\lambda^\prime$
since the sky subtractions 
where calculated using the same noisy sky-fibers.
These correlations for $\lambda=\lambda^\prime$ generate small
correlations for $\lambda\ne\lambda^\prime$ through continuum
fitting, as discussed in Sect. \ref{xiestimationsec}.
Because the correlations are 
primarily for  $\lambda=\lambda^\prime$
they contribute a spurious $\xif(\rperp,\rpar=0)$
for pairs of pixels on the same half-plate.

Errors on the calibration vector $A_q$ are also Poissonian in that
they are due to the fluctuations in the imperfect modeling of $\sim10$
calibration stars per half-plate.
The mean over all spectra
of the imperfections are visible in the mean ``transmission''
of the unabsorbed parts of quasar spectra shown in Fig.~\ref{meanfluxfig}.
As mentioned  in Section \ref{samplesec},  
the mean imperfections are removed by dividing
forest spectra by $R(\lambda)$ from Fig.~\ref{meanfluxfig}.
However, fluctuations from exposure
to exposure remain because of the spectral variations of 
the small number of calibration stars used.
The statistical properties of these variations have been estimated
by comparing the $A_q$ measured with random sub-samples of stars
and, independently,
by comparing the $A_q$ on different half-plates for the same exposure.
Unlike the correlations due to sky-subtraction, which are confined
to $\lambda=\lambda^\prime$, the errors in the calibration vectors
are strongly correlated
over the characteristic wavelength range of stellar
spectral features, $\Delta\lambda\sim1$~nm.

We constructed a model for pipeline-induced correlations that uses the
measured statistical properties of the $A_q(\lambda)$ and $B_q(\lambda)$.
The contribution of
the sky  noise was estimated using a fit of random realizations of
the signal in the sky fibers (using a degree 2 polynomial fit as a
function of the fiber number).
The realizations were based on the  statistical uncertainty
estimation  from the pipeline (which we have found to be accurate at
the 5\% level, see Fig. \ref{fig:rms_new_prod}, bottom panel).
A normalization
factor was applied to account for the average difference of calibration between
the sky and target fibers (a $\sim10\%$ correction).
Performing a
fit allowed us to capture accurately the resulting correlations as a
function of fiber separation.

The contribution of the calibration was estimated using the fact that
because the flux calibration of the two spectrographs is determined
independently, the difference of those two calibration solutions
for each exposure provides us with an estimate of their statistical
fluctuation (up to a normalization factor $\sqrt{2}$) while retaining
the correlation of those fluctuations as a function of wavelength.
Those differences for all DR12 plates was computed. Their average
per observation run was subtracted in order to account for the actual
difference of throughput of the two spectrographs and their evolution
during the course of the survey. This data set provided us with a
library of calibration uncertainties that was used to calculate
calibration-induced correlations between spectra.

This model was used to calculate the expected
correlation between pixels in the \Lya forest of one quasar
and pixels in the CIV forest ($142<\lamrf<152$~nm) of another quasar.
The physical correlation of the two forests is due mostly 
to the auto-correlation of the weak  CIV absorption in the two forests,
but any pipeline-induced correlations should be present at full strength.
Since this cross-correlation is designed to isolate pipeline-induced effects,
we denote this 
correlation function as $\xipl(\lambda_2-\lambda_1,\theta)$ where
$(\lambda_1,\lambda_2)$ are the wavelengths in the two
forests and $\theta$ is the angular separation. We also distinguish
between  $\xipl$ and correlations between flux pairs on the same half-plate,
$\xipl({hp=hp^\prime})$, 
and correlations between flux pairs on 
different half-plates,
$\xipl({hp\neq hp^\prime})$.
While these correlations are naturally given as a function of wavelength
and angular separations,
for convenience, the $\xipl$ will be given as functions of pseudo-separations
$(\rperppseudo,\rparpseudo)$ calculated using the \Lya rest-frame
wavelength to define redshifts in both spectra.
With this approach, $\rparpseudo=0$ corresponds to absorption at
the same observed wavelength.

\begin{figure}[t]
\begin{center}
\includegraphics[width=.49\textwidth]{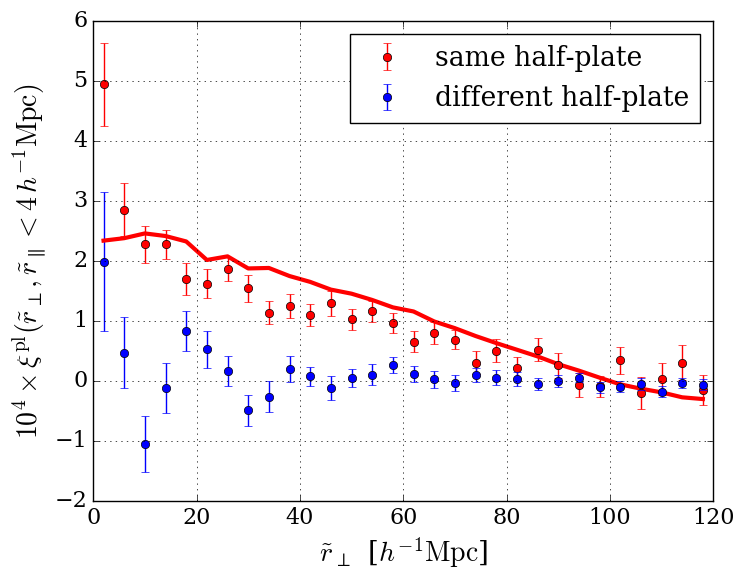}
\caption{
  Correlation between pixels in the \lya-forest and pixels in the
  CIV-forest, $\xipl(\lambda_2-\lambda_1,\theta)$ for small wavelength
  differences.  The wavelength difference and angular separation have
  been transformed to pseudo-separations
$(\rperppseudo,\rparpseudo<4~\hMpc)$
  assuming \Lya absorption
  in both forests.
  The red points show the correlation for pairs on the same halfplate
$\xipl_{hp=hp^\prime}(\rperppseudo,\rparpseudo<4~\hMpc)$.
The much smaller correlations for different half-plates are the blue points.
The red line represents the prediction for our model of calibration
and sky  noise.
}
\label{xihzlzrpeq10_dif_same_halfplates}
\end{center}
\end{figure}

This correlation for same-half-plate pairs (measured using the
techniques of Sect. \ref{xiestimationsec})
is shown with the red points  in Fig.~\ref{xihzlzrpeq10_dif_same_halfplates}
for the first $\rparpseudo$ bin ($\rparpseudo<4~\hMpc$).
Superimposed  on the data is the prediction of the model of
the pipeline-induced correlations.
For the first $\rparpseudo$ bin, the correlation is dominated by
those induced by to the sky-subtraction model.
There is good agreement between this
simple model and the observed correlations.

The pipeline-induced  correlations that we have
considered  do not contribute to correlations
between pixels observed on different half-plates.
The different-half-plates correlation functions
for the $\rparpseudo<4~\hMpc$ bin is shown by the blue points
in Fig.~\ref{xihzlzrpeq10_dif_same_halfplates}.
The correlations for different half-plates (blue points) are clearly
far less than those for same half-plates (red points).
For $\rparpseudo<4~\hMpc,\rperppseudo<~60~\hMpc$, 
$\chi^2=28.6$ ($N_{\rm dof}=15$) for the no-correlation hypothesis. 

\begin{figure}[t]
\begin{center}
\includegraphics[width=.49\textwidth]{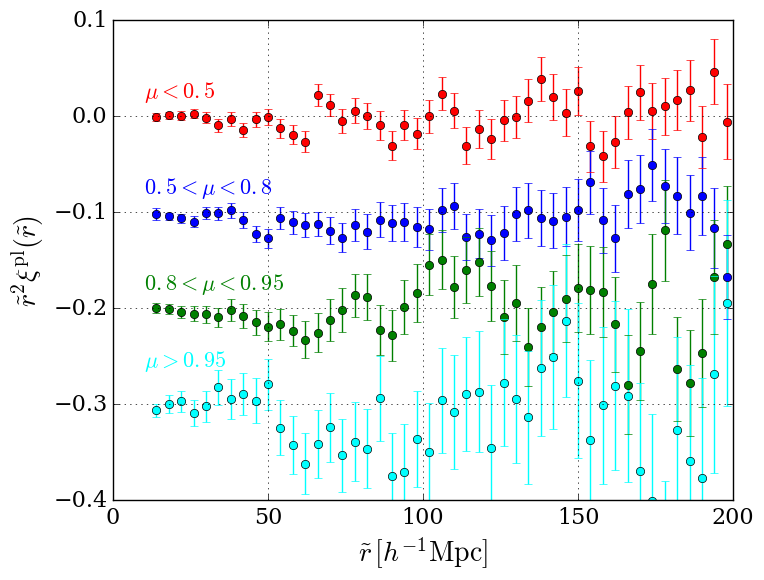}
\caption{
  Null test for pipeline-induced correlations.
   As in Fig. \ref{xihzlzrpeq10_dif_same_halfplates}, the 
correlation between the \lya-forest pixels and the CIV-forest pixels,
$\xipl(\rparpseudo,\rperppseudo)$ is shown, but now 
for four angular ranges as labeled and offset by 0.1 successively for clarity.
Pixel pairs with $\rparpseudo<4~\hMpc$ are excluded.
There is no evidence for pipeline-induced systematics with
the $\chi^2$ for vanishing correlation
for $12<\tilde{r}<160~\hMpc$ is $(32.5,\,31.2,\,22.7,\,40.0)$ for the
four ranges, each with 38 data points.
  }
\label{crosscorrhizlzfig}
\end{center}
\end{figure}

For the $\rparpseudo>4~\hMpc$ bins, 
the model predicts much smaller correlations. 
In particular, the sky-subtraction model noise induces non-zero 
correlations only because of the continuum fit which distorts 
the original correlation function as described in 
Section \ref{xiestimationsec}.
This behavior is illustrated in 
Fig.~\ref{crosscorrhizlzfig} showing $\xipl$ in four ranges
of $\mu$.
Because of the low level of absorption fluctuations redward of \Lya emission,
the variance of the points is a factor $\sim7$ smaller
than the corresponding variance for the forest-forest correlation
function.

The model for pipeline-induced correlations assumes
that the sky-subtraction errors are entirely due to
Poisson statistics of photo-electron counts,
neglecting possible wavelength- and position-dependent systematic
mis-estimates of the sky flux.
While
the changes applied to the pipeline (Appendix \ref{sec:new_pipeline}) 
considerably reduced the systematic sky residuals, 
significant residuals remain on bright sky lines due to an 
imperfect PSF model, PSF variations and displacement of spectral 
traces (due to changes of temperature and variations of the gravity 
load of the spectrographs which are at the Cassegrain focus of the telescope). 
We tested the effect of those sky residuals by computing the 
correlation function of fake \Lya forests consisting of the residual 
signal in the nearest sky fiber in the CCD  divided by the quasar continuum 
model (we made sure to use only once each sky fiber in this process to 
avoid introducing an artificial correlation). The measured correlation signal 
could be entirely explained by the
Poissonian sky model noise described above. 
We hence conclude that the systematic sky residuals induce a negligible 
contamination to the \Lya correlation function. 

This analysis presented above evaluates the contamination 
due to any additive signal to the \Lya forest that leads to a 
correlated signal in the CCD. It includes the systematic sky residuals, 
their fluctuations from plate to plate, but also the potential effect of 
scattered light in the spectrograph.

We have used our model of pipeline-induced correlations to
calculate their effect on the determination of the BAO peak
position.  Since the model contains no scale near the BAO scale,
it is not surprising that it predicts no measurable
influence, as reported in Table \ref{table:pipeline_systematics}
of Sect. \ref{systsec}.
Furthermore,
to facilitate the comparison of the measured correlation function
with the physical model developed in the next section,
we do not use pairs of pixels on the same
exposure that would contribute to the $\rperp<4~\hMpc$ bins of
$\xif(\rperp,\rpar)$.

\begin{figure*}[t]
\centering
\includegraphics[width=\columnwidth]{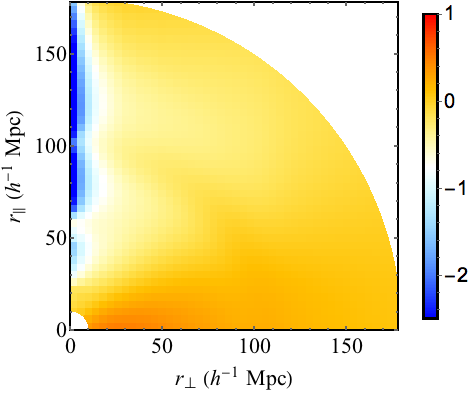}
\includegraphics[width=\columnwidth]{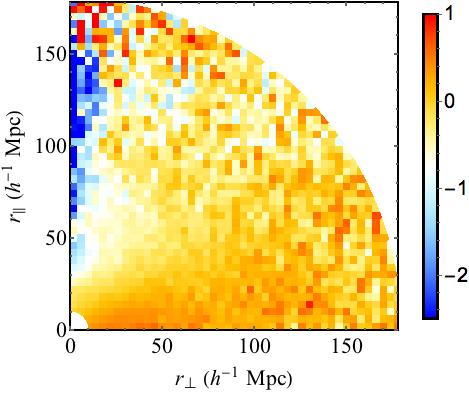}
\caption{
  Two-dimensional representation of 
  $r^2\xi(\rperp,\rpar)$ in units of $(\hMpc)^2$.
  The right panel shows the measurement
  and the left panel the best-fit model, $\xicosmo$,
  modified by
  the distortion matrix via
  eqn. (\ref{xitheoryeq}).
  The BAO feature is at
  $r\sim100~\hMpc$.  The effects of metal-\Lya correlations are seen
  in the lowest $\rperp$ bin, in particular the peak at $50<\rpar<70~\hMpc$
  due to SiIIa and SiIIb.}
\label{2dmodelfig}
\end{figure*}

\begin{table}
  \centering
  \caption{
    Parameters of the physical model for the correlation function
    and their best-fit values for the data fit over
    the range $10<r<180~\hMpc$.
    The parameters
    include the BAO peak-position parameters, $(\aperp,\apar)$,
    the bias parameters $(b,\beta)$ at redshift $z=2.3$
    for the
    various components, the length scale, $\Lhcds$, for unidentified HCDs,
    and the bias parameter, $b_\Gamma$ for UV fluctuations.
    The biases $\bhcds$ and $b_\Gamma$ are relative to $\blya$.
    Parameters without uncertainties are fixed in the fit.
    These include the UV absorber response bias, $b_a^\prime$, the UV
    attenuation length, $\lambda_{\rm UV}$,
    the BAO amplitude parameter, $\Apeak$, the bias redshift-evolution
    parameter, $\gamma$, and the peak-broadening parameters $\vec{\Sigma}$.
  } 
\label{paramtable}
\begin{tabular}{l l l}
  parameters & best fit & best fit \\
  \hline
 $\apar,\aperp$ &  $  1.053  \pm  0.036 $ &  $ 0.965  \pm  0.055 $   \\ 
 $b(1+\beta),\beta\;$ \Lya &  $  -0.325  \pm  0.004 $ &  $ 1.663  \pm  0.085 $   \\ 
 $b,\beta\;$ Si3 &  $  -0.0033  \pm  0.0013 $ &  $ 0.500   $   \\ 
 $b,\beta\;$ Si2a &  $  -0.0044  \pm  0.0009 $ &  $ 0.500   $   \\ 
 $b,\beta\;$ Si2b &  $  -0.0035  \pm  0.0009 $ &  $ 0.500   $   \\ 
 $b,\beta\;$ Si2c &  $  -0.0015  \pm  0.0012 $ &  $ 0.500   $   \\ 
 $b,\beta\;$ CIV &  $  -0.0233  \pm  0.0114 $ &  $ 0.500   $   \\ 
 $b,\beta\;$ HCDs &  $  -0.0288  \pm  0.0043 $ &  $ 0.681  \pm  0.175 $   \\ 
 $\Lhcds$ &  $  24.3410  \pm  1.1308 $  \\ 
 $b_\Gamma,\,b^\prime_a$  &  $  0.1125  \pm  0.0512 $ & $-2/3$   \\ 
 $\lamuv [\hMpc] $  &  $  300.00   $  \\ 
 $\Apeak$  &  $  1.00   $  \\ 
 $\gamma$  &  $  2.90   $  \\ 
 $\Sigma_\perp,\Sigma_\parallel\;$ &  $  3.26     $ &  $ 6.41   $   \\ 
\end{tabular}
\end{table}

\section{Fits for the BAO peak position}
\label{fitssec}

To determine the position of the BAO peak in the transverse
and radial directions, we fit the measured $\xif(\rperp,\rpar)$
to functions that describe the underlying
large-scale-structure correlations.
These correlations are primarily due to \Lya absorption
in the intergalactic medium (IGM), but
we also include absorption by  
metals in the IGM and neutral hydrogen in high-column-density systems (HCDs).
The physical correlations are corrected for
distortions that are 
introduced by the procedure for determining the quasar continuum
(Sect. \ref{distortionsec}).
The fitting routine we use  gives results consistent with
those found using the publicly available
baofit\footnote{http://darkmatter.ps.uci.edu/baofit/}
package  \citep{2013JCAP...03..024K,2015JCAP...11..034B} modified to
include the effects described in this Section.
The parameters of the fits are described below and
in Table \ref{paramtable}.
The best-fit model correlation function 
is shown in Fig. \ref{2dmodelfig}.

\begin{figure*}[tb]
\centering
\includegraphics[width=\columnwidth]{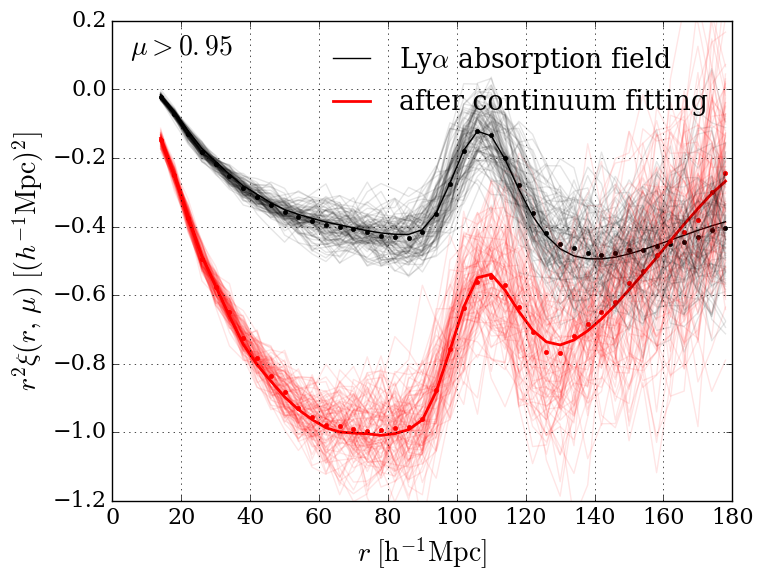}
\includegraphics[width=\columnwidth]{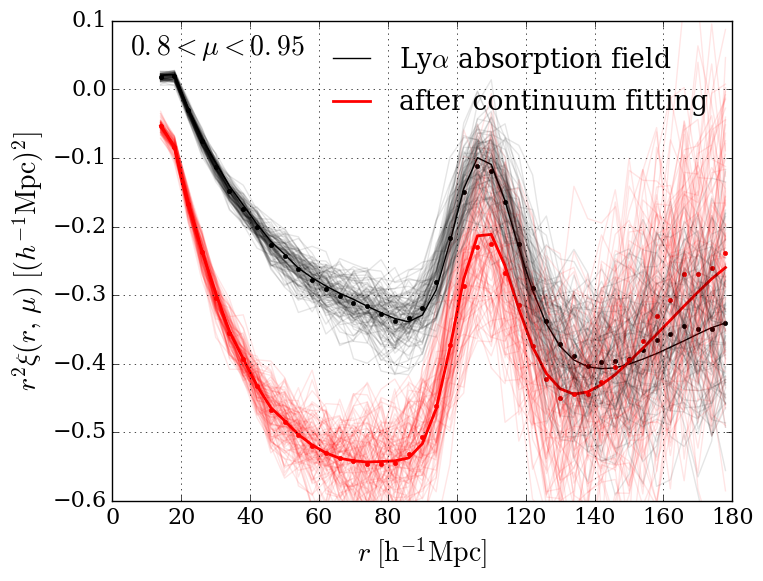}
\includegraphics[width=\columnwidth]{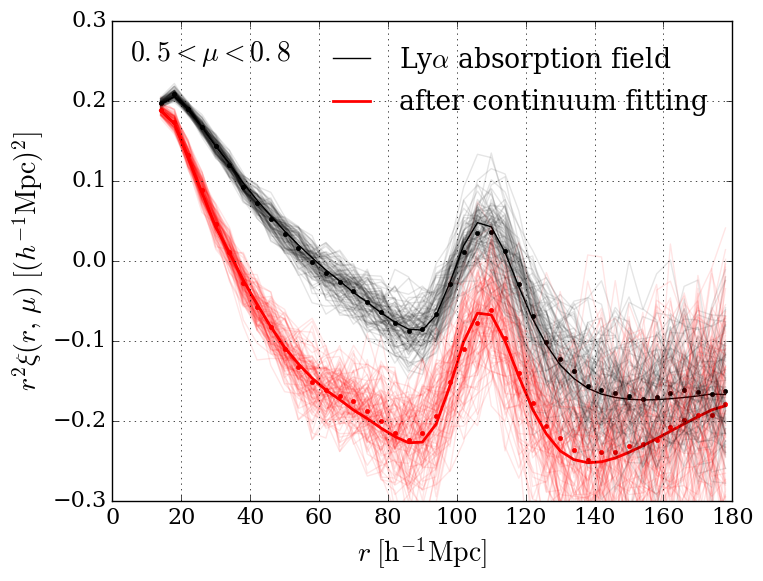}
\includegraphics[width=\columnwidth]{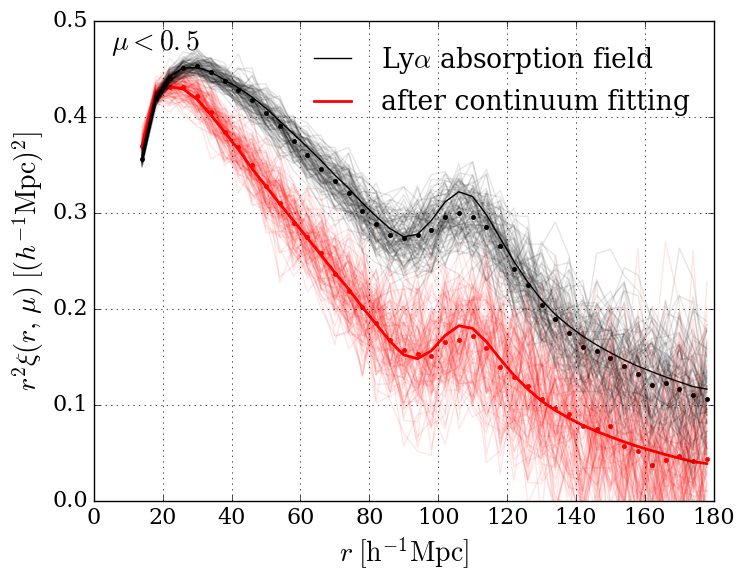}
\caption{
Correlation function for the metal-free mocks in four
ranges of $\mu$.
The black  points and curves correspond to mocks with \Lya absorption
but without the addition of a quasar continuum.
The red points and curves correspond to mocks with the
addition of a continuum.
The points correspond to stacks of 100 mocks and the light
curves to individual mocks.  The heavy curves correspond
to the input model of Table \ref{modtable}
(after distortion
by the  matrix $D_{AA^\prime}$ (eqn. \ref{DABeq}) for the red curve). 
}
\label{stackedmocksfig}
\end{figure*}

\begin{figure*}[tb]
\centering
\includegraphics[width=\columnwidth]{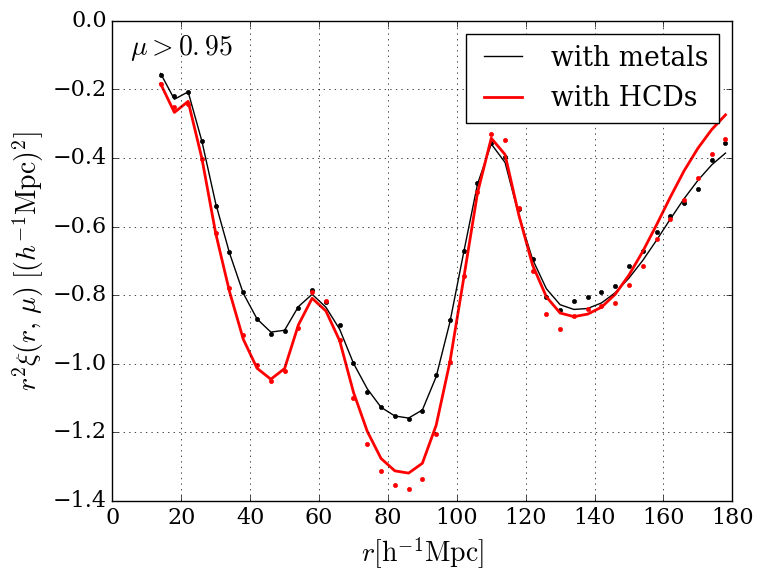}
\includegraphics[width=\columnwidth]{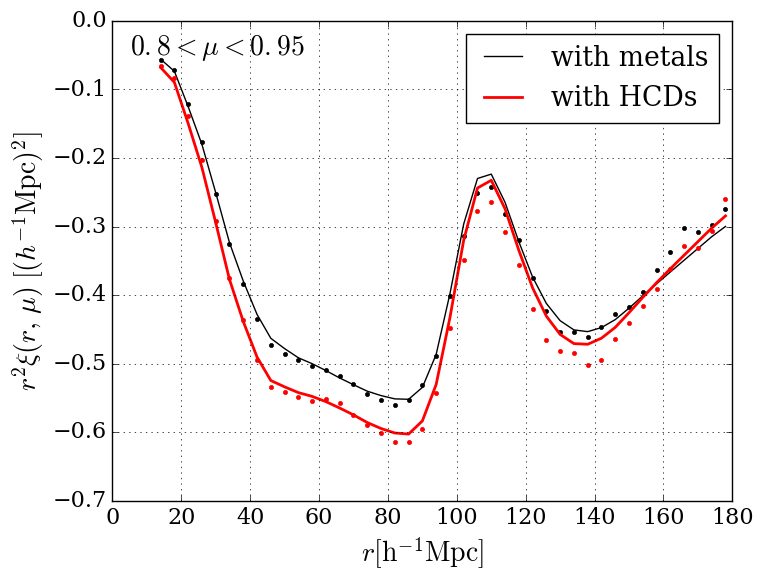}
\includegraphics[width=\columnwidth]{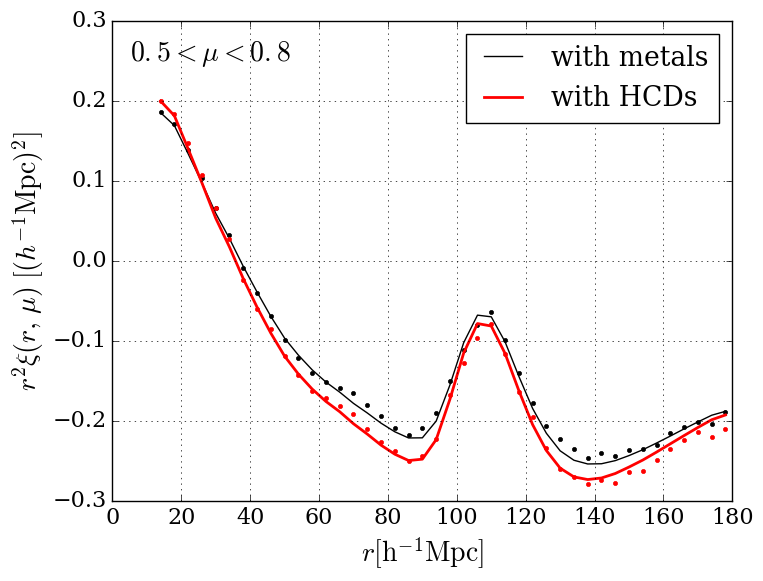}
\includegraphics[width=\columnwidth]{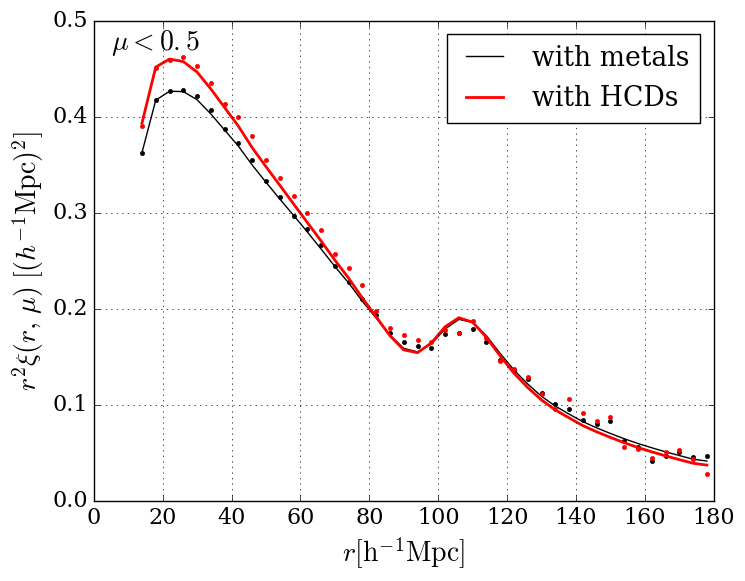}
\caption{
Correlation function for the stack of the 100 mocks in
four ranges of $\mu$.
The red points represent
the measured correlation function of the mocks with metals
and the red curves shows the best fit.
The black points  indicate
the correlation function after including, in addition to metals,  HCDs
unmasked for $\NHI<10^{20}{\rm cm^{-2}}$
and the black curve the best fit.
The peak at $r\sim60~\hMpc$ due to SiIIa and SiIIb is apparent in the range
$\mu>0.95$ but not in the range $0.8<\mu<0.95$.
}
\label{stackedmockswithmetalsfig}
\end{figure*}

\begin{figure*}[tb]
\includegraphics[width=\columnwidth]{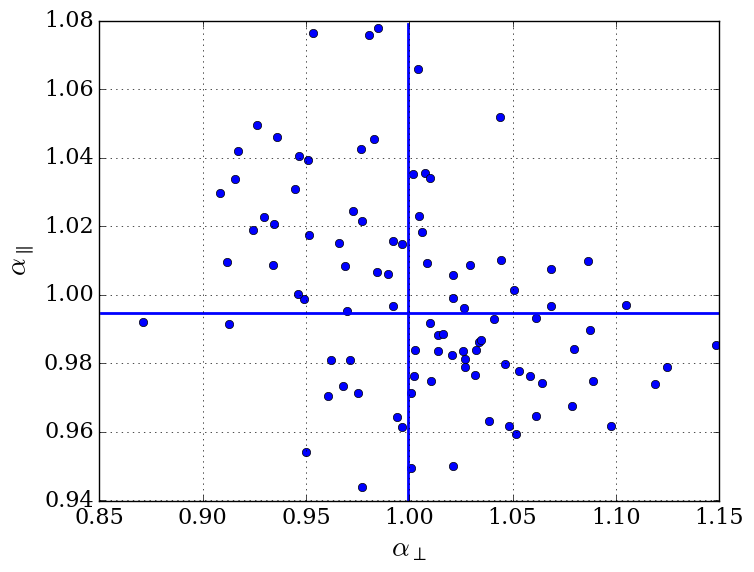}
\includegraphics[width=\columnwidth]{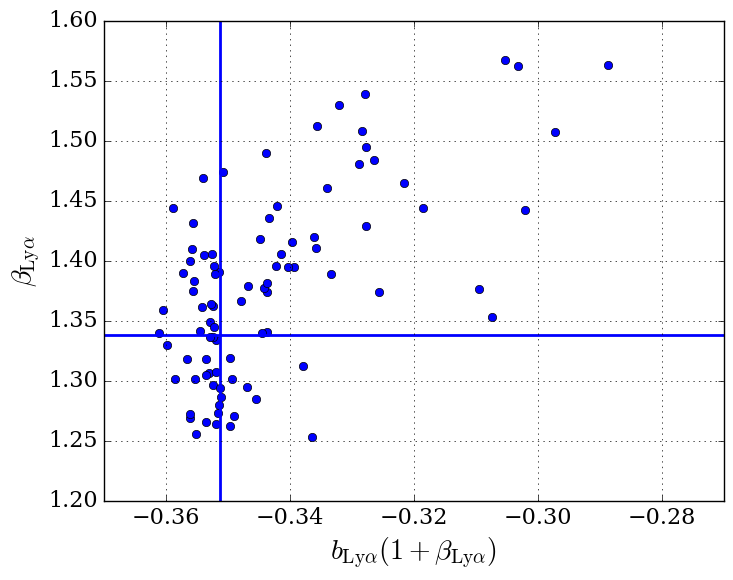}
\caption{
Measured $\aperp$ and $\apar$ (left) and  $\betalya$ and $\blya(1+\betalya)$
(right) for the 100 mock catalogs including HCDs with $\NHI>10^{20}{\rm cm^2}$
masked.
There are four outliers on the left plot and two on the right.
The horizontal and vertical blue lines show the weighted means
of the distributions, also given in Table \ref{fitmocktable}.
The distribution and mean values of $(\aperp,\apar)$ indicates
no significant bias in the
reconstruction of the BAO peak-position parameters. 
}
\label{mocksoutputfig} 
\end{figure*}

\subsection{The model of the correlation function}
\label{fitmodelsec}

The general
form for the model correlation function $\xif_A$
of the $(\rperp,\rpar)$ bin $A$
is a distorted sum of the cosmological, or ``physical''  correlation function, $\xicosmo_A$,
and a slowly varying function, $B_A$, used to test
for systematics:
\begin{equation}
\xif_A  = \sum_{A^\prime}D_{AA^\prime}
\left[ \xicosmo_{A^\prime}
\,+\, B_{A^\prime} \right]
\label{xitheoryeq}
\end{equation}
Here, $D_{AA^\prime}$ is the distortion matrix (eqn. \ref{DABeq})
that models the effects of continuum fitting.

The physical component of the model is dominated by the auto-correlation
due to \Lya absorption.  It is assumed to be a biased version of
the  total matter auto-correlation of the appropriate flat-\lcdm model
(Table \ref{modtable}) modified to free the position of the BAO
peak \citep{2013JCAP...03..024K}:
\begin{equation}
  \xilya(\rperp,\rpar,\aperp,\apar) =
  \xismooth(\rperp,\rpar) + \xipeak(\aperp\rperp,\apar\rpar)
\end{equation}
where the BAO peak-position parameters to be fit are
\begin{equation}
\apar = \frac { \left[\DHub(\bar z)/r_d\right] }
{\left[D_H(\bar z)/r_d\right]_{\rm fid}}
\hspace*{3mm}{\rm and}\hspace*{5mm}
\aperp = \frac { \left[\DM(\bar z)/r_d\right] }
{\left[\DM(\bar z)/r_d\right]_{\rm fid}} ~,
\label{eq:alpha}
\end{equation}
and where the subscript ``fid'' refers to the fiducial cosmological
model from Table \ref{modtable} used to transform angle differences
and redshift differences to $(\rperp,\rpar)$.
The  nominal correlation function,
$\xilya(\rperp,\rpar,\aperp=\apar=1)$, is derived from
its Fourier transform
\begin{equation}
  P_{\rm Ly\alpha}(\vec{k},z)=P_{\rm QL}(\vec{k},z)
\blya^2(1+\betalya\mu_{\vec{k}}^2)^2
F_{\rm NL}(\vec{k})
G(\vec{k})
\label{lyapowerspectrumeqn}
\end{equation}
where $P_{\rm QL}$ is the (quasi) linear power spectrum decomposed
into a smooth component and a peak component corrected for non-linear
broadening of the BAO peak:
\begin{equation}
  P_{\rm QL}(\vec{k},z)= P_{\rm smooth}(k,z)\; +\,
 e^{-k^2\Sigma^2(\mu_k)/2}
   \Apeak P_{\rm peak}(k,z)
\label{peaksmooth}
\end{equation}
The smooth component is derived from the CAMB-calculated linear
power spectrum, $P_{\rm L}$,
via the side-band technique \citep{2013JCAP...03..024K} 
and $P_{\rm peak}=P_{\rm L}-P_{\rm smooth}$.
The correction for non-linear broadening of the BAO peak is parameterized by
$\Sigma^2=\mu_k^2\Sigma_\parallel^2+(1-\mu_k^2)\Sigma_\perp^2$.
The nominal values used
are $\Sigma_\parallel=6.41~\hMpc$ and $\Sigma_\perp=3.26~\hMpc$
\citep{2013JCAP...03..024K}.

In equation \ref{lyapowerspectrumeqn},
the bias, $\blya$, is assumed to have a redshift dependence 
$b_n\propto(1+z)^\gamma$ with $\gamma=2.9$
(so that $P(k,z)\blya(z)^2\propto (1+z)^{3.8}$),
and $\betalya$ is assumed redshift independent.
The function $F_{\rm NL}$ corrects for
non-linear effects at large $k$ due to
the isotropic enhancement of power due to non-linear growth,
the isotropic suppression of power due to gas pressure, and
the suppression of power due to line-of-sight non-linear
peculiar velocity and thermal broadening.
We use the form given by equation 21 and Table 1 of
\citet{2003ApJ...585...34M}.
The forms proposed by \citet{2015JCAP...12..017A}
produce nearly identical results for the range of $(\rperp,\rpar)$
used in this study.

  The last term in equation \ref{lyapowerspectrumeqn},
  $G(\vec{k})=G_\parallel(k_\parallel)G_\perp(k_\perp)$, accounts for
  the binning in $(\rperp,\rpar)$ space which effectively
  averages the correlation function over a bin.
(The large width of these bins renders unnecessary a term that
  accounts for the spectrometer resolution.)
  If the distribution of observed $(\rperp,\rpar)$ were uniform
  in a bin, $G(\vec{k})$ would be the Fourier transform of the bin.
  The distribution is approximately uniform
in the radial direction, which implies
$G_\parallel={\rm sinc}^2(R_\parallel k_\parallel/2)$ where
$R_\parallel=4~\hMpc$ (the bin width).
In the perpendicular direction,
the distribution is approximately proportional to $\rperp$
so
$G_\perp=[(J_1(k_\perp r_+)-J_1(k_\perp r_-))/k_\perp\Delta r]^2$
where $r_+$ and $r_-$ are the extrema of a given $\rperp$ bin
and $\Delta r=r_+-r_-$ is the bin width.
The expression is bin-dependent and performing a Fourier transform
on each bin
would be too time-consuming. We therefore
replace $G_\perp$ by
${\rm sinc^2}(k_\perp \Delta r)$ and evaluate the
correlation function at the average $\rperp$ of the pairs in a bin.
This procedure produces a sufficiently accurate correlation function.

We allow for
fluctuations of ionizing UV radiation 
\citep{2014PhRvD..89h3010P,2014MNRAS.442..187G} which lead to
a scale-dependent bias of \Lya absorption 
given by eqn. 12 of \citet{2014MNRAS.442..187G}.
The effect of the fluctuations is to increase $\blya$ from
its nominal value at small scale to a different value at
large scale.  The transition scale is determined by the
UV photon mean-free-path which we set to a comoving value of $300~\hMpc$
\citep{2013ApJ...769..146R}.
We then fit for one parameter, $b_\Gamma$ corresponding
to the $b_\Gamma(b_s-b_a)$ of \citet{2014MNRAS.442..187G}; it
determines the change
in $\blya$ between large and small scales.
A second bias, $b_a^\prime$, that determines the precise dependence
of the bias on scale, is set to the nominal value of $-2/3$
used by \citet{2014MNRAS.442..187G}.

In our previous investigations, $\xilya$ was assumed to be a sufficiently
accurate approximation for the correlation function.
In this study, we include absorption
by unidentified High-Column-Density Systems (HCDs) and by metals.
The total physical correlation function is then the sum
of the auto- and cross-correlations of the various components:
\begin{equation}
\xicosmo_A\, =\, \sum_{m,n}\xi^{mn}_A
\end{equation}
Each component, $n$ or $m$,
has its own bias and redshift-space distortion parameter.

The correlation function for HCDs and their cross-correlation
with \Lya absorption 
is derived from a power spectrum as in
eqn. \ref{lyapowerspectrumeqn}
but with different bias parameters \citep{2012JCAP...07..028F}.
While
these absorbers are  expected to trace the 
underlying density field
their effect on the flux-transmission field depends on whether 
they are identified and given the special treatment
described in Sect. \ref{samplesec}.
If they are correctly identified with the total absorption region masked
and the wings correctly modeled, they can be expected to have no
significant effect on the field.
Conversely, if they are not identified, the measured
correlation function will be modified because
their absorption is spread  along the radial direction.
We model this by multiplying the bias for
unidentified HCDs, $\bhcds$, by a function
of $k_\parallel$,
$F_{\rm HCD}(k_\parallel)$, that reflects the typical
column densities of the unidentified HCDs.
Following the studies with the mock spectra described in Sect. \ref{mocksec},
we adopted the
form $F_{\rm HCD}(k_\parallel)={\rm sinc}(\Lhcds k_\parallel)$,
where $\Lhcds$ is a free parameter.
The parameter $\betahcds$ is poorly determined by the fit
and we impose a Gaussian prior $\betahcds=0.5\pm 0.2$

Because there is little
absorption by metals
the treatment of metal components is simplified without
the separation into peak and smooth components.
The fiducial correlation function is directly used
to calculate in real space both metal-metal and metal-\Lya
correlations.  The parameters $(b,\beta)$ for each
species (Table \ref{paramtable})
are included by using a multipole expansion.

Absorption by metals is complicated
by the fact that in the data analysis
the \Lya transition is naturally assumed when transforming
wavelength differences to position differences.
Because of this procedure,
for the cross-correlation between absorption by transition $n$
and transition $m$,
the nominal $(\rperp,\rpar)$ of the bin $A$ is not equal to
the true separation coordinates $(\rperp^{mn},\rpar^{mn})$.
The model correlation, $\xi^{mn}_A$  for the bin $A$ is thus
evaluated at the true, rather than nominal, separation.

Because amplitudes for SiII and SiIII are mostly determined
by the excess correlation at $(\rperp\sim0, \rpar\neq0)$,
the $\beta$ for each metal is poorly determined.
We therefore fix their value to $\beta=0.5$
corresponding to host halos with bias of
2, the value found for DLAs \citep{2012JCAP...07..028F}
which is also typical
of star-forming galaxies.

  The bias, $\bciv$, of foreground ($z\sim1.65$) CIV absorption
  is poorly determined by the
  fit of the \lya-forest auto correlation. This is because 
  the large wavelength difference between the \Lya and CIV
  transitions  places the peak corresponding to
  vanishing metal-\Lya separation outside the forest.
We have therefore measured $\bciv$ by
correlating the flux in the CIV forest 
with quasars at $z\sim1.65$.
We compared
these correlations with the analogous ones between the $\dqlam$
in the \Lya forest with $z\sim2.3$ quasars.
The resulting $\mu$-averaged correlation ratio is
\begin{equation}
  \frac{\xi^{\rm q-CIV}(z=1.65)}{\xi^{\rm q-Ly\alpha}(z=2.3)}
= 0.08 \pm 0.02
\hspace*{10mm} r=10\,\hMpc
\label{lyacivratio}
\end{equation}
Because the quasar-CIV flux correlation is linear in $\bciv$ while
the contribution to the flux auto-correlation is proportional
to $\bciv^2$,
this result implies a negligible ($<1\%$) contamination
of the \lya-forest auto-correlation by CIV auto-correlation.

\begin{table*}[ht]
\begin{center}
\caption{
  Weighted mean of fit parameters and
  mean uncertainties for six sets of 100 mocks
  of increasing realism: \Lya only; including a quasar continuum; 
  including metal absorption (for Met1 or Met2); including 
  HCDs (and Met1) and masking
  those with $\NHI>10^{20}{\rm cm^{-2}}$ or $>10^{21}{\rm cm^{-2}}$.
  Fits are over the range $10<r<180~\hMpc$.
  The bias parameter, $\blya$, refers to  the reference redshift $z=2.25$.
  The input values for the mock generation were $b(1+\beta)_{\rm Ly\alpha}=-0.336$
  and $\betalya=1.4$.
  }
\begin{tabular}{l c c c c c}
mock set  
    & $\overline{\apar}\hspace*{2mm}(\overline{\sigma_{\apar}})$ 
    & $\overline{\aperp}\hspace*{2mm}(\overline{\sigma_{\aperp}})$  
    & $\overline{b(1 \! + \! \beta)_{\rm Ly\alpha}}\hspace*{2mm}(\overline{\sigma_{b(1+\beta)}})$
    & $\overline{\betalya}\hspace*{2mm}(\overline{\sigma_{\betalya}})$
    & $\overline{\chi^2_{\rm min}}/DOF$\hspace*{3mm} $\overline{Prob.}$  \\
\noalign{\smallskip} 
\hline
\noalign{\smallskip}
Ly$\alpha$ only & $ 0.998 \; ( 0.014 ) $ & $ 1.002 \; ( 0.020 ) $ & $ -0.341 \; ( 0.001 ) $ & $ 1.360 \; ( 0.017 ) $ & $ 1632.5 /( 1590 - 4 ) \hspace*{2mm}p= 0.292 $ \\ 
+continuum & $ 1.000 \; ( 0.023 ) $ & $ 0.993 \; ( 0.040 ) $ & $ -0.338 \; ( 0.002 ) $ & $ 1.339 \; ( 0.034 ) $ & $ 1589.1 /( 1590 - 4 ) \hspace*{2mm}p= 0.487 $ \\ 
+metals (Met1) & $ 0.998 \; ( 0.025 ) $ & $ 0.993 \; ( 0.040 ) $ & $ -0.337 \; ( 0.002 ) $ & $ 1.334 \; ( 0.038 ) $ & $ 1583.8 /( 1590 - 12 ) \hspace*{2mm}p= 0.473 $ \\ 
(or) +metals (Met2) & $ 1.002 \; ( 0.023 ) $ & $ 0.992 \; ( 0.039 ) $ & $ -0.339 \; ( 0.002 ) $ & $ 1.334 \; ( 0.037 ) $ & $ 1582.8 /( 1590 - 12 ) \hspace*{2mm}p= 0.480 $ \\ 
+HCDs (20) & $ 0.995 \; ( 0.032 ) $ & $ 0.999 \; ( 0.058 ) $ & $ -0.351 \; ( 0.035 ) $ & $ 1.338 \; ( 0.172 ) $ & $ 1590.0 /( 1590 - 15 ) \hspace*{2mm}p= 0.429 $ \\ 
(or) +HCDs (21) & $ 0.995 \; ( 0.032 ) $ & $ 0.995 \; ( 0.056 ) $ & $ -0.359 \; ( 0.026 ) $ & $ 1.321 \; ( 0.145 ) $ & $ 1594.1 /( 1590 - 15 ) \hspace*{2mm}p= 0.417 $ \\ 
\label{fitmocktable}
\end{tabular}
\end{center}
\end{table*}

To transform the ratio  (\ref{lyacivratio}) into a
constraint on $\bciv$ that is needed for the fits, we equate 
it to the appropriate
function of bias parameters:
\begin{equation}
  \frac{\xi^{\rm q-CIV}_{z=1.65}}{\xi^{\rm q-Ly\alpha}_{z=2.3}} =
  \frac{[\bciv\bqso\xi_{\rm L}]_{z=1.65}}{[\blya\bqso\xi_{\rm L}]_{z=2.3}}
  \frac{1+(\betaciv+\betaq)/3 + \betaciv\betaq/5}
       {1+(\betalya+\betaq)/3 + \betalya\betaq/5}
\end{equation}
where $\xi_{\rm L}$ is the linear correlation function at $10~\hMpc$.
To determine $\bciv(1.65)$ we use the values
$(\betalya,\betaq,\betaciv)=(1.4,0.27,0.5)$,
a growth factor $\xi_{\rm L}(1.65)/\xi_{\rm L}(2.3)=1.5$,
and  the redshift evolution of the quasar
bias from \citet{2005MNRAS.356..415C}, $\bqso(1.65)]/\bqso(2.3)=0.70$.
The resulting value of $\bciv(1.65)/\blya(2.3)$
is multiplied by the measured value of $\blya$ (Table \ref{fitautotable})
and then used as a Gaussian prior in the fits:
$\bciv(2.3)=-0.0224\pm 0.0112$ (and $\bciv<0$), where we have conservatively
doubled the uncertainty in (\ref{lyacivratio}).

To test for systematic errors, the model in
equation (\ref{xitheoryeq})
includes the broadband term $B_A$, a smoothly varying function of
$(\rperp,\rpar)$:
\begin{equation}
B(r,\mu) =
\sum_{j=0}^{\jmax} \sum_{i=\imin}^{\imax} a_{ij} \frac{L_{j}(\mu)}{r^i}
\hspace*{5mm}\; (j\,\rm{even}) ,
\label{xibroadbandeq}
\end{equation}
where the $L_{j}$ are Legendre polynomials.
In our previous studies, $B_A$  had a central role because we did not
attempt to model distortions due to continuum fitting
as we now do with $D_{AA^\prime}$.
We include $B_A$ now as an option to study how the 
fit position of the BAO peak depends on the assumed
form of the smooth component of the correlation function.
By including this function, we can assure that the measurement of the BAO
parameters
is due only to the peak position, and is not significantly
influenced by the assumed form of the smooth component.

\subsection{Fits with the mock data sets}
\label{mockanalysissec}

The 100 mock sets were used to test the fitting procedure and to 
search for possible biases in the fit BAO parameters.
Studies were made on individual mock sets and for the
stack of the 100 sets.  The former test our ability
to reliably extract $(\aperp,\apar)$ in data sets similar
to the BOSS observations.  The latter test the accuracy with
which we model  the analysis procedure including the effects of  weighting,
$(\rperp,\rpar)$ binning, and continuum fitting.

Four types of mocks
were produced and analyzed.  The results are summarized in
table \ref{fitmocktable} and Figs.
\ref{stackedmocksfig} and \ref{stackedmockswithmetalsfig}.
The four types
and the analysis procedure are as follows:
\begin{enumerate}
\item Only \Lya absorption:  in the fits
  the distortion matrix is set  to the identity matrix and
  only the parameters $(\aperp,\apar,\blya, \betalya)$ are fit.
\item A quasar continuum is added:  the
  fits use the distortion matrix and 
  the parameters $(\aperp,\apar,\blya, \betalya)$ are fit.
\item Metal absorption is added: the
  additional parameters are the $(b_i,\beta_i)$ for each metal transition, $i$.
\item High column density systems are added:
  three additional parameters $(\bhcds,\betahcds,\Lhcds)$ are fit.
\end{enumerate}

\begin{figure*}[ht]
\centering
\includegraphics[width=\columnwidth]{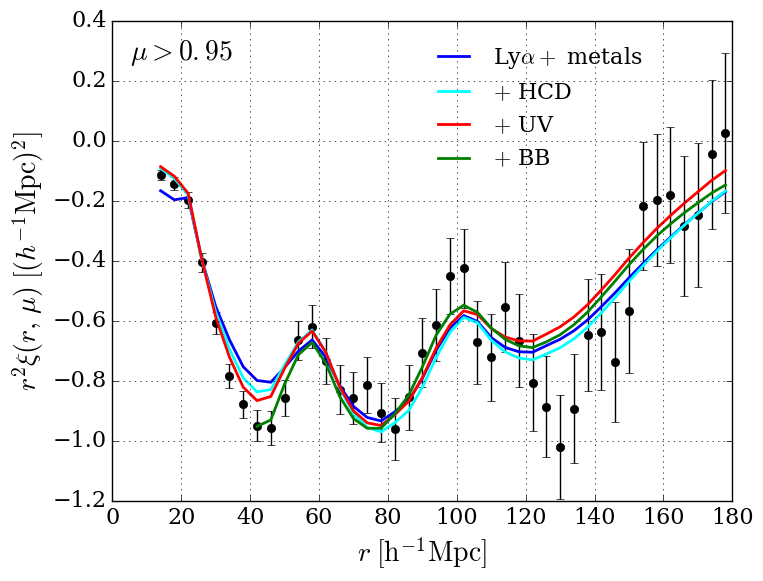}
\includegraphics[width=\columnwidth]{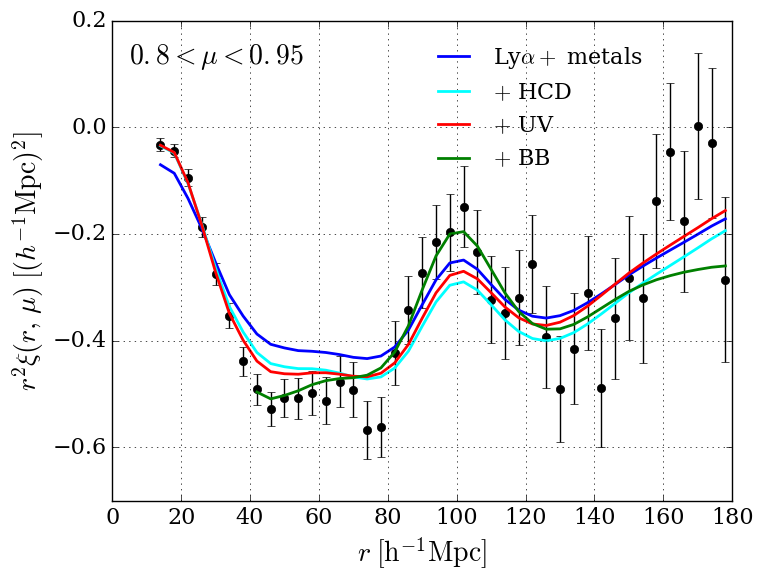}
\includegraphics[width=\columnwidth]{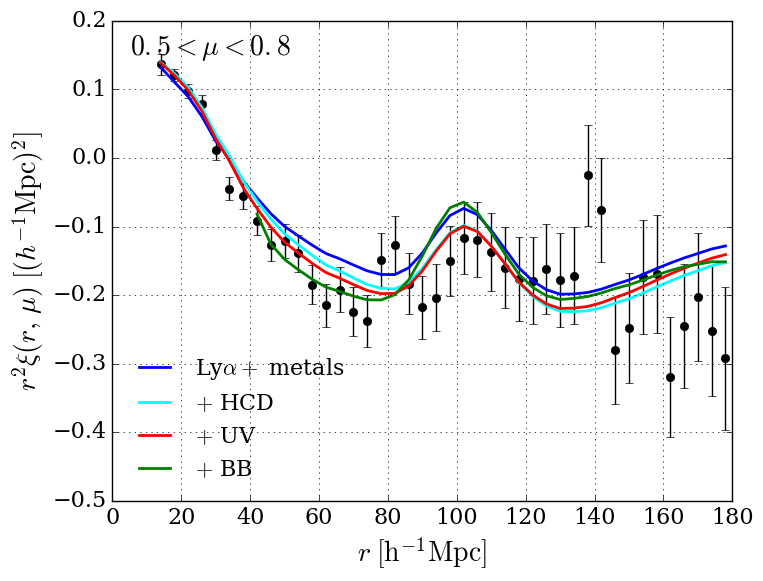}
\includegraphics[width=\columnwidth]{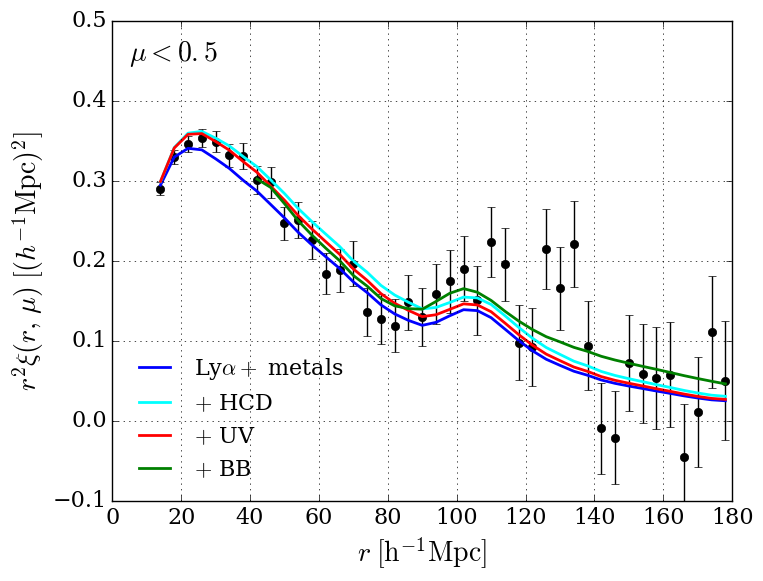}
\caption{Measured correlation function
in four ranges of $\mu$.  The most radial range (top-left $\mu>0.95$) 
has, in addition to the BAO peak at $\sim100~\hMpc$,
a peak at $\sim60~\hMpc$ due to correlated absorption by
\Lya and SiIII(119.0,119.3) at the same physical position.
In the next radial bin (top-right, $0.8<\mu<0.95$) only the BAO
peak is visible.
The lines show fits including successively \Lya and metal absorption,
unidentified HCDs absorption,  UV flux fluctuations, and
a $(\imin,\imax,\jmax)=(0,2,6)$ broadband (BB).
}
\label{wedgesfig}
\end{figure*}

Figure~\ref{stackedmocksfig} shows the correlation
function in four ranges of $\mu$.
The black points are the stack of the
measured correlation function for type 1 mocks
(only \Lya absorption).
The red points are the stacked measurements for type 2 mocks
(quasar continuum added).
The difference between the black and red points shows the effect
of continuum-fitting.
The heavy curves are the best fit model for the cosmology used
to generate the mocks.
The good agreement between the red curve and points indicates
the accuracy of the modeling of the continuum-fitting distortion.

Fits with type 3 mocks (including metals) were used to search for
possible influences of metal absorption on derived BAO parameters.
Metal absorption is mostly confined to the small $\rperp$ region
and this is seen in the two top panels of Fig.~\ref{stackedmockswithmetalsfig}.
The most radial sector, $\mu>0.95$, displays the effects of
metal absorption, most prominently the peak due to SiIIa,b absorption
around $r\sim60~\hMpc$.  The peak is not visible in the next sector,
$0.8<\mu<0.95$.

Two sets of fits for type 4 mocks (HCDs) were also performed.
In one,
HCDs with $\NHI<10^{20}{\rm cm^{-2}}$ were not given the DLA treatment
described in Section \ref{deltaestimationsec}. 
This cut approximates the  threshold
for DLA detection either automatically or visually.
In the second set, only HCDs with $\NHI>10^{21}~{\rm cm^{-2}}$ were
treated specially, 
testing our sensitivity to an underestimation of 
the efficiency to flag DLAs.

Table \ref{fitmocktable} lists the mean fit values for the 100 mocks
and for the 4 types.
Fig.~\ref{mocksoutputfig}
presents the corresponding distributions for type 4 mocks.
For all types, the mean of the best-fit values of  $(\aperp,\apar)$ 
is consistent with unity at the sub-percent level.
Since the precision of our measurement with the data is of order three percent,
this means that the mocks do not indicate any significant bias in
the measurement of the BAO peak
position.

\subsection{Fits of the data with physical correlation function}
\label{dataanalysissec}

\begin{table*}[t]
\centering
\caption{
Results of fits of the data assuming
increasingly complicated forms for $\xicosmo(\rperp,\rpar)$: \Lya absorption only;
including  metal absorption;
including unidentified HCDs; 
and including fluctuations in the ionizing UV flux
(i.e. the complete physical model whose complete set of
parameters is given in Table \ref{paramtable}).
Fits are over the range $10<r<180~\hMpc$.
The bias parameter, $\blya$, refers to  the reference redshift $z=2.3$.
}
\begin{tabular}{l c c c c c}
analysis  
    & $\apar$ 
    & $\aperp$  
    & $\blya(1+\betalya)$
    & $\betalya$
    & $\chi^2_{\rm min}/DOF$, prob  \\
\noalign{\smallskip} 
\hline
\noalign{\smallskip}
\Lya  & $ 1.040  \pm 0.033  $ & $ 0.975  \pm 0.056  $ & $ -0.326  \pm 0.002  $ & $ 1.246  \pm 0.044  $ & $ 1763.8  /(  1590 - 4 )\;\;p=0.001  $  \\ 
+metals  & $ 1.050  \pm 0.035  $ & $ 0.967  \pm 0.054  $ & $ -0.330  \pm 0.002  $ & $ 1.275  \pm 0.045  $ & $ 1644.5  /(  1590 - 9 )\;\;p=0.130  $  \\ 
+HCD  & $ 1.053  \pm 0.036  $ & $ 0.962  \pm 0.054  $ & $ -0.321  \pm 0.003  $ & $ 1.656  \pm 0.086  $ & $ 1561.4  /(  1590 - 12 )\;\;p=0.612  $  \\ 
+UV  & $ 1.053  \pm 0.036  $ & $ 0.965  \pm 0.055  $ & $ -0.326  \pm 0.003  $ & $ 1.666  \pm 0.085  $ & $ 1556.5  /(  1590 - 13 )\;\;p=0.639  $  \\ 
\label{fitautotable}
\end{tabular}
\end{table*}%

\begin{table*}
\begin{center}
\caption{
  Best-fit values of $(\aperp,\apar)$  for fits including metals, HCDs and UV fluctuations,
  with and without power-law broadbands (BBs) of the form
  $(\imin,\imax,\jmax)=(0,2,6)$.
  The fit without broadbands is over $10<r<180~\hMpc$ and those with broadbands
  are over $40<r<180~\hMpc$.
  The broadband fit with  ``physical priors'' requires that the best-fit values
  for parameters
  be within two standard deviations of those found without the broadband
  and imposes $\Apeak=1$.
  The broadband fit with ``no additional priors'' adds no constraints
  beyond fixing the parameters that are fixed in the fit without broadband
  (see Table \ref{paramtable}).
}
\begin{tabular}{l c c c c }
analysis  
  & $\apar$ 
    & $\aperp$  
    & $\apar^{0.7}\aperp^{0.3}$
    & $\chi^2_{\rm min}/DOF$, prob.  \\
\noalign{\smallskip} 
\hline
\noalign{\smallskip}
\vspace*{3mm}
no BB  & $  1.053  ^{+  0.037 \; +  0.077 }_{- 0.035 \; - 0.071 } $ & $  0.965  ^{+  0.061 \; +  0.146 }_{- 0.051 \; - 0.102 } $ & $  1.026  ^{+ 0.025 \; +  0.052 }_{- 0.024 \; - 0.047 }  $ & $  1556.5 /( 1590 - 13 )\;  0.639 $ \\ 
\vspace*{3mm}
BB, physical priors  & $  1.057  ^{+  0.039 \; +  0.081 }_{- 0.037 \; - 0.073 } $ & $  0.962  ^{+  0.063 \; +  0.155 }_{- 0.053 \; - 0.106 } $ & $  1.028  ^{+ 0.026 \; +  0.054 }_{- 0.025 \; - 0.049 }  $ & $  1475.8 /( 1515 - 25 )\;  0.598 $ \\ 
\vspace*{3mm}
BB, no additional priors  & $  1.043  ^{+  0.035 \; +  0.075 }_{- 0.033 \; - 0.068 } $ & $  1.000  ^{+  0.095 \; +  0.346 }_{- 0.060 \; - 0.114 } $ & $  1.030  ^{+ 0.023 \; +  0.063 }_{- 0.021 \; - 0.043 }  $ & $  1472.6 /( 1515 - 26 )\;  0.614 $ \\ 
\label{fitbroadbandtable}
\end{tabular}
\end{center}
\end{table*}%

The data were fit over the range
$10<r<180~\hMpc$
under various hypotheses of increasing complexity
concerning the form of $\xicosmo(\rperp,\rpar)$.
The results are summarized in 
Table \ref{fitautotable}.
The first fit assumes only \Lya absorption, the second
includes absorption by metals, and the third absorption
by unidentified HCDs.
The fourth fit also assumes fluctuation in the flux of
ionizing UV radiation.
Inclusion of each effect yields a significantly better
$\chi^2$;  however, the  values of $(\aperp,\apar)$ do
not change significantly with each step.
The data and fit correlation functions  are presented
for the four angular ranges in Fig.~\ref{wedgesfig}.

  In what follows, we will focus our attention on the fourth
  fit that will be referred to as the ``complete physical model''.
The BAO peak-position parameters for this model are
\begin{equation}
 \apar= 1.053  \pm  0.036  \hspace*{5mm}\Rightarrow \frac{\DHub(z=2.33)}{r_d}= 9.07  \pm  0.31
\label{aparresult}
\end{equation}
\begin{equation}
 \aperp= 0.965  \pm  0.055  \hspace*{5mm}\Rightarrow \frac{\DM(z=2.33)}{r_d}= 37.77  \pm  2.13
\; .
\label{aperpresult}
\end{equation}
where we have used the Planck values from Table \ref{modtable}
that were used in the (angle,redshift) to $(\rperp,\rpar)$ transformation.
The one, two and three standard deviation contours
(corresponding to $\Delta\chi^2=2.3,6.18,11.83$)
are shown as the red contours
in Fig.~\ref{alphadatafig}.
Our result is within one standard deviation of the prediction
of flat-\lcdm model of Table \ref{modtable} that is
in agreement with the Planck data.

  In the next Section we will study the robustness of the determination
  of the BAO-peak position parameters by
  adding power-law broadband terms to the complete physical model.
  We can also estimate
the significance of the BAO peak by modifying this  model by
 setting $\Apeak=0$.
 This degrades the quality of the fit by $\Delta\chi^2=27.7$ corresponding
 to a BAO peak detection of $5.2\sigma$.  Adding broadband terms reduces
 this to $\Delta\chi^2=21.8$ ($4.7\sigma$).

Other than $(\aperp,\apar)$, the  parameters of the model listed in
Table \ref{paramtable} have no direct cosmological implications.
However, the fact that
the best-fit values are physically reasonable
reinforces our confidence in the model and, therefore, 
on the cosmological conclusions.
Most importantly,
the best-fit values of $(\blya,\betalya)$ in table \ref{fitautotable}
are both relatively stable and near the expected values.
For example, the model of \citet{2015JCAP...12..017A}
predicts $\betalya\sim1.4$ and $|\blya|(1+\betalya)$ in the
range 0.25 to 0.33.
In Appendix \ref{biassec} we discuss in more detail the comparison
of our measurement of $(\blya,\betalya)$ with theory and
with previous measurements.

The best-fit values of most of the other parameters
in Table \ref{paramtable} are near those found in the mock fits.
For the metal biases, this only means that we have placed a reasonable
amount of metal absorption in the mock spectra.
Of more significance is 
the value of $\Lhcds=(24.3\pm1.13)~\hMpc$ which is, as expected, roughly the
width of a DLA of $\NHI=10^{20}~{\rm cm^2}$, corresponding
to our estimated threshold for good efficiency
in identifying DLAs.  In the mocks, best-fit values
of $\Lhcds$ when masking only those DLAs with $\NHI>10^{20}~{\rm cm^2}$
cluster near $\Lhcds\sim25$ in about half the fits.
For the other half, the fits find $\Lhcds\sim0$, corresponding to the
addition of a second \Lya component but with a different $\beta$.
This indicates that more than one way exists to model
sufficiently well the correlation
function with unidentified HCDs.
The bias parameter found in the data, $\bhcds=-0.0288\pm0.0043$,
is about a factor of 7 higher than that found in those mocks
that find $\Lhcds\sim25$.  This indicates that
there are unidentified HCDs in the data with $\NHI>10^{20}{\rm cm^2}$
 and or that the HCDs that
 we have placed in the mocks are less biased than in reality.

 We have not generated mocks with fluctuations in the flux of
 ionizing radiation.  However,
 we note that the bias parameter $b_\Gamma=0.1125\pm0.0511$ is near
 the value of $0.13$ suggested by \citet{2014MNRAS.442..187G}.

\subsection{Fits of the data with power-law broadbands added}

In addition to statistical uncertainties,
it is important to determine to what extent
the derived values of $(\aperp,\apar)$
are influenced by the assumed form of $\xicosmo$.
The amplitude and form of the metal
components are well understood and  constrained by
the low-$\rperp$ bins.
However,
the HCD and UV parameters 
have a more diffuse effect on the correlation
function and are not strongly constrained by
independent data.
For example, 
the UV background fluctuation model
we have used includes only fluctuations that follow the
underlying density field.
Additional correlations \citep{2014MNRAS.442..187G}
due to the discrete nature of sources of UV radiation
would modify the absorption
in neighboring forests since they are influenced by the same
discrete sources.
This ``shot-noise''
term would give an additive, isotropic and positive contribution
to the \Lya auto-correlation function.
Its shape and amplitude as a function of comoving separation are
however poorly known as they depend on the quasar luminosity distribution
and variability.

To study the effects of possible
uncertainties in the form of the non-BAO component, $\xicosmo_{\rm smooth}$,
we have performed fits adding to the complete physical model
broadband functions in the form
of power laws (\ref{xibroadbandeq}) defined by $(\imin,\imax,\jmax)$. 
With the addition of new components of $\xicosmo_{\rm smooth}$, the
connection between the BAO peak amplitude and the smooth components
is lost so we also performed fits
releasing the constraint $\Apeak=1$.

The parameters of slowly varying broadband forms are
necessarily somewhat degenerate with the parameters of the
slowly varying part of the physical model: 
$(b,\beta)$ and the parameters describing HCDs and UV fluctuations.
As such, fits with power-law broadband terms can produce unphysical
best-fit values of these parameters.
Conversely, we expect that a small number of power-law terms
will not significantly affect the BAO-peak parameters.
Fits with the mock spectra indicate that this is the case
if the power-law broadbands are restricted to the ranges
$-2\leq\imax\leq3$, $(\imax-\imin)\leq3$
and $\jmax\leq6$.
Of course in the case of the mocks we know precisely the form
of the physical correlation function while for the data we do not.
Shifts of ($\aperp,\apar$) in the data when using
power-law terms could indicate that the underlying
physical correlation function has not been modeled with sufficient
precision.

We want to ensure that the power-law terms model variations of
the slowly-varying part of the correlation function under the
BAO peak.
We therefore perform these fits only over the restricted range
$40<r<180~\hMpc$, avoiding undue influence of the $10<r<40~\hMpc$ range
on the amplitudes of the power laws.
Using this range, the fit with the complete physical
model without power-law
broadbands yields
BAO parameters indistinguishable from those found over
the full range $10<r<180~\hMpc$  (Table \ref{fitautotable}).

\begin{figure}[tb] 
\includegraphics[width=\columnwidth]{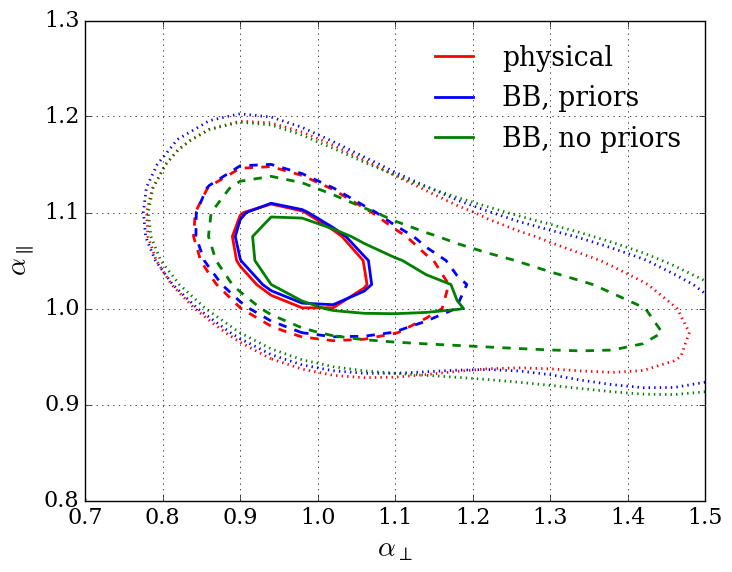}
\caption{
  Contours for $(\aperp,\apar)$ at the (68.3, 95.45, 99.73)\% CL
  (solid, dashed, dotted).
  The red contours are for the physical model with HCDs and UV fluctuations
  from Tables \ref{paramtable} and \ref{fitautotable}.
  The blue (green) contours are for the $(\imin,\imax,\jmax)=(0,2,6)$
  power-law broadband with (without) priors on the 
  parameters of the physical correlation function.
}
\label{alphadatafig}
\end{figure}

To choose appropriate values of $(\imin,\imax,\jmax)$, we note that
apart from the BAO peak, $r^2\xi$ does not vary wildly
over the range  $40<r<180~\hMpc$. A reasonable choice
is therefore $(\imin,\imax)=(0,2)$ corresponding to a parabola
and which amounts to freeing the
value and first two derivatives of $r^2\xi_{\rm smooth}$ underneath the BAO peak.
A reasonable choice for $\jmax$ is $\jmax=6$ corresponding
to approximately independent broadbands in each of the four angular ranges in
Fig. \ref{wedgesfig}.
The best-fit correlation function for 
$(\imin,\imax,\jmax)=(0,2,6)$ and $\Apeak$ free 
is shown in the figure and the
BAO-peak parameters on the third line of  Table \ref{fitbroadbandtable}.
The best-fit value of $\apar$ is $\sim1\%$ (i.e. $\sim0.3\sigma$)
lower than the 
value for the complete physical model  and uncertainty is unchanged.
The value of $\aperp$ is about 3\% higher
than the standard value with the error increasing slightly
from $\sim0.055$ to $\sim0.075$.
The one- and two-standard deviation contours are shown in green
in Fig. \ref{alphadatafig}. 
We emphasize that these results are obtained without placing
any priors on the physical parameters, other than those already
used in the standard fits.

We have experimented with other combinations of $(\imin,\imax,\jmax)$ in the
range $\jmax\leq 6$, $-2<\imin<0$, $-1<\imax<3$.
While most combinations yield values of  $(\aperp,\apar)$ that
are close to the nominal values, there are
cases where there are significant changes.
These cases always correspond to fits yielding unphysical
values for $(b,\beta)$ of one or more components.
In all cases, the large changes in $(\aperp,\apar)$ can be
avoided by
placing a ``physical priors'' on the $(b,\beta)$ that requires them
to take values within two standard deviations of
those found in the standard fits
over the range $10<r<180~\hMpc$ and imposing $\Apeak=1$.
The results for the $(0,2,6)$ model are
listed oneline two of Table \ref{fitbroadbandtable} and the contours
shown in blue in Fig. \ref{alphadatafig}.

\subsection{Summary of data fits}

Table \ref{fitbroadbandtable} and Fig. \ref{alphadatafig} summarizes
the BAO parameters found for fits with our most complete physical
model of the correlation function, with and without power-law broadbands.
The results of the fits with physical priors imply  that the
uncertainty in the form of $\xicosmo_{\rm smooth}$ does not
translate into a significant uncertainty in $(\aperp,\apar)$.
Because of the priors placed on the physical parameters,
this conclusion follows only if our basic physical model is not
too far removed from reality.
Without such priors, the uncertainty of $\aperp$ can be significantly
increased (line 3 of Table \ref{fitbroadbandtable}).

The limits on $\apar$ are relatively stable with the inclusion
of broadbands.
A quantity that is more robust than
either $\apar$ or $\aperp$ is $\apar^\gamma\aperp^{1-\gamma}$ with
the exponent $\gamma$ chosen to minimize the variance.
For the observed correlation and variances of $(\aperp,\apar)$,
one calculates $\gamma\sim0.7$ with
\begin{equation}
  \apar^{0.7}\aperp^{0.3}=1.028  \pm 0.026 \;,
\end{equation}
where the uncertainty is chosen to cover the one-standard-deviations
of the three fits in Table \ref{fitbroadbandtable}.
Using the Planck model values from Table \ref{modtable}, we deduce
\begin{equation}
  \frac{\DHub^{0.7}\DM^{0.3}}{r_d} =13.94  \pm 0.35 
  \hspace*{5mm}{\rm at}\;z=2.33\;.
\end{equation}
This result is largely independent of the addition of power-law
broadbands.  It contains most of the cosmological constraints
of our measurement if used in combination with constraints
from the forest-quasar cross-correlation, since the latter
has more constraining power on $\DM/r_d$ than the auto-correlation
result.

\subsection{Comparison with previous results}

We can compare the results (\ref{aparresult}) and (\ref{aperpresult})
with those of \citet{2015A&A...574A..59D}: 
$\DHub(2.34)/r_d=9.18\pm0.28$,
$\DM/r_d=37.67\pm2.17$
and
$\DHub^{0.7}\DM^{0.3}/r_d=14.02\pm0.3$.
The change in $\DHub/r_d$ is $\sim0.5\sigma$.
To determine whether this change
is due to the change in analysis procedure
or in the data set we performed fits using only data
used in the DR11 analysis.
The results for a HCDS,UV analysis are
$\DHub(2.33)/r_d=9.14\pm0.28$ and
$\DM(2.34)/r_d=37.54\pm2.17$,
within $0.15\sigma$ and $0.06\sigma$ of the
results of \citet{2015A&A...574A..59D}.
This result implies that most of the change is due to the change
in the data set.

The $0.5\sigma$ change in  $\DHub/r_d$
is typical of the variations we can expect when increasing the sample
size from DR11 to DR12.  We confirmed this expectations
by using jackknife samples
of DR12 to simulate DR11 data sets and observing changes in $\DHub/r_d$.
Roughly 30\%
of the jackknife samples had a change
as large as that observed for DR11 and we therefore conclude that the DR12-DR11
difference is primarily statistical.

\begin{table}
\begin{center}
  \caption{Systematic shifts of the \Lya bias parameters
    and BAO peak-position parameters due to various spurious
  correlations discussed in Sect. \ref{instcorrsec}.
Individual shifts in $(\aperp,\apar)$  are less than $\sigma/10$ and
  the total shifts are  consistent with zero.
  Their uncertainty is dominated by the statistical uncertainty on the sky
  model.}
  \begin{tabular}{l|cccc}
  & $\betalya$ & $ b(1+\beta) $ & $\apar$ & $\aperp$\\
\hline
Sky model noise & $-0.026$ & $-0.002$ & $<0.001$ & $<0.001$ \\
Calibration noise & $+0.047$ & $+0.002$ & $<0.001$ & $+0.001$ \\
Fiber cross-talk & $+0.003$ & $<0.001$ & $<0.001$ & $<0.001$ \\
ISM absorption & $+0.003$ & $<0.001$ & $<0.001$ & $<0.001$ \\
\hline
Sum & $+0.027$ & $<0.001$ & $+0.001$ & $<0.001$ \\
\hline
Quadratic sum & $+0.054$ & $+0.002$ & $<0.001$ & $+0.001$ \\
\end{tabular}
    \label{table:pipeline_systematics}
\end{center}
\end{table}

\begin{figure}
\centering
\includegraphics[width=\columnwidth]{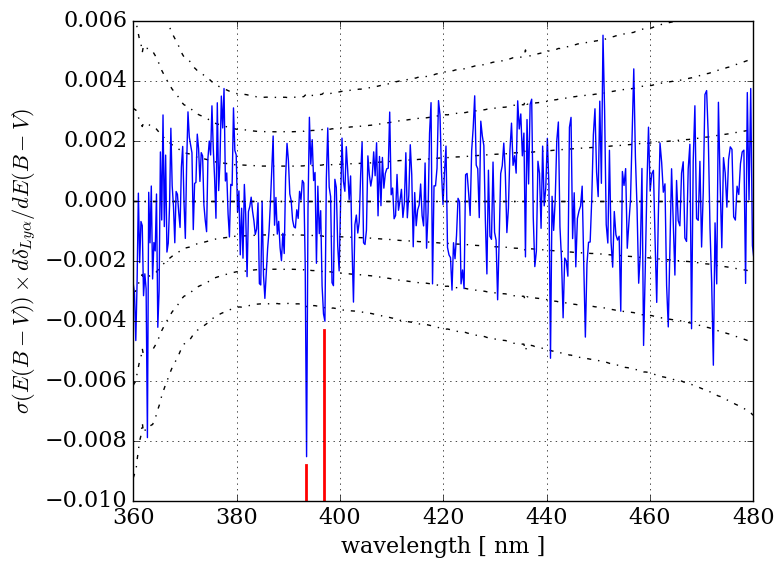}
\caption{Derivative of  the normalized \lya-forest flux-decrement,
  $\delta_{LyaF}$, with respect to the
  Milky Way dust column density normalized to its
  dispersion. 
The dashed curves represent the 0,1,2 and 3-$\sigma$ uncertainties.
The red vertical lines mark the positions of the calcium H and K lines.
\label{fig:lya_delta_vs_extinction}}
\end{figure}

\section{Systematic Errors in the BAO-peak position}
\label{systsec}

Systematic errors in $(\aperp,\apar)$ can  
result if the measured correlations function
is not sufficiently well modeled by the fitting
function (\ref{xitheoryeq}).
Such effects can be introduced by the pipeline and
analysis procedures, by improper modeling for the flux-transmission
field, and by astrophysical absorption not related to cosmological
large-scale structure.

Spurious correlations
introduced by the pipeline
were discussed in 
Section \ref{instcorrsec}.  The most important correlations
were confined to the lowest $\rpar$ bin with negligible
effects on the fitting of the correlation function.
No evidence was found for any additional instrumental
correlations that could have an influence of the measurement
of  $(\aperp,\apar)$.
Since our model of pipeline-induced correlations introduces
no scales near the BAO scale, it is not surprising that
it predicts negligible effects on the BAO scale.
Upper limits on the pipeline-induced shifts
in $(\aperp,\apar)$ are  given
in Table \ref{table:pipeline_systematics}.

Biases could also be introduced at the analysis level, for example
through the small correlations between weights and the measured
$\dqlam$.
These correlations are estimated to have a negligible
impact on the BAO parameters.
If they did not, they would have produced biased estimates of $(\aperp,\apar)$
in the mock data sets, while no such biases are seen.

The second type of systematic error  would result from
incorrect modeling of the various contributions to the
physical correlations.
The physical correlations are parameterized by the $(b_i,\beta_i)$
for each transition.  These  parameters are marginalized over and
their uncertainties therefore contribute to the reported statistical
errors on $(\aperp,\apar)$.

Modeling of the contribution
of unidentified HCDs and of UV fluctuations also contain free-parameters
that are marginalized over.
We have verified that the best-fit BAO parameters are insensitive
to assumption about non-linearities and redshift evolution
of the bias parameters.
We have also investigated  possible influence of foreground
absorbers in addition to CIV.
Finally, we have included power-law broadband
parameters to place limits on our sensitivity to unmodeled
slowly-varying contributions.

In our physical model of the correlation function
we have ignored two effects that could potentially bias our results
at some level.
First, we have assumed that the lines of sight sample random positions
of the Universe, while in a real survey we only have information
in pixels in front of a quasar.
This makes our measured correlation sensitive to the three-point
function of quasar-lya-lya,
but given that this is averaged over pixel pairs at
different separations from the quasars we expect this
contamination to be very smooth on BAO scales. 
The second effect is that the transmitted flux fraction
could be affected by the relative velocity between baryons and
dark matter in the early Universe \citep{2010PhRvD..82h3520T}.
These relative velocities are quite small in the redshift range of interest,
but it is possible that the different velocities at early times left an
imprint in the distribution of neutral hydrogen even at low redshift,
similar to what has been suggested in the case of galaxy clustering
\citep{2010JCAP...11..007D,2011JCAP...07..018Y,2016PhRvL.116l1303B}.
In order to study this effect in the Lyman alpha forest we would
require detailed hydrodynamical simulations, and this is beyond
the scope of this paper. Given that the density of quasars should
have a different dependency on the relative velocity, differences
in the BAO scale measured from the auto-correlation and from the
cross-correlation with quasars \citep{2014JCAP...05..027F}
could be
used to constraint the amplitude of this effect.

The third type of systematic
error would be due to correlations of non-cosmological origin.
One possibility is
atomic or molecular absorption in the Galactic interstellar medium (ISM). 
The average ISM absorption is included in the fit mean flux transmission, 
or the subsequent subtraction of the average $\dqlam$.
However, residuals correlated across the sky might still be present. 
Current studies show that the ISM
 absorption is highly correlated with dust column density. 
We hence use the estimated Milky-Way dust extinction 
as a proxy for potential ISM absorption and fit for the 
correlation of 
$\dqlam$
with this dust
extinction parameter.
The resulting correlation is shown in Fig.~\ref{fig:lya_delta_vs_extinction}. 
The Ca H and K lines are the only known ISM lines to appear significantly. 
Therefore ISM absorption 
has a negligible impact on the \Lya correlation function. 
We have verified this conclusion
by computing the auto-correlation function of the quantity 
$E(B-V) \times d\dqlam/d E(B-V)$.

A final source of correlations that we have considered would originate
in our use of a unique quasar-continuum template which we adapt
to individual quasars by the $(a_q,b_q)$ parameters of
equation \ref{templatewarpeq}.
Templates could be improved by making them depend on
a non-forest variable, like
quasar redshift or CIV emission-line strength.
These improved templates could change the derived flux correlation
function if the non-forest variables of the quasars are themselves
correlated  at large distances.
This is not expected to be the case of CIV line strength since
this strength is believed to be due to local conditions and
therefore not correlated at large distances.
We have tested this by using two templates, one for strong CIV
emitters (relative to side-bands) and one for weak emitters.
This slightly improves the pixel variance (by 3\%) but
makes no significant change in the measured correlation
function and less than $0.1\%$ difference in the  derived $(\aperp,\apar)$.

\begin{figure}[t] 
\includegraphics[width=.95\columnwidth]{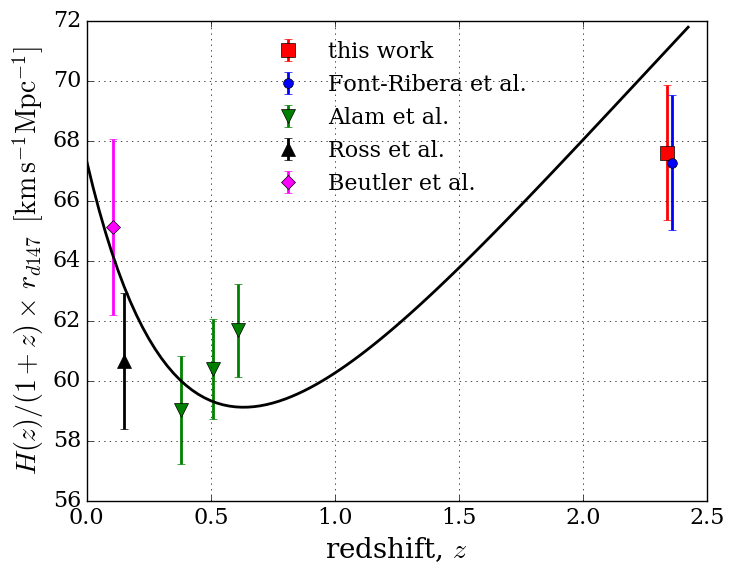}
\caption{Comoving expansion
  rate, $r_dH(z)/(1+z)$, measured with BAO, with $r_d$ in units
of 147.3~Mpc.
The red square is the present measurement at $z=2.33$.
Also shown are measurements using
the \lya-quasar cross correlation at
$z=2.36$ \citep{2014JCAP...05..027F},
and  galaxy correlations at
$0.35<z<0.65$ \citep{2016arXiv160703155A},
at $z=0.15$ \citep{2015MNRAS.449..835R}, and
at $z=0.11$ \citep{2011MNRAS.416.3017B}. 
The points at $z=0.11$  
and $z=0.15$ use
the SNIa measurement of $q_0$ \citep{2014A&A...568A..22B}
to convert the measured $\DV(z)\propto \DM^{2/3}\DHub^{1/3}$ to $D_H(z)$.
The black line is the prediction of the Planck flat \lcdm model
(table \ref{modtable}).
}
\label{adotfig}
\end{figure}

\begin{figure}[tb]
\includegraphics[width=\columnwidth]{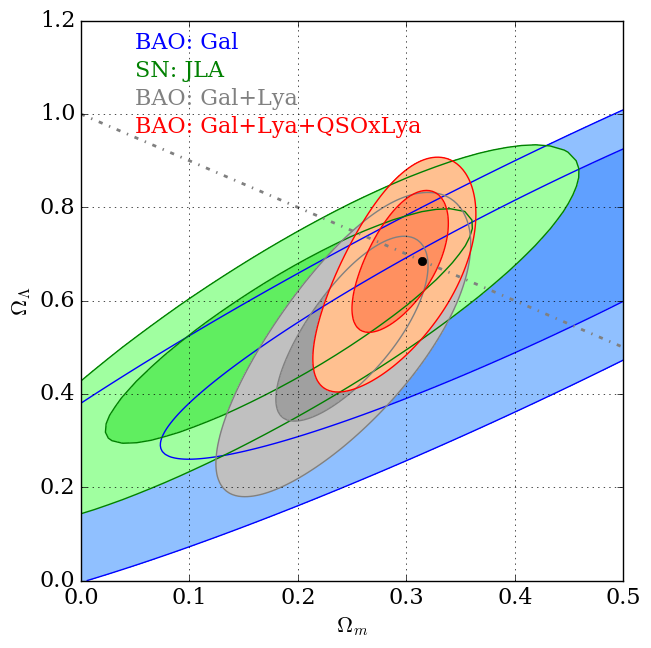}
\caption{One- and two-standard deviation
  constraints on the o\lcdm parameters
  $(\om,\ol)$ using BAO measurements of
  $\DM(z)/r_d$ and $\DHub(z)/r_d$ without CMB measurements, i.e.
  without imposing the CMB value of $r_d$.
  The red contours use
  the work presented here
  (line 1 of Table \ref{fitbroadbandtable}),
  the $z<0.8$ galaxy
  data \citep{2016arXiv160703155A,2011MNRAS.416.3017B,2015MNRAS.449..835R}
  and the quasar-forest cross-correlation \citep{2014JCAP...05..027F}.
  The gray contours exclude the quasar-forest cross-correlation.
  The blue contours use only the $z<0.8$ galaxy data and the 
  green contours show the constraints derived from the SNIa
  Hubble diagram \citep{2014A&A...568A..22B}.
  The black dot is the position of the flat-\lcdm model that
  describes CMB data \citep{2016A&A...594A..13P}.
}
\label{omolfig}
\end{figure}

A continuum that evolves with redshift could be more problematic
because the correlation function measurement  necessarily requires quasar
pairs that are near in redshift.
We searched for the effect of such evolution by using a
quasar continuum template
that evolves linearly in redshift.
The results
do not differ significantly from those for the redshift-independent
template.

We have searched for unsuspected systematics
by dividing the data into two redshift bins, $z<2.295$ and $z>2.295$.
The 95\%CL ranges on   $\apar$  in the high and low redshift
ranges are
$0.90<\apar<1.094$ and $1.02<\apar<1.15$.
For $\aperp$ the ranges are
$0.74<\aperp<1.5$ and $0.86<\aperp<1.01$.
As with the DR11 results of \citet{2015A&A...574A..59D},
the results are consistent
with the expected absence of  evolution at the two-standard-deviation level.

\section{Cosmological implications}
\label{cosmosec}

Our measurements of $\DM/r_d$ and $\DHub/r_d$ at $z=2.33$
are in good agreement with the values predicted by
the flat-\lcdm model consistent with the  CMB data.
The precision is, however,
less than that of the BOSS galaxy BAO data \citep{2016arXiv160703155A},
and
for simple cosmological models,
parameter constraints based on combinations of CMB and BAO data
are not strongly changed by our results.
Our constraints  however,
can provide tests of more general models with
complicated expansion dynamics \citep{2015PhRvD..92l3516A}.

When combined with  other BAO data, our measurement  
provides interesting results using only low-redshift data.
For example, the low-redshift expansion history can be
directly measured through the measurements of $\DHub(z)/r_d$.
The Hubble distance  (\ref{aparresult}) can be written as
\begin{equation}
H(z=2.33) = (224.\pm8.)~{\rm km\,s^{-1}Mpc^{-1}}\frac{147.33~{\rm Mpc}}{r_d}
\end{equation}
Figure~\ref{adotfig}  displays this result along with those of
\citet{2016arXiv160703155A,2011MNRAS.416.3017B,2015MNRAS.449..835R}
and \citet{2014JCAP...05..027F}.
The expected
phases of acceleration at $z<0.7$ and deceleration at
$z>0.7$ are striking.

Combining $\DHub/r_d$ and  $\DM/r_d$ measurements,
the cosmological parameters can be measured 
independently of CMB data.
Non-flat \lcdm models (o\lcdm) have three parameters $(\om,\ol,H_0r_d)$
that determine $\DM/r_d$ and $\DHub/r_d$  and
constraints on $(\om,\ol)$ can be derived by marginalizing
over $H_0r_d$. 
This procedure is equivalent to the use of SNIa to
measure  $(\om,\ol)$ by marginalizing over the parameters
that determine supernova luminosities.
Fig.~\ref{omolfig} shows the constraints derived using various
combinations of BAO results.
By themselves, the low-redshift BAO data
provide no strong  evidence for $\ol>0$.
The inclusion of the \lya-forest data results in $\sim10\%$ precision
on $(\om,\ol)$:
\begin{equation}
\om=0.296 \pm 0.029  \hspace*{10mm}
\ol=0.699 \pm 0.100
\end{equation}
consistent with spatial flatness:
\begin{equation}
\Omega_k = -0.002 \pm 0.119 \;.
\end{equation}
The fit has $\chi^2=12.27$ for 12 data points and three parameters.
The values of $(\om,\ol)$ agree well with the flat-\lcdm
values found using CMB anisotropies $(\om,\ol)=(0.315,0.685)$
\citep{2016A&A...594A..13P}.

\section{Conclusions}
\label{conclusionssec}

In this paper, we present an analysis of the complete SDSS-3
(DR12) set of \Lya forests. The position of the BAO peak in the
flux correlation function is in good agreement with that expected
in flat-\lcdm models favored by CMB anisotropies.

Our analysis represents a significant improvement over our previous
work, both in the understanding of instrumental effects on  the
flux transmission and in the data analysis.
The modeling of the distortion of the correlation function due
to continuum fitting allows fitting the
entire correlation function using a physical model without
resorting to arbitrary ``broadband terms''.
In the future, this approach
may allow alternative ways to study the cosmological parameters, for
instance by eliminating the need to separate the correlation
function into ``smooth'' and ``peak'' components.

Now that we have direct access to the full correlation function,
further improvements in the analysis will profit from better
modeling of absorption mechanisms.  In particular, it will be important
to fully use external constraints on HCD and metal absorption.

The improvements in modeling and data analysis that we have implemented
here will clearly be important for future projects with
reduced statistical errors.  For example, the
DESI project \citep{2016arXiv161100036D} will have up to 15 times
the number of forest-pixel pairs resulting in a significant decrease
of statistical errors of the correlation function.
The understanding the physics of the
broadband correlation function may be the limiting factor in
the extraction of BAO parameters.

The cosmological information from the \lya-forest auto-correlation function
studied here is complemented by the study of the cross-correlation
between the forest and neighboring quasars.
Results and implications of a new measurement of this function will
be presented in a forthcoming publication
(du Mas des Bourboux et al., in preparation).

\begin{acknowledgements}
Funding for SDSS-III has been provided by the Alfred~P. Sloan Foundation, the Participating Institutions, the National Science Foundation, and the U.S. Department of Energy Office of Science. The SDSS-III web site is http://www.sdss3.org/.

SDSS-III is managed by the Astrophysical Research Consortium for the Participating Institutions of the SDSS-III Collaboration including the University of Arizona, the Brazilian Participation Group, Brookhaven National Laboratory, Carnegie Mellon University, University of Florida, the French Participation Group, the German Participation Group, Harvard University, the Instituto de Astrofisica de Canarias, the Michigan State/Notre Dame/JINA Participation Group, Johns Hopkins University, Lawrence Berkeley National Laboratory, Max Planck Institute for Astrophysics, Max Planck Institute for Extraterrestrial Physics, New Mexico State University, New York University, Ohio State University, Pennsylvania State University, University of Portsmouth, Princeton University, the Spanish Participation Group, University of Tokyo, University of Utah, Vanderbilt University, University of Virginia, University of Washington, and Yale University.

The French Participation Group of SDSS-III was supported by the
Agence Nationale de la
Recherche under contracts ANR-08-BLAN-0222 and ANR-12-BS05-0015-01.
NB and JG acknowledge support from the French Programme National Cosmologie et Galaxies.
MB, MMP and IP were supported by the A*MIDEX project (ANR- 11-IDEX-0001-02) funded by the “Investissements d’Avenir” French Government program, managed by the French National Research Agency (ANR), and by ANR under contract ANR-14-ACHN-0021.
The work of JB and KD was supported in part by the U.S. Department of Energy,
Office of Science, Office of High Energy Physics,
under Award Number DE-SC0009959.
N.P.R acknowledges support from the S.T.F.C. and the Ernest Rutherford
Fellowship scheme.

\end{acknowledgements}

\appendix

\section{Modifications applied to the BOSS spectral extraction pipeline}
\label{sec:new_pipeline}

Each of the two  BOSS spectrograph slitheads consists of 500 fibers organized in 25 aligned blocks (or bundles) of 20 fibers (see~\citealt{2013AJ....146...32S} for a description of the BOSS spectrographs). Fibers in a bundle are separated by 260~$\mu$m whereas the separation between fibers of adjacent bundles is 624~$\mu$m.  
This design allows one to treat independently each fiber bundle in the data reduction:  whereas the overlap of light coming from fibers of the same bundle must be accounted for, one can safely ignore
interbundle contamination.

The spectral processing consists in a row-by-row extraction from each CCD,
where the signal of a row of pixels illuminated by a fiber bundle is fit independently from the rest of the data (wavelength are dispersed approximately along CCD columns).
The extraction consists in minimizing the following quantity 
$$ \chi^2 = \sum_{pixel\ i} w_{i} \left( D_{i} - \sum_{fiber\ f} \hat{N}_{f} \, \hat{P}_{f,i} - \sum_k a_k i^k \right)^2$$
where $D$ are the pixel values (after preprocessing, i.e. after the subtraction of the bias, conversion from digital units to electrons, and flat-fielding), $w$ the weights assigned to each pixel, $P$ the true cross-dispersion PSF profile
of a fiber and $\hat{P}$ the profile used (a Gaussian).
The quantity $\hat{N}$ is the number of collected electrons from a fiber, to be estimated. The last term is a low-order polynomial contribution used to fit any residual light in the row; it  compensates both for scattered light and the imperfect modeling of the cross-dispersion profile tails.

For the extraction to be optimal, one must have $w_{i} = var_{i}^{-1}$, where  $var_{i}$ is the variance of $D_{i}$. 
This pixel variance is, however, not known a priori.
In the BOSS DR12 pipeline, the pixel data were directly used as an estimate of the expected signal: 
$var_{i} = var_{read} + D_{i}$ if $D_{i}>0$ and $var = var_{read}$ if not, where $var_{read}$ is the readout noise variance, and
 $D_{i}$ is here an estimate of the Poisson noise.

This use of the data to determine the variance results in a biased
estimate of the flux.
As an illustration, if one considers a single fiber extraction, 
without the background polynomial
contribution, the $\chi^2$ minimization yields
$$ \hat{N} = \frac{ \sum_{i} w_{i} D_{i} \hat{P}_{i} }{ \sum_{i} w_{i} \hat{P}_{i}^2 }$$
To first order, assuming $D_i= N P_i + \delta_i >0$, and noting $\bar{w}_i = 1/var(\delta_{i})$ the expectation value of $w_i$,  one has
$w_{i} \simeq \bar{w}_{i} (1 - \bar{w}_{i} \delta_i)$
and
\begin{equation}
\hat{N} \simeq  N \frac{ \sum_{i} \bar{w}_{i} P_i \hat{P}_{i} }{ \sum_{i} \bar{w}_{i} \hat{P}_{i}^2 } - \frac{ \sum_{i} \bar{w}_i \hat{P}_{i} }{ \sum_{i} \bar{w}_{i} \hat{P}_{i}^2} \label{eq:bias}
\end{equation}

The first term produces an unbiased estimate of $N$ if $\hat{P}=P$,
which is not strictly the case. This bias is in principle corrected for 
by the calibration, but since the weights $\bar{w}$ depend on the 
amount of signal, the calibration obtained with bright spectra does 
not perfectly correct the bias of faint spectra. The second term 
is a systematic underestimation of the signal; it can be interpreted 
as an effective number of pixels contributing to the measurement 
(if the profile is normalized, with $\sum_i P_i = 1$).

The fractional bias is more important at low fluxes, and as a consequence  
i) this bias has a non trivial wavelength dependence correlated with 
bright sky emission lines and ii) its amplitude relative to the calibrated 
flux depends on the throughput and hence the observation conditions. 
Those two effects could lead to correlated biases in the \Lya forest.

 
We have rerun the data processing using $w_i=1/var_{read}$ 
(this weight is in practice not a constant quantity per CCD amplifier 
because of the flat-field correction). We also improved the outlier 
rejection procedure that discards unmasked pixels affected by cosmic rays.

This new approach to extraction is unbiased but less optimal.
Expressing this procedure using
a matrix representation, if we define $X$ as the parameters 
of the model (fluxes and polynomial corrections) such that 
we can interpret the data as $D = H X + \delta$ (where $\delta$ is the noise), 
the optimal extraction results in a covariance of the parameters 
$C_X = ( \sum_i w_i H_i H_i^T)^{-1}$, whereas the sub-optimal extraction 
produces a larger covariance $C_X' = C_X (\sum w_i^2 var_i H_i H_i^T) C_X $.
We use this latter expression in the subsequent steps of the pipeline, 
using the estimated flux (and not directly the pixel values) to evaluate 
the Poisson noise contribution to $var_i$.

\begin{figure}[tb]
\centering
\includegraphics[width=\columnwidth]{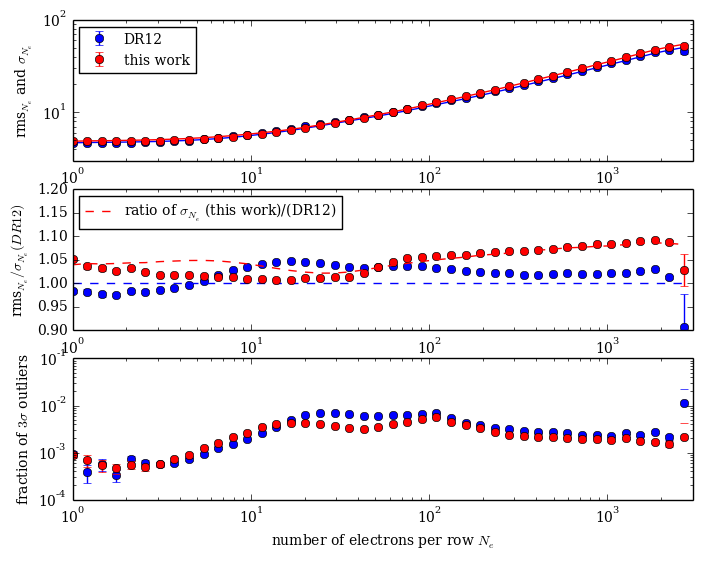}
\caption{Comparison of flux measurement uncertainties between the DR12 data 
reduction (in blue) and this new extraction (in red). This analysis is 
based on the data from camera b1, plate 7339, where many observations 
of the same targets were performed. The top panel shows the empirical 
r.m.s. (circles) among the various observations, and the corresponding 
predicted uncertainty (lines) of the number of electrons per CCD row 
($N_e$) as a function of $N_e$. The second panel is the same quantity 
normalized by the predicted uncertainty of DR12. The bottom panel displays
the fraction of 3~$\sigma$ outliers that were discarded in the two 
other panels. As expected, there is a larger dispersion with the new, 
less optimal reduction at high flux. The larger dispersion at low flux 
is not easy to understand; it is  probably a consequence of the biased 
estimator in DR12, where negative statistical fluctuations are given
a high weight.\label{fig:rms_new_prod}}
\end{figure}

The ratio of uncertainties does not exceed 5\% over the dynamic range of 
BOSS observations of quasars. It reaches a higher value of 10\% for the
brightest measurements (see Fig.~\ref{fig:rms_new_prod}).

Subsequent steps of the data processing (fiber flat-field, sky spectrum model, 
photometric calibration) also involve the weighted average of the data.
Similarly to the extraction, using directly the flux inverse variances 
as weights results in a bias because
of the flux-inverse-variance correlation.
We hence took 
care of replacing the inverse variances of spectra by their average over 
many wavelength (using a spline fit) to mitigate this bias and
to simultaneously
minimize the loss of optimality. Those changes have been 
incorporated in the pipeline used for the
SDSS data release DR13 \citep{2016arXiv160802013S}.

\section{The bias parameters of the \Lya forest}
\label{biassec}

   The physical model of the \Lya forest autocorrelation function involves
two independent parameters for the linear biasing: the density bias factor
$\blya$, and the redshift distortion factor $\betalya$. The values for our best fit
with errors are given in Table 5. As usual, we give the value of
$\blya(1+\betalya)$
because its error is smaller than the error for $\blya$ and less correlated with
$\betalya$ (see \citet{2011JCAP...09..001S}).
Some variations of these parameters occur
as the model is improved to include the effects of metal lines, and possible
effects of HCDs and ionizing intensity fluctuations. As expected, there is
a degeneracy in the values of various parameters of the model that affect
broadband terms. In particular, the value of $\betalya$ shows a substantial
change
from 1.2 to 1.7 depending on the model that is used. The first model,
assuming
that only \Lya forest correlation is present with a constant ionizing
intensity, does not yield a good fit. Taking into account the effect of
metal
lines is the main improvement of the model that allows for an acceptable
fit,
and does not change appreciably the value of $\betalya$ although errors are
substantially increased. The addition of the effects of HCDs and UV
fluctuations further improves the model and increases substantially the best
fit value of $\betalya$.

   To compare our values of $b$ and $\betalya$ with the previous determination
   from the DR11 data from  \citet{2015JCAP...11..034B},
     we first translate their
result
to the cosmological model we use in this paper, taking into account that the
measurement of the \Lya autocorrelation depends only on $b\sigma_8(\bar z)$,
where $\bar z=2.33$ is the mean redshift of our \Lya forest correlation
measurement. \citet{2015JCAP...11..034B} measured $\betalya=1.39 \pm 0.11$, and
$\blya(1+\betalya)=-0.374\pm 0.007$, using a model with $\Omega_m=0.27$ and
$\sigma_8=0.7877$,
implying $\sigma_8(\bar z)=0.3072$. The model used in this paper has
$\Omega_m=0.3147$ and $\sigma_8=0.8298$,
implying $\sigma_8(\bar{z})=0.3131$.
Therefore the \Lya bias of \citet{2015JCAP...11..034B}
translated to our model is
$\blya(1+\betalya)=-0.367 \pm 0.007$.
This should be compared with our result on the first row of Table 5,
because \citet{2015JCAP...11..034B} did not consider the effects from metals,
HCDs or UV fluctuations

   The main reason for the difference with the values of $\blya(1+\betalya)$ and
$\betalya$ we obtain here for our \Lya forest model applied to DR12 given in
Table 5 is the different radial range for the fit: \citet{2015JCAP...11..034B}
restricted their fit to $r>40~\hMpc$, whereas we use $r>10~\hMpc$. There are
other details that contribute to the difference, e.g., the continuum
distortion and bin smoothing modeling, but the radial range of the fit is
most important. As mentioned earlier, the $\chi^2$ value shows that our
\Lya fit is not good, reflecting the fact that the shape of the correlation
cannot be properly fit over our broad radial range. This is why
the values of $\|\blya\|\,(1+\betalya)$ and $\betalya$ decrease when extending
the fit to
smaller separations. The better fits obtained for the more complete models
in Table 5 show that $\blya(1+\betalya)$ is relatively stable, but the value of
$\betalya$ can be more model dependent.

   These results can also be compared to theoretical predictions from
hydrodynamic simulations of the intergalactic medium, which were
compared to the observational measurement of \citet{2015JCAP...11..034B} in
Section 7 of \citet{2015JCAP...12..017A}. There, it was noticed that
while the predicted value of $\betalya$ is roughly in agreement with the
observational determination, the value of $\|\blya\|\,(1+\betalya)$ was
predicted to
be lower than observed. Now, the lower value of $\|\blya\|\, (1+\betalya)$ we
measure
from DR12, correcting for the effect of metals and extending the fit
to smaller separations, is in much better agreement with the range of
the expected theoretical values. Despite the considerable uncertainties
that are still present both in the predictions from hydrodynamic
simulations of the \Lya forest (as discussed in \citet{2015JCAP...12..017A}),
and in the observational determinations
that are affected by the modeling of metals, HCDs and
ionizing intensity fluctuations, we think that the rough agreement we
now find of two independently predicted linear bias factors of the \Lya
forest is fairly remarkable. In the future, we expect to be able to
better measure these linear factors and their redshift evolution.

\bibliographystyle{aa}
\bibliography{baodr12}

\begin{thebibliography}{71}
\expandafter\ifx\csname natexlab\endcsname\relax\def\natexlab#1{#1}\fi

\bibitem[{{Alam} {et~al.}(2015){Alam}, {Albareti}, {Allende Prieto}, {Anders},
  {Anderson}, {Anderton}, {Andrews}, {Armengaud}, {Aubourg}, {Bailey}, \&
  et~al.}]{2015ApJS..219...12A}
{Alam}, S., {Albareti}, F.~D., {Allende Prieto}, C., {et~al.} 2015, \apjs, 219,
  12

\bibitem[{{Alam} {et~al.}(2016){Alam}, {Ata}, {Bailey}, {Beutler}, {Bizyaev},
  {Blazek}, {Bolton}, {Brownstein}, {Burden}, {Chuang}, {Comparat}, {Cuesta},
  {Dawson}, {Eisenstein}, {Escoffier}, {Gil-Mar{\'{\i}}n}, {Grieb}, {Hand},
  {Ho}, {Kinemuchi}, {Kirkby}, {Kitaura}, {Malanushenko}, {Malanushenko},
  {Maraston}, {McBride}, {Nichol}, {Olmstead}, {Oravetz}, {Padmanabhan},
  {Palanque-Delabrouille}, {Pan}, {Pellejero-Ibanez}, {Percival}, {Petitjean},
  {Prada}, {Price-Whelan}, {Reid}, {Rodr{\'{\i}}guez-Torres}, {Roe}, {Ross},
  {Ross}, {Rossi}, {Rubi{\~n}o-Mart{\'{\i}}n}, {S{\'a}nchez}, {Saito},
  {Salazar-Albornoz}, {Samushia}, {Satpathy}, {Sc{\'o}ccola}, {Schlegel},
  {Schneider}, {Seo}, {Simmons}, {Slosar}, {Strauss}, {Swanson}, {Thomas},
  {Tinker}, {Tojeiro}, {Vargas Maga{\~n}a}, {Vazquez}, {Verde}, {Wake}, {Wang},
  {Weinberg}, {White}, {Wood-Vasey}, {Y{\`e}che}, {Zehavi}, {Zhai}, \&
  {Zhao}}]{2016arXiv160703155A}
{Alam}, S., {Ata}, M., {Bailey}, S., {et~al.} 2016, ArXiv e-prints
  [\eprint[arXiv]{1607.03155}]

\bibitem[{{Anderson} {et~al.}(2014{\natexlab{a}}){Anderson}, {Aubourg},
  {Bailey}, {Beutler}, {Bhardwaj}, {Blanton}, {Bolton}, {Brinkmann},
  {Brownstein}, {Burden}, {Chuang}, {Cuesta}, {Dawson}, {Eisenstein},
  {Escoffier}, {Gunn}, {Guo}, {Ho}, {Honscheid}, {Howlett}, {Kirkby}, {Lupton},
  {Manera}, {Maraston}, {McBride}, {Mena}, {Montesano}, {Nichol}, {Nuza},
  {Olmstead}, {Padmanabhan}, {Palanque-Delabrouille}, {Parejko}, {Percival},
  {Petitjean}, {Prada}, {Price-Whelan}, {Reid}, {Roe}, {Ross}, {Ross}, {Sabiu},
  {Saito}, {Samushia}, {S{\'a}nchez}, {Schlegel}, {Schneider}, {Scoccola},
  {Seo}, {Skibba}, {Strauss}, {Swanson}, {Thomas}, {Tinker}, {Tojeiro},
  {Maga{\~n}a}, {Verde}, {Wake}, {Weaver}, {Weinberg}, {White}, {Xu},
  {Y{\`e}che}, {Zehavi}, \& {Zhao}}]{2014MNRAS.441...24A}
{Anderson}, L., {Aubourg}, {\'E}., {Bailey}, S., {et~al.} 2014{\natexlab{a}},
  \mnras, 441, 24

\bibitem[{{Anderson} {et~al.}(2014{\natexlab{b}}){Anderson}, {Aubourg},
  {Bailey}, {Beutler}, {Bolton}, {Brinkmann}, {Brownstein}, {Chuang}, {Cuesta},
  {Dawson}, {Eisenstein}, {Ho}, {Honscheid}, {Kazin}, {Kirkby}, {Manera},
  {McBride}, {Mena}, {Nichol}, {Olmstead}, {Padmanabhan},
  {Palanque-Delabrouille}, {Percival}, {Prada}, {Ross}, {Ross}, {S{\'a}nchez},
  {Samushia}, {Schlegel}, {Schneider}, {Seo}, {Strauss}, {Thomas}, {Tinker},
  {Tojeiro}, {Verde}, {Wake}, {Weinberg}, {Xu}, \&
  {Yeche}}]{2014MNRAS.439...83A}
{Anderson}, L., {Aubourg}, E., {Bailey}, S., {et~al.} 2014{\natexlab{b}},
  \mnras, 439, 83

\bibitem[{{Anderson} {et~al.}(2012){Anderson}, {Aubourg}, {Bailey}, {Bizyaev},
  {Blanton}, {Bolton}, {Brinkmann}, {Brownstein}, {Burden}, {Cuesta}, {da
  Costa}, {Dawson}, {de Putter}, {Eisenstein}, {Gunn}, {Guo}, {Hamilton},
  {Harding}, {Ho}, {Honscheid}, {Kazin}, {Kirkby}, {Kneib}, {Labatie},
  {Loomis}, {Lupton}, {Malanushenko}, {Malanushenko}, {Mandelbaum}, {Manera},
  {Maraston}, {McBride}, {Mehta}, {Mena}, {Montesano}, {Muna}, {Nichol},
  {Nuza}, {Olmstead}, {Oravetz}, {Padmanabhan}, {Palanque-Delabrouille}, {Pan},
  {Parejko}, {P{\^a}ris}, {Percival}, {Petitjean}, {Prada}, {Reid}, {Roe},
  {Ross}, {Ross}, {Samushia}, {S{\'a}nchez}, {Schlegel}, {Schneider},
  {Sc{\'o}ccola}, {Seo}, {Sheldon}, {Simmons}, {Skibba}, {Strauss}, {Swanson},
  {Thomas}, {Tinker}, {Tojeiro}, {Maga{\~n}a}, {Verde}, {Wagner}, {Wake},
  {Weaver}, {Weinberg}, {White}, {Xu}, {Y{\`e}che}, {Zehavi}, \&
  {Zhao}}]{2012MNRAS.427.3435A}
{Anderson}, L., {Aubourg}, E., {Bailey}, S., {et~al.} 2012, \mnras, 427, 3435

\bibitem[{{Arinyo-i-Prats} {et~al.}(2015){Arinyo-i-Prats},
  {Miralda-Escud{\'e}}, {Viel}, \& {Cen}}]{2015JCAP...12..017A}
{Arinyo-i-Prats}, A., {Miralda-Escud{\'e}}, J., {Viel}, M., \& {Cen}, R. 2015,
  \jcap, 12, 017

\bibitem[{{Aubourg} {et~al.}(2015){Aubourg}, {Bailey}, {Bautista}, {Beutler},
  {Bhardwaj}, {Bizyaev}, {Blanton}, {Blomqvist}, {Bolton}, {Bovy},
  {Brewington}, {Brinkmann}, {Brownstein}, {Burden}, {Busca}, {Carithers},
  {Chuang}, {Comparat}, {Croft}, {Cuesta}, {Dawson}, {Delubac}, {Eisenstein},
  {Font-Ribera}, {Ge}, {Le Goff}, {Gontcho}, {Gott}, {Gunn}, {Guo}, {Guy},
  {Hamilton}, {Ho}, {Honscheid}, {Howlett}, {Kirkby}, {Kitaura}, {Kneib},
  {Lee}, {Long}, {Lupton}, {Maga{\~n}a}, {Malanushenko}, {Malanushenko},
  {Manera}, {Maraston}, {Margala}, {McBride}, {Miralda-Escud{\'e}}, {Myers},
  {Nichol}, {Noterdaeme}, {Nuza}, {Olmstead}, {Oravetz}, {P{\^a}ris},
  {Padmanabhan}, {Palanque-Delabrouille}, {Pan}, {Pellejero-Ibanez},
  {Percival}, {Petitjean}, {Pieri}, {Prada}, {Reid}, {Rich}, {Roe}, {Ross},
  {Ross}, {Rossi}, {Rubi{\~n}o-Mart{\'{\i}}n}, {S{\'a}nchez}, {Samushia},
  {Santos}, {Sc{\'o}ccola}, {Schlegel}, {Schneider}, {Seo}, {Sheldon},
  {Simmons}, {Skibba}, {Slosar}, {Strauss}, {Thomas}, {Tinker}, {Tojeiro},
  {Vazquez}, {Viel}, {Wake}, {Weaver}, {Weinberg}, {Wood-Vasey}, {Y{\`e}che},
  {Zehavi}, {Zhao}, \& {BOSS Collaboration}}]{2015PhRvD..92l3516A}
{Aubourg}, {\'E}., {Bailey}, S., {Bautista}, J.~E., {et~al.} 2015, \prd, 92,
  123516

\bibitem[{{Bautista} {et~al.}(2015){Bautista}, {Bailey}, {Font-Ribera},
  {Pieri}, {Busca}, {Miralda-Escud{\'e}}, {Palanque-Delabrouille}, {Rich},
  {Dawson}, {Feng}, {Ge}, {Gontcho}, {Ho}, {Le Goff}, {Noterdaeme},
  {P{\^a}ris}, {Rossi}, \& {Schlegel}}]{2015JCAP...05..060B}
{Bautista}, J.~E., {Bailey}, S., {Font-Ribera}, A., {et~al.} 2015, \jcap, 5, 60

\bibitem[{{Becker} {et~al.}(1995){Becker}, {White}, \&
  {Helfand}}]{1995ApJ...450..559B}
{Becker}, R.~H., {White}, R.~L., \& {Helfand}, D.~J. 1995, \apj, 450, 559

\bibitem[{{Betoule} {et~al.}(2014){Betoule}, {Kessler}, {Guy}, {Mosher},
  {Hardin}, {Biswas}, {Astier}, {El-Hage}, {Konig}, {Kuhlmann}, {Marriner},
  {Pain}, {Regnault}, {Balland}, {Bassett}, {Brown}, {Campbell}, {Carlberg},
  {Cellier-Holzem}, {Cinabro}, {Conley}, {D'Andrea}, {DePoy}, {Doi}, {Ellis},
  {Fabbro}, {Filippenko}, {Foley}, {Frieman}, {Fouchez}, {Galbany}, {Goobar},
  {Gupta}, {Hill}, {Hlozek}, {Hogan}, {Hook}, {Howell}, {Jha}, {Le Guillou},
  {Leloudas}, {Lidman}, {Marshall}, {M{\"o}ller}, {Mour{\~a}o}, {Neveu},
  {Nichol}, {Olmstead}, {Palanque-Delabrouille}, {Perlmutter}, {Prieto},
  {Pritchet}, {Richmond}, {Riess}, {Ruhlmann-Kleider}, {Sako}, {Schahmaneche},
  {Schneider}, {Smith}, {Sollerman}, {Sullivan}, {Walton}, \&
  {Wheeler}}]{2014A&A...568A..22B}
{Betoule}, M., {Kessler}, R., {Guy}, J., {et~al.} 2014, \aap, 568, A22

\bibitem[{{Beutler} {et~al.}(2011){Beutler}, {Blake}, {Colless}, {Jones},
  {Staveley-Smith}, {Campbell}, {Parker}, {Saunders}, \&
  {Watson}}]{2011MNRAS.416.3017B}
{Beutler}, F., {Blake}, C., {Colless}, M., {et~al.} 2011, \mnras, 416, 3017

\bibitem[{{Blake} {et~al.}(2011){Blake}, {Davis}, {Poole}, {Parkinson},
  {Brough}, {Colless}, {Contreras}, {Couch}, {Croom}, {Drinkwater}, {Forster},
  {Gilbank}, {Gladders}, {Glazebrook}, {Jelliffe}, {Jurek}, {Li}, {Madore},
  {Martin}, {Pimbblet}, {Pracy}, {Sharp}, {Wisnioski}, {Woods}, {Wyder}, \&
  {Yee}}]{2011MNRAS.415.2892B}
{Blake}, C., {Davis}, T., {Poole}, G.~B., {et~al.} 2011, \mnras, 415, 2892

\bibitem[{{Blazek} {et~al.}(2016){Blazek}, {McEwen}, \&
  {Hirata}}]{2016PhRvL.116l1303B}
{Blazek}, J.~A., {McEwen}, J.~E., \& {Hirata}, C.~M. 2016, Physical Review
  Letters, 116, 121303

\bibitem[{{Blomqvist} {et~al.}(2015){Blomqvist}, {Kirkby}, {Bautista},
  {Arinyo-i-Prats}, {Busca}, {Miralda-Escud{\'e}}, {Slosar}, {Font-Ribera},
  {Margala}, {Schneider}, \& {Vazquez}}]{2015JCAP...11..034B}
{Blomqvist}, M., {Kirkby}, D., {Bautista}, J.~E., {et~al.} 2015, \jcap, 11, 034

\bibitem[{{Bolton} {et~al.}(2012){Bolton}, {Schlegel}, {Aubourg}, {Bailey},
  {Bhardwaj}, {Brownstein}, {Burles}, {Chen}, {Dawson}, {Eisenstein}, {Gunn},
  {Knapp}, {Loomis}, {Lupton}, {Maraston}, {Muna}, {Myers}, {Olmstead},
  {Padmanabhan}, {P{\^a}ris}, {Percival}, {Petitjean}, {Rockosi}, {Ross},
  {Schneider}, {Shu}, {Strauss}, {Thomas}, {Tremonti}, {Wake}, {Weaver}, \&
  {Wood-Vasey}}]{2012AJ....144..144B}
{Bolton}, A.~S., {Schlegel}, D.~J., {Aubourg}, {\'E}., {et~al.} 2012, \aj, 144,
  144

\bibitem[{{Bovy} {et~al.}(2011){Bovy}, {Hennawi}, {Hogg}, {Myers},
  {Kirkpatrick}, {Schlegel}, {Ross}, {Sheldon}, {McGreer}, {Schneider}, \&
  {Weaver}}]{2011ApJ...729..141B}
{Bovy}, J., {Hennawi}, J.~F., {Hogg}, D.~W., {et~al.} 2011, \apj, 729, 141

\bibitem[{{Busca} {et~al.}(2013){Busca}, {Delubac}, {Rich}, {Bailey},
  {Font-Ribera}, {Kirkby}, {Le Goff}, {Pieri}, {Slosar}, {Aubourg}, {Bautista},
  {Bizyaev}, {Blomqvist}, {Bolton}, {Bovy}, {Brewington}, {Borde}, {Brinkmann},
  {Carithers}, {Croft}, {Dawson}, {Ebelke}, {Eisenstein}, {Hamilton}, {Ho},
  {Hogg}, {Honscheid}, {Lee}, {Lundgren}, {Malanushenko}, {Malanushenko},
  {Margala}, {Maraston}, {Mehta}, {Miralda-Escud{\'e}}, {Myers}, {Nichol},
  {Noterdaeme}, {Olmstead}, {Oravetz}, {Palanque-Delabrouille}, {Pan},
  {P{\^a}ris}, {Percival}, {Petitjean}, {Roe}, {Rollinde}, {Ross}, {Rossi},
  {Schlegel}, {Schneider}, {Shelden}, {Sheldon}, {Simmons}, {Snedden},
  {Tinker}, {Viel}, {Weaver}, {Weinberg}, {White}, {Y{\`e}che}, \&
  {York}}]{2013A&A...552A..96B}
{Busca}, N.~G., {Delubac}, T., {Rich}, J., {et~al.} 2013, \aap, 552, A96

\bibitem[{{Chuang} \& {Wang}(2012)}]{2012MNRAS.426..226C}
{Chuang}, C.-H. \& {Wang}, Y. 2012, \mnras, 426, 226

\bibitem[{{Cole} {et~al.}(2005){Cole}, {Percival}, {Peacock}, {Norberg},
  {Baugh}, {Frenk}, {Baldry}, {Bland-Hawthorn}, {Bridges}, {Cannon}, {Colless},
  {Collins}, {Couch}, {Cross}, {Dalton}, {Eke}, {De Propris}, {Driver},
  {Efstathiou}, {Ellis}, {Glazebrook}, {Jackson}, {Jenkins}, {Lahav}, {Lewis},
  {Lumsden}, {Maddox}, {Madgwick}, {Peterson}, {Sutherland}, \&
  {Taylor}}]{2005MNRAS.362..505C}
{Cole}, S., {Percival}, W.~J., {Peacock}, J.~A., {et~al.} 2005, \mnras, 362,
  505

\bibitem[{{Croft} {et~al.}(2016){Croft}, {Miralda-Escud{\'e}}, {Zheng},
  {Bolton}, {Dawson}, {Peterson}, {York}, {Eisenstein}, {Brinkmann},
  {Brownstein}, {Cen}, {Delubac}, {Font-Ribera}, {Hamilton}, {Lee}, {Myers},
  {Palanque-Delabrouille}, {P{\^a}ris}, {Petitjean}, {Pieri}, {Ross}, {Rossi},
  {Schlegel}, {Schneider}, {Slosar}, {Vazquez}, {Viel}, {Weinberg}, \&
  {Y{\`e}che}}]{2016MNRAS.457.3541C}
{Croft}, R.~A.~C., {Miralda-Escud{\'e}}, J., {Zheng}, Z., {et~al.} 2016,
  \mnras, 457, 3541

\bibitem[{{Croom} {et~al.}(2005){Croom}, {Boyle}, {Shanks}, {Smith}, {Miller},
  {Outram}, {Loaring}, {Hoyle}, \& {da {\^A}ngela}}]{2005MNRAS.356..415C}
{Croom}, S.~M., {Boyle}, B.~J., {Shanks}, T., {et~al.} 2005, \mnras, 356, 415

\bibitem[{{Dalal} {et~al.}(2010){Dalal}, {Pen}, \&
  {Seljak}}]{2010JCAP...11..007D}
{Dalal}, N., {Pen}, U.-L., \& {Seljak}, U. 2010, \jcap, 11, 007

\bibitem[{{Dawson} {et~al.}(2013){Dawson}, {Schlegel}, {Ahn}, {Anderson},
  {Aubourg}, {Bailey}, {Barkhouser}, {Bautista}, {Beifiori}, {Berlind},
  {Bhardwaj}, {Bizyaev}, {Blake}, {Blanton}, {Blomqvist}, {Bolton}, {Borde},
  {Bovy}, {Brandt}, {Brewington}, {Brinkmann}, {Brown}, {Brownstein}, {Bundy},
  {Busca}, {Carithers}, {Carnero}, {Carr}, {Chen}, {Comparat}, {Connolly},
  {Cope}, {Croft}, {Cuesta}, {da Costa}, {Davenport}, {Delubac}, {de Putter},
  {Dhital}, {Ealet}, {Ebelke}, {Eisenstein}, {Escoffier}, {Fan}, {Filiz Ak},
  {Finley}, {Font-Ribera}, {G{\'e}nova-Santos}, {Gunn}, {Guo}, {Haggard},
  {Hall}, {Hamilton}, {Harris}, {Harris}, {Ho}, {Hogg}, {Holder}, {Honscheid},
  {Huehnerhoff}, {Jordan}, {Jordan}, {Kauffmann}, {Kazin}, {Kirkby}, {Klaene},
  {Kneib}, {Le Goff}, {Lee}, {Long}, {Loomis}, {Lundgren}, {Lupton}, {Maia},
  {Makler}, {Malanushenko}, {Malanushenko}, {Mandelbaum}, {Manera}, {Maraston},
  {Margala}, {Masters}, {McBride}, {McDonald}, {McGreer}, {McMahon}, {Mena},
  {Miralda-Escud{\'e}}, {Montero-Dorta}, {Montesano}, {Muna}, {Myers},
  {Naugle}, {Nichol}, {Noterdaeme}, {Nuza}, {Olmstead}, {Oravetz}, {Oravetz},
  {Owen}, {Padmanabhan}, {Palanque-Delabrouille}, {Pan}, {Parejko},
  {P{\^a}ris}, {Percival}, {P{\'e}rez-Fournon}, {P{\'e}rez-R{\`a}fols},
  {Petitjean}, {Pfaffenberger}, {Pforr}, {Pieri}, {Prada}, {Price-Whelan},
  {Raddick}, {Rebolo}, {Rich}, {Richards}, {Rockosi}, {Roe}, {Ross}, {Ross},
  {Rossi}, {Rubi{\~n}o-Martin}, {Samushia}, {S{\'a}nchez}, {Sayres}, {Schmidt},
  {Schneider}, {Sc{\'o}ccola}, {Seo}, {Shelden}, {Sheldon}, {Shen}, {Shu},
  {Slosar}, {Smee}, {Snedden}, {Stauffer}, {Steele}, {Strauss}, {Streblyanska},
  {Suzuki}, {Swanson}, {Tal}, {Tanaka}, {Thomas}, {Tinker}, {Tojeiro},
  {Tremonti}, {Vargas Maga{\~n}a}, {Verde}, {Viel}, {Wake}, {Watson}, {Weaver},
  {Weinberg}, {Weiner}, {West}, {White}, {Wood-Vasey}, {Yeche}, {Zehavi},
  {Zhao}, \& {Zheng}}]{2013AJ....145...10D}
{Dawson}, K.~S., {Schlegel}, D.~J., {Ahn}, C.~P., {et~al.} 2013, \aj, 145, 10

\bibitem[{{Delubac} {et~al.}(2015){Delubac}, {Bautista}, {Busca}, {Rich},
  {Kirkby}, {Bailey}, {Font-Ribera}, {Slosar}, {Lee}, {Pieri}, {Hamilton},
  {Aubourg}, {Blomqvist}, {Bovy}, {Brinkmann}, {Carithers}, {Dawson},
  {Eisenstein}, {Gontcho}, {Kneib}, {Le Goff}, {Margala}, {Miralda-Escud{\'e}},
  {Myers}, {Nichol}, {Noterdaeme}, {O'Connell}, {Olmstead},
  {Palanque-Delabrouille}, {P{\^a}ris}, {Petitjean}, {Ross}, {Rossi},
  {Schlegel}, {Schneider}, {Weinberg}, {Y{\`e}che}, \&
  {York}}]{2015A&A...574A..59D}
{Delubac}, T., {Bautista}, J.~E., {Busca}, N.~G., {et~al.} 2015, \aap, 574, A59

\bibitem[{{DESI Collaboration} {et~al.}(2016){DESI Collaboration}, {Aghamousa},
  {Aguilar}, {Ahlen}, {Alam}, {Allen}, {Allende Prieto}, {Annis}, {Bailey},
  {Balland}, \& et~al.}]{2016arXiv161100036D}
{DESI Collaboration}, {Aghamousa}, A., {Aguilar}, J., {et~al.} 2016, ArXiv
  e-prints [\eprint[arXiv]{1611.00036}]

\bibitem[{{Eisenstein} {et~al.}(2011){Eisenstein}, {Weinberg}, {Agol},
  {Aihara}, {Allende Prieto}, {Anderson}, {Arns}, {Aubourg}, {Bailey},
  {Balbinot}, \& et~al.}]{2011AJ....142...72E}
{Eisenstein}, D.~J., {Weinberg}, D.~H., {Agol}, E., {et~al.} 2011, \aj, 142, 72

\bibitem[{{Eisenstein} {et~al.}(2005){Eisenstein}, {Zehavi}, {Hogg},
  {Scoccimarro}, {Blanton}, {Nichol}, {Scranton}, {Seo}, {Tegmark}, {Zheng},
  {Anderson}, {Annis}, {Bahcall}, {Brinkmann}, {Burles}, {Castander},
  {Connolly}, {Csabai}, {Doi}, {Fukugita}, {Frieman}, {Glazebrook}, {Gunn},
  {Hendry}, {Hennessy}, {Ivezi{\'c}}, {Kent}, {Knapp}, {Lin}, {Loh}, {Lupton},
  {Margon}, {McKay}, {Meiksin}, {Munn}, {Pope}, {Richmond}, {Schlegel},
  {Schneider}, {Shimasaku}, {Stoughton}, {Strauss}, {SubbaRao}, {Szalay},
  {Szapudi}, {Tucker}, {Yanny}, \& {York}}]{2005ApJ...633..560E}
{Eisenstein}, D.~J., {Zehavi}, I., {Hogg}, D.~W., {et~al.} 2005, \apj, 633, 560

\bibitem[{{Font-Ribera} {et~al.}(2013){Font-Ribera}, {Arnau},
  {Miralda-Escud{\'e}}, {Rollinde}, {Brinkmann}, {Brownstein}, {Lee}, {Myers},
  {Palanque-Delabrouille}, {P{\^a}ris}, {Petitjean}, {Rich}, {Ross},
  {Schneider}, \& {White}}]{2013JCAP...05..018F}
{Font-Ribera}, A., {Arnau}, E., {Miralda-Escud{\'e}}, J., {et~al.} 2013, \jcap,
  5, 18

\bibitem[{{Font-Ribera} {et~al.}(2014){Font-Ribera}, {Kirkby}, {Busca},
  {Miralda-Escud{\'e}}, {Ross}, {Slosar}, {Rich}, {Aubourg}, {Bailey},
  {Bhardwaj}, {Bautista}, {Beutler}, {Bizyaev}, {Blomqvist}, {Brewington},
  {Brinkmann}, {Brownstein}, {Carithers}, {Dawson}, {Delubac}, {Ebelke},
  {Eisenstein}, {Ge}, {Kinemuchi}, {Lee}, {Malanushenko}, {Malanushenko},
  {Marchante}, {Margala}, {Muna}, {Myers}, {Noterdaeme}, {Oravetz},
  {Palanque-Delabrouille}, {P{\^a}ris}, {Petitjean}, {Pieri}, {Rossi},
  {Schneider}, {Simmons}, {Viel}, {Yeche}, \& {York}}]{2014JCAP...05..027F}
{Font-Ribera}, A., {Kirkby}, D., {Busca}, N., {et~al.} 2014, \jcap, 5, 27

\bibitem[{{Font-Ribera} {et~al.}(2012){Font-Ribera}, {McDonald}, \&
  {Miralda-Escud{\'e}}}]{2012JCAP...01..001F}
{Font-Ribera}, A., {McDonald}, P., \& {Miralda-Escud{\'e}}, J. 2012, \jcap, 1,
  1

\bibitem[{{Font-Ribera} \& {Miralda-Escud{\'e}}(2012)}]{2012JCAP...07..028F}
{Font-Ribera}, A. \& {Miralda-Escud{\'e}}, J. 2012, \jcap, 7, 28

\bibitem[{{Gontcho A Gontcho} {et~al.}(2014){Gontcho A Gontcho},
  {Miralda-Escud{\'e}}, \& {Busca}}]{2014MNRAS.442..187G}
{Gontcho A Gontcho}, S., {Miralda-Escud{\'e}}, J., \& {Busca}, N.~G. 2014,
  \mnras, 442, 187

\bibitem[{{Gunn} {et~al.}(1998){Gunn}, {Carr}, {Rockosi}, {Sekiguchi}, {Berry},
  {Elms}, {de Haas}, {Ivezi{\' c}}, {Knapp}, {Lupton}, {Pauls}, {Simcoe},
  {Hirsch}, {Sanford}, {Wang}, {York}, {Harris}, {Annis}, {Bartozek},
  {Boroski}, {Bakken}, {Haldeman}, {Kent}, {Holm}, {Holmgren}, {Petravick},
  {Prosapio}, {Rechenmacher}, {Doi}, {Fukugita}, {Shimasaku}, {Okada}, {Hull},
  {Siegmund}, {Mannery}, {Blouke}, {Heidtman}, {Schneider}, {Lucinio}, \&
  {Brinkman}}]{1998AJ....116.3040G}
{Gunn}, J.~E., {Carr}, M., {Rockosi}, C., {et~al.} 1998, \aj, 116, 3040

\bibitem[{{Gunn} {et~al.}(2006){Gunn}, {Siegmund}, {Mannery}, {Owen}, {Hull},
  {Leger}, {Carey}, {Knapp}, {York}, {Boroski}, {Kent}, {Lupton}, {Rockosi},
  {Evans}, {Waddell}, {Anderson}, {Annis}, {Barentine}, {Bartoszek}, {Bastian},
  {Bracker}, {Brewington}, {Briegel}, {Brinkmann}, {Brown}, {Carr},
  {Czarapata}, {Drennan}, {Dombeck}, {Federwitz}, {Gillespie}, {Gonzales},
  {Hansen}, {Harvanek}, {Hayes}, {Jordan}, {Kinney}, {Klaene}, {Kleinman},
  {Kron}, {Kresinski}, {Lee}, {Limmongkol}, {Lindenmeyer}, {Long}, {Loomis},
  {McGehee}, {Mantsch}, {Neilsen}, {Neswold}, {Newman}, {Nitta}, {Peoples},
  {Pier}, {Prieto}, {Prosapio}, {Rivetta}, {Schneider}, {Snedden}, \&
  {Wang}}]{2006AJ....131.2332G}
{Gunn}, J.~E., {Siegmund}, W.~A., {Mannery}, E.~J., {et~al.} 2006, \aj, 131,
  2332

\bibitem[{{Kirkby} {et~al.}(2013){Kirkby}, {Margala}, {Slosar}, {Bailey},
  {Busca}, {Delubac}, {Rich}, {Bautista}, {Blomqvist}, {Brownstein},
  {Carithers}, {Croft}, {Dawson}, {Font-Ribera}, {Miralda-Escud{\'e}}, {Myers},
  {Nichol}, {Palanque-Delabrouille}, {P{\^a}ris}, {Petitjean}, {Rossi},
  {Schlegel}, {Schneider}, {Viel}, {Weinberg}, \&
  {Y{\`e}che}}]{2013JCAP...03..024K}
{Kirkby}, D., {Margala}, D., {Slosar}, A., {et~al.} 2013, \jcap, 3, 24

\bibitem[{{Kirkpatrick} {et~al.}(2011){Kirkpatrick}, {Schlegel}, {Ross},
  {Myers}, {Hennawi}, {Sheldon}, {Schneider}, \&
  {Weaver}}]{2011ApJ...743..125K}
{Kirkpatrick}, J.~A., {Schlegel}, D.~J., {Ross}, N.~P., {et~al.} 2011, \apj,
  743, 125

\bibitem[{{Lawrence} {et~al.}(2007){Lawrence}, {Warren}, {Almaini}, {Edge},
  {Hambly}, {Jameson}, {Lucas}, {Casali}, {Adamson}, {Dye}, {Emerson},
  {Foucaud}, {Hewett}, {Hirst}, {Hodgkin}, {Irwin}, {Lodieu}, {McMahon},
  {Simpson}, {Smail}, {Mortlock}, \& {Folger}}]{2007MNRAS.379.1599L}
{Lawrence}, A., {Warren}, S.~J., {Almaini}, O., {et~al.} 2007, \mnras, 379,
  1599

\bibitem[{{Lewis} {et~al.}(2000){Lewis}, {Challinor}, \&
  {Lasenby}}]{2000ApJ...538..473L}
{Lewis}, A., {Challinor}, A., \& {Lasenby}, A. 2000, \apj, 538, 473

\bibitem[{{Margala} {et~al.}(2016){Margala}, {Kirkby}, {Dawson}, {Bailey},
  {Blanton}, \& {Schneider}}]{2016ApJ...831..157M}
{Margala}, D., {Kirkby}, D., {Dawson}, K., {et~al.} 2016, \apj, 831, 157

\bibitem[{{Martin} {et~al.}(2005){Martin}, {Fanson}, {Schiminovich},
  {Morrissey}, {Friedman}, {Barlow}, {Conrow}, {Grange}, {Jelinsky},
  {Milliard}, {Siegmund}, {Bianchi}, {Byun}, {Donas}, {Forster}, {Heckman},
  {Lee}, {Madore}, {Malina}, {Neff}, {Rich}, {Small}, {Surber}, {Szalay},
  {Welsh}, \& {Wyder}}]{2005ApJ...619L...1M}
{Martin}, D.~C., {Fanson}, J., {Schiminovich}, D., {et~al.} 2005, \apjl, 619,
  L1

\bibitem[{{McDonald}(2003)}]{2003ApJ...585...34M}
{McDonald}, P. 2003, \apj, 585, 34

\bibitem[{{McDonald} \& {Eisenstein}(2007)}]{2007PhRvD..76f3009M}
{McDonald}, P. \& {Eisenstein}, D.~J. 2007, \prd, 76, 063009

\bibitem[{{Mehta} {et~al.}(2012){Mehta}, {Cuesta}, {Xu}, {Eisenstein}, \&
  {Padmanabhan}}]{2012MNRAS.427.2168M}
{Mehta}, K.~T., {Cuesta}, A.~J., {Xu}, X., {Eisenstein}, D.~J., \&
  {Padmanabhan}, N. 2012, \mnras, 427, 2168

\bibitem[{{Noterdaeme} {et~al.}(2012){Noterdaeme}, {Petitjean}, {Carithers},
  {P{\^a}ris}, {Font-Ribera}, {Bailey}, {Aubourg}, {Bizyaev}, {Ebelke},
  {Finley}, {Ge}, {Malanushenko}, {Malanushenko}, {Miralda-Escud{\'e}},
  {Myers}, {Oravetz}, {Pan}, {Pieri}, {Ross}, {Schneider}, {Simmons}, \&
  {York}}]{2012A&A...547L...1N}
{Noterdaeme}, P., {Petitjean}, P., {Carithers}, W.~C., {et~al.} 2012, \aap,
  547, L1

\bibitem[{{Padmanabhan} {et~al.}(2012){Padmanabhan}, {Xu}, {Eisenstein},
  {Scalzo}, {Cuesta}, {Mehta}, \& {Kazin}}]{2012MNRAS.427.2132P}
{Padmanabhan}, N., {Xu}, X., {Eisenstein}, D.~J., {et~al.} 2012, \mnras, 427,
  2132

\bibitem[{{P{\^a}ris} {et~al.}(2012){P{\^a}ris}, {Petitjean}, {Aubourg},
  {Bailey}, {Ross}, {Myers}, {Strauss}, {Anderson}, {Arnau}, {Bautista},
  {Bizyaev}, {Bolton}, {Bovy}, {Brandt}, {Brewington}, {Browstein}, {Busca},
  {Capellupo}, {Carithers}, {Croft}, {Dawson}, {Delubac}, {Ebelke},
  {Eisenstein}, {Engelke}, {Fan}, {Filiz Ak}, {Finley}, {Font-Ribera}, {Ge},
  {Gibson}, {Hall}, {Hamann}, {Hennawi}, {Ho}, {Hogg}, {Ivezi{\'c}}, {Jiang},
  {Kimball}, {Kirkby}, {Kirkpatrick}, {Lee}, {Le Goff}, {Lundgren}, {MacLeod},
  {Malanushenko}, {Malanushenko}, {Maraston}, {McGreer}, {McMahon},
  {Miralda-Escud{\'e}}, {Muna}, {Noterdaeme}, {Oravetz},
  {Palanque-Delabrouille}, {Pan}, {Perez-Fournon}, {Pieri}, {Richards},
  {Rollinde}, {Sheldon}, {Schlegel}, {Schneider}, {Slosar}, {Shelden}, {Shen},
  {Simmons}, {Snedden}, {Suzuki}, {Tinker}, {Viel}, {Weaver}, {Weinberg},
  {White}, {Wood-Vasey}, \& {Y{\`e}che}}]{2012A&A...548A..66P}
{P{\^a}ris}, I., {Petitjean}, P., {Aubourg}, {\'E}., {et~al.} 2012, \aap, 548,
  A66

\bibitem[{{P{\^a}ris} {et~al.}(2014){P{\^a}ris}, {Petitjean}, {Aubourg},
  {Ross}, {Myers}, {Streblyanska}, {Bailey}, {Hall}, {Strauss}, {Anderson},
  {Bizyaev}, {Borde}, {Brinkmann}, {Bovy}, {Brandt}, {Brewington},
  {Brownstein}, {Cook}, {Ebelke}, {Fan}, {Filiz Ak}, {Finley}, {Font-Ribera},
  {Ge}, {Hamann}, {Ho}, {Jiang}, {Kinemuchi}, {Malanushenko}, {Malanushenko},
  {Marchante}, {McGreer}, {McMahon}, {Miralda-Escud{\'e}}, {Muna},
  {Noterdaeme}, {Oravetz}, {Palanque-Delabrouille}, {Pan}, {Perez-Fournon},
  {Pieri}, {Riffel}, {Schlegel}, {Schneider}, {Simmons}, {Viel}, {Weaver},
  {Wood-Vasey}, {Y{\`e}che}, \& {York}}]{2014A&A...563A..54P}
{P{\^a}ris}, I., {Petitjean}, P., {Aubourg}, {\'E}., {et~al.} 2014, \aap, 563,
  A54

\bibitem[{{P{\^a}ris} {et~al.}(2016){P{\^a}ris}, {Petitjean}, {Ross}, {Myers},
  {Aubourg}, {Streblyanska}, {Bailey}, {Armengaud}, {Palanque-Delabrouille},
  {Y{\`e}che}, {Hamann}, {Strauss}, {Albareti}, {Bovy}, {Bizyaev}, {Brandt},
  {Brusa}, {Buchner}, {Comparat}, {Croft}, {Dwelly}, {Fan}, {Font-Ribera},
  {Ge}, {Georgakakis}, {Hall}, {Jian}, {Kinemuchi}, {Malanushenko},
  {Malanushenko}, {McMahon}, {Menzel}, {Merloni}, {Nandra}, {Noterdaeme},
  {Oravetz}, {Pan}, {Pieri}, {Prada}, {Salvato}, {Schlegel}, {Schneider},
  {Simmons}, {Viel}, {Weinberg}, \& {Zhu}}]{2016arXiv160806483P}
{P{\^a}ris}, I., {Petitjean}, P., {Ross}, N.~P., {et~al.} 2016, ArXiv e-prints
  [\eprint[arXiv]{1608.06483}]

\bibitem[{{Peebles} \& {Yu}(1970)}]{1970ApJ...162..815P}
{Peebles}, P.~J.~E. \& {Yu}, J.~T. 1970, \apj, 162, 815

\bibitem[{{Percival} {et~al.}(2007){Percival}, {Cole}, {Eisenstein}, {Nichol},
  {Peacock}, {Pope}, \& {Szalay}}]{2007MNRAS.381.1053P}
{Percival}, W.~J., {Cole}, S., {Eisenstein}, D.~J., {et~al.} 2007, \mnras, 381,
  1053

\bibitem[{{Percival} {et~al.}(2010){Percival}, {Reid}, {Eisenstein}, {Bahcall},
  {Budavari}, {Frieman}, {Fukugita}, {Gunn}, {Ivezi{\'c}}, {Knapp}, {Kron},
  {Loveday}, {Lupton}, {McKay}, {Meiksin}, {Nichol}, {Pope}, {Schlegel},
  {Schneider}, {Spergel}, {Stoughton}, {Strauss}, {Szalay}, {Tegmark},
  {Vogeley}, {Weinberg}, {York}, \& {Zehavi}}]{2010MNRAS.401.2148P}
{Percival}, W.~J., {Reid}, B.~A., {Eisenstein}, D.~J., {et~al.} 2010, \mnras,
  401, 2148

\bibitem[{{Pieri} {et~al.}(2010){Pieri}, {Frank}, {Weinberg}, {Mathur}, \&
  {York}}]{2010ApJ...724L..69P}
{Pieri}, M.~M., {Frank}, S., {Weinberg}, D.~H., {Mathur}, S., \& {York}, D.~G.
  2010, \apjl, 724, L69

\bibitem[{{Pieri} {et~al.}(2014){Pieri}, {Mortonson}, {Frank}, {Crighton},
  {Weinberg}, {Lee}, {Noterdaeme}, {Bailey}, {Busca}, {Ge}, {Kirkby},
  {Lundgren}, {Mathur}, {P{\^a}ris}, {Palanque-Delabrouille}, {Petitjean},
  {Rich}, {Ross}, {Schneider}, \& {York}}]{2014MNRAS.441.1718P}
{Pieri}, M.~M., {Mortonson}, M.~J., {Frank}, S., {et~al.} 2014, \mnras, 441,
  1718

\bibitem[{{Planck Collaboration} {et~al.}(2014){Planck Collaboration}, {Ade},
  {Aghanim}, {Armitage-Caplan}, {Arnaud}, {Ashdown}, {Atrio-Barandela},
  {Aumont}, {Baccigalupi}, {Banday}, \& et~al.}]{2014A&A...571A..16P}
{Planck Collaboration}, {Ade}, P.~A.~R., {Aghanim}, N., {et~al.} 2014, \aap,
  571, A16

\bibitem[{{Planck Collaboration} {et~al.}(2016){Planck Collaboration}, {Ade},
  {Aghanim}, {Arnaud}, {Ashdown}, {Aumont}, {Baccigalupi}, {Banday},
  {Barreiro}, {Bartlett}, \& et~al.}]{2016A&A...594A..13P}
{Planck Collaboration}, {Ade}, P.~A.~R., {Aghanim}, N., {et~al.} 2016, \aap,
  594, A13

\bibitem[{{Pontzen}(2014)}]{2014PhRvD..89h3010P}
{Pontzen}, A. 2014, \prd, 89, 083010

\bibitem[{{Prochaska} {et~al.}(2005){Prochaska}, {Herbert-Fort}, \&
  {Wolfe}}]{2005ApJ...635..123P}
{Prochaska}, J.~X., {Herbert-Fort}, S., \& {Wolfe}, A.~M. 2005, \apj, 635, 123

\bibitem[{{Ross} {et~al.}(2015){Ross}, {Samushia}, {Howlett}, {Percival},
  {Burden}, \& {Manera}}]{2015MNRAS.449..835R}
{Ross}, A.~J., {Samushia}, L., {Howlett}, C., {et~al.} 2015, \mnras, 449, 835

\bibitem[{{Ross} {et~al.}(2012){Ross}, {Myers}, {Sheldon}, {Y{\`e}che},
  {Strauss}, {Bovy}, {Kirkpatrick}, {Richards}, {Aubourg}, {Blanton}, {Brandt},
  {Carithers}, {Croft}, {da Silva}, {Dawson}, {Eisenstein}, {Hennawi}, {Ho},
  {Hogg}, {Lee}, {Lundgren}, {McMahon}, {Miralda-Escud{\'e}},
  {Palanque-Delabrouille}, {P{\^a}ris}, {Petitjean}, {Pieri}, {Rich}, {Roe},
  {Schiminovich}, {Schlegel}, {Schneider}, {Slosar}, {Suzuki}, {Tinker},
  {Weinberg}, {Weyant}, {White}, \& {Wood-Vasey}}]{2012ApJS..199....3R}
{Ross}, N.~P., {Myers}, A.~D., {Sheldon}, E.~S., {et~al.} 2012, \apjs, 199, 3

\bibitem[{{Rudie} {et~al.}(2013){Rudie}, {Steidel}, {Shapley}, \&
  {Pettini}}]{2013ApJ...769..146R}
{Rudie}, G.~C., {Steidel}, C.~C., {Shapley}, A.~E., \& {Pettini}, M. 2013,
  \apj, 769, 146

\bibitem[{{SDSS Collaboration} {et~al.}(2016){SDSS Collaboration}, {Albareti},
  {Allende Prieto}, {Almeida}, {Anders}, {Anderson}, {Andrews},
  {Aragon-Salamanca}, {Argudo-Fernandez}, {Armengaud}, \&
  et~al.}]{2016arXiv160802013S}
{SDSS Collaboration}, {Albareti}, F.~D., {Allende Prieto}, C., {et~al.} 2016,
  ArXiv e-prints [\eprint[arXiv]{1608.02013}]

\bibitem[{{Slosar} {et~al.}(2011){Slosar}, {Font-Ribera}, {Pieri}, {Rich}, {Le
  Goff}, {Aubourg}, {Brinkmann}, {Busca}, {Carithers}, {Charlassier},
  {Cort{\^e}s}, {Croft}, {Dawson}, {Eisenstein}, {Hamilton}, {Ho}, {Lee},
  {Lupton}, {McDonald}, {Medolin}, {Muna}, {Miralda-Escud{\'e}}, {Myers},
  {Nichol}, {Palanque-Delabrouille}, {P{\^a}ris}, {Petitjean}, {Pi{\v s}kur},
  {Rollinde}, {Ross}, {Schlegel}, {Schneider}, {Sheldon}, {Weaver}, {Weinberg},
  {Yeche}, \& {York}}]{2011JCAP...09..001S}
{Slosar}, A., {Font-Ribera}, A., {Pieri}, M.~M., {et~al.} 2011, \jcap, 9, 1

\bibitem[{{Slosar} {et~al.}(2013){Slosar}, {Ir{\v s}i{\v c}}, {Kirkby},
  {Bailey}, {Busca}, {Delubac}, {Rich}, {Aubourg}, {Bautista}, {Bhardwaj},
  {Blomqvist}, {Bolton}, {Bovy}, {Brownstein}, {Carithers}, {Croft}, {Dawson},
  {Font-Ribera}, {Le Goff}, {Ho}, {Honscheid}, {Lee}, {Margala}, {McDonald},
  {Medolin}, {Miralda-Escud{\'e}}, {Myers}, {Nichol}, {Noterdaeme},
  {Palanque-Delabrouille}, {P{\^a}ris}, {Petitjean}, {Pieri}, {Pi{\v s}kur},
  {Roe}, {Ross}, {Rossi}, {Schlegel}, {Schneider}, {Suzuki}, {Sheldon},
  {Seljak}, {Viel}, {Weinberg}, \& {Y{\`e}che}}]{2013JCAP...04..026S}
{Slosar}, A., {Ir{\v s}i{\v c}}, V., {Kirkby}, D., {et~al.} 2013, \jcap, 4, 26

\bibitem[{{Smee} {et~al.}(2013){Smee}, {Gunn}, {Uomoto}, {Roe}, {Schlegel},
  {Rockosi}, {Carr}, {Leger}, {Dawson}, {Olmstead}, {Brinkmann}, {Owen},
  {Barkhouser}, {Honscheid}, {Harding}, {Long}, {Lupton}, {Loomis}, {Anderson},
  {Annis}, {Bernardi}, {Bhardwaj}, {Bizyaev}, {Bolton}, {Brewington}, {Briggs},
  {Burles}, {Burns}, {Castander}, {Connolly}, {Davenport}, {Ebelke}, {Epps},
  {Feldman}, {Friedman}, {Frieman}, {Heckman}, {Hull}, {Knapp}, {Lawrence},
  {Loveday}, {Mannery}, {Malanushenko}, {Malanushenko}, {Merrelli}, {Muna},
  {Newman}, {Nichol}, {Oravetz}, {Pan}, {Pope}, {Ricketts}, {Shelden},
  {Sandford}, {Siegmund}, {Simmons}, {Smith}, {Snedden}, {Schneider},
  {SubbaRao}, {Tremonti}, {Waddell}, \& {York}}]{2013AJ....146...32S}
{Smee}, S.~A., {Gunn}, J.~E., {Uomoto}, A., {et~al.} 2013, \aj, 146, 32

\bibitem[{{Sunyaev} \& {Zeldovich}(1970)}]{1970Ap&SS...7....3S}
{Sunyaev}, R.~A. \& {Zeldovich}, Y.~B. 1970, \apss, 7, 3

\bibitem[{{Suzuki}(2006)}]{2006ApJS..163..110S}
{Suzuki}, N. 2006, \apjs, 163, 110

\bibitem[{{Suzuki} {et~al.}(2005){Suzuki}, {Tytler}, {Kirkman}, {O'Meara}, \&
  {Lubin}}]{2005ApJ...618..592S}
{Suzuki}, N., {Tytler}, D., {Kirkman}, D., {O'Meara}, J.~M., \& {Lubin}, D.
  2005, \apj, 618, 592

\bibitem[{{Tseliakhovich} \& {Hirata}(2010)}]{2010PhRvD..82h3520T}
{Tseliakhovich}, D. \& {Hirata}, C. 2010, \prd, 82, 083520

\bibitem[{{Xu} {et~al.}(2013){Xu}, {Cuesta}, {Padmanabhan}, {Eisenstein}, \&
  {McBride}}]{2013MNRAS.431.2834X}
{Xu}, X., {Cuesta}, A.~J., {Padmanabhan}, N., {Eisenstein}, D.~J., \&
  {McBride}, C.~K. 2013, \mnras, 431, 2834

\bibitem[{{Y{\`e}che} {et~al.}(2010){Y{\`e}che}, {Petitjean}, {Rich},
  {Aubourg}, {Busca}, {Hamilton}, {Le Goff}, {Paris}, {Peirani}, {Pichon},
  {Rollinde}, \& {Vargas-Maga{\~n}a}}]{2010A&A...523A..14Y}
{Y{\`e}che}, C., {Petitjean}, P., {Rich}, J., {et~al.} 2010, \aap, 523, A14

\bibitem[{{Yoo} {et~al.}(2011){Yoo}, {Dalal}, \&
  {Seljak}}]{2011JCAP...07..018Y}
{Yoo}, J., {Dalal}, N., \& {Seljak}, U. 2011, \jcap, 7, 018

\end{thebibliography}

\end{document}